\documentclass[twocolumn,showpacs,preprintnumbers,amsmath,amssymb,pra,aps,superscriptaddress]{revtex4-1}

\usepackage{graphicx}
\usepackage{amsmath}
\usepackage{verbatim}
\usepackage{bm}
\usepackage{siunitx}
\usepackage[space]{grffile}
\usepackage{xr} 
\usepackage{multirow}
\usepackage[section]{placeins} 
\usepackage{layouts} 
\usepackage{ifthen} 
\usepackage{helvet} 
\usepackage{enumerate}  
\usepackage{float}

\usepackage{bibunits}
\newcommand{\Np}{N_{\mathrm{p}}}
\newcommand{\Npavg}{\langle N_{\mathrm{p}} \rangle}
\newcommand{\MT}{\mathrm{M}_{\mathrm{T}}}
\newcommand{\MB}{\mathrm{M}_{\mathrm{B}}}

\newcommand{\DM}{\mathrm{D}_{\mathrm{M}}}

\newcommand{\DT}{\mathrm{D}_{\mathrm{T}}}
\newcommand{\DB}{\mathrm{D}_{\mathrm{B}}}

\newcommand{\be}{\begin{equation}}
\newcommand{\ee}{\end{equation}}
\newcommand{\bea}{\begin{eqnarray}}
\newcommand{\eea}{\end{eqnarray}}

\newcommand{\ket}[1]{\left\lvert #1 \right\rangle}

\newcommand{\K}{\mathrm{K}}
\newcommand{\mK}{\mathrm{mK}}
\newcommand{\kHz}{\mathrm{kHz}}
\newcommand{\MHz}{\mathrm{MHz}}
\newcommand{\GHz}{\mathrm{GHz}}
\newcommand{\us}{\mu\mathrm{s}}
\newcommand{\ns}{\mathrm{ns}}

\newcommand{\mm}{\mathrm{mm}}

\newcommand{\nm}{\mathrm{nm}}

\newcommand{\dB}{\mathrm{dB}}

\newcommand{\QM}{\mathrm{Q}_{\mathrm{M}}}

\newcommand{\QT}{\mathrm{Q}_{\mathrm{T}}}

\newcommand{\QA}{\mathrm{Q}_{\mathrm{A}}}
\newcommand{\QB}{\mathrm{Q}_{\mathrm{B}}}

\newcommand{\Tone}{T_{1}}
\newcommand{\Ttwostar}{T_{2}^{\ast}}
\newcommand{\Ttwoecho}{T_{2}^{\mathrm{E}}}

\newcommand{\Hethree}{^3\mathrm{He}}
\newcommand{\Hefour}{^4\mathrm{He}}

\newcommand{\Rmnum}[1]{\expandafter\@slowromancap\romannumeral #1@}

\newcommand{\figlabel}[1]{{\usefont{T1}{phv}{b}{n}\footnotesize(#1)}}

\defaultbibliographystyle{apsrev4-1}
\defaultbibliography{Asaad_References}

\begin{document}

\title{Independent, extensible control of same-frequency superconducting qubits by selective broadcasting}
\author{S.~Asaad}
\author{C.~Dickel}
\author{S.~Poletto}
\author{A.~Bruno}
\author{N.~K.~Langford}
\author{M.~A.~Rol}
\affiliation{QuTech, Delft University of Technology, P.O. Box 5046, 2600 GA Delft, The Netherlands}
\affiliation{Kavli Institute of Nanoscience, Delft University of Technology, P.O. Box 5046, 2600 GA Delft, The Netherlands}
\author{D.~Deurloo}
\affiliation{QuTech, Delft University of Technology, P.O. Box 5046, 2600 GA Delft, The Netherlands}
\affiliation{Netherlands Organisation for Applied Scientific Research (TNO), P.O. Box 155, 2600 AD Delft, Netherlands}
\author{L.~DiCarlo}
\affiliation{QuTech, Delft University of Technology, P.O. Box 5046, 2600 GA Delft, The Netherlands}
\affiliation{Kavli Institute of Nanoscience, Delft University of Technology, P.O. Box 5046, 2600 GA Delft, The Netherlands}

\date{\today}

\begin{abstract}
A critical ingredient for realizing large-scale quantum information processors will be the ability to make economical use of qubit control hardware.
We demonstrate an extensible strategy for reusing control hardware on same-frequency transmon qubits in a circuit QED chip with surface-code-compatible connectivity.
A vector switch matrix enables selective broadcasting of input pulses to multiple transmons with individual tailoring of pulse quadratures for each, as required to minimize the effects of leakage on weakly anharmonic qubits.
Using randomized benchmarking, we compare multiple broadcasting strategies that each pass the surface-code error threshold for single-qubit gates.
In particular, we introduce a selective-broadcasting control strategy using five pulse primitives, which allows independent, simultaneous Clifford gates on arbitrary numbers of qubits.
\end{abstract}

\maketitle

\begin{bibunit}

Building a fault-tolerant quantum computer requires the ability to efficiently address and control individual qubits in a large-scale system.
Many leading experimental quantum information platforms, among them trapped ions~\cite{Monroe13}, electronic spins in impurities and quantum dots~\cite{Awschalom13} and superconducting circuits~\cite{Devoret13}, employ qubits with level transitions in the microwave frequency domain.
Addressing these transitions often involves expensive microwave electronics scaling linearly with the number of qubits.
To move beyond the state of the art in microwave-frequency quantum processors, such as those recently used for small-scale quantum error correction in superconducting circuits~\cite{Kelly15,Corcoles15,Riste15}, it will already be beneficial to have a hardware-efficient control strategy that harnesses economies of scale.
One approach is to use microwave pulses from a single control source for multiple qubits~\cite{Hornibrook15}, requiring frequency-matched qubits and high-speed routing of pulses to separate control lines.
The linear scaling of control equipment could then be reduced to a constant overhead for the most expensive resources.

Using control equipment for multiple qubits has previously been demonstrated for optical addressing in atomic systems, where qubits naturally have the same frequency~\cite{Knoernschild10,Weitenberg11,Crain14,Xia15}.
Such frequency reuse also becomes possible in circuit quantum electrodynamics (cQED)~\cite{Blais04} in the context of fault-tolerant computation strategies~\cite{Divincenzo09,Helmer09cavity,Ghosh12} which rely only on local interactions between qubits mediated by bus resonators.
The natural isolation between different lattice sites allows the use of repeating patterns of qubit frequencies with selectivity provided by spatial separation.
A tileable unit cell with a handful of qubit frequencies~\cite{Gambetta2014frequency} could therefore provide a promising route towards scalability.
Crucially, this also solves the frequency-crowding problem which arises when trying to fit many distinct-frequency qubits within the finite useful bandwith of the circuit-based devices, particularly for designs based on weakly anharmonic qubits where higher levels must also be avoided~\cite{Schutjens13,Vesterinen14}.
While no qubit experiments have yet shown the viability of this approach, Hornibrook \emph{et al.}\ have recently demonstrated a cryogenic switching matrix for pulse distribution operating at $20 ~\mK$, triggered by a field-programmable gate array at $4~\K$~\cite{Hornibrook15}.
Cryogenic control equipment may shorten feedback latency and reduce wiring complexity across temperature stages, but the isolation and operational frequency range achieved are currently insufficient for typical cQED experiments.

In this Letter, we demonstrate frequency reuse in an extensible solid-state multiqubit architecture.
Specifically, we show independent simultaneous control of two same-frequency qubits with a home-built room-temperature vector switch matrix (VSM).
The VSM allows tailoring of control pulses to individual qubit properties, and routing of the pulses to either one or both of the qubits using fast digital markers.
We develop several different approaches to selective pulse broadcasting, including a simple scheme for implementing independent Clifford control on an arbitrary number of qubits with a constant overhead in time.
The device for this experiment is designed to allow testing in a circuit with the correct connectivity of a relevant surface-code lattice~\cite{Bravyi98, Fowler12}.
Using randomized benchmarking (RB), we show that all control schemes exceed the fidelity threshold for surface code and are dominated by qubit relaxation.
We also develop a method for measuring leakage to the second excited state directly within the context of RB~\cite{chasseur15,Epstein14}.
We characterize the limitations of our system and find no major obstacles to scaling up to larger implementations.

\begin{figure}
  \centering
  \includegraphics[width=\linewidth]{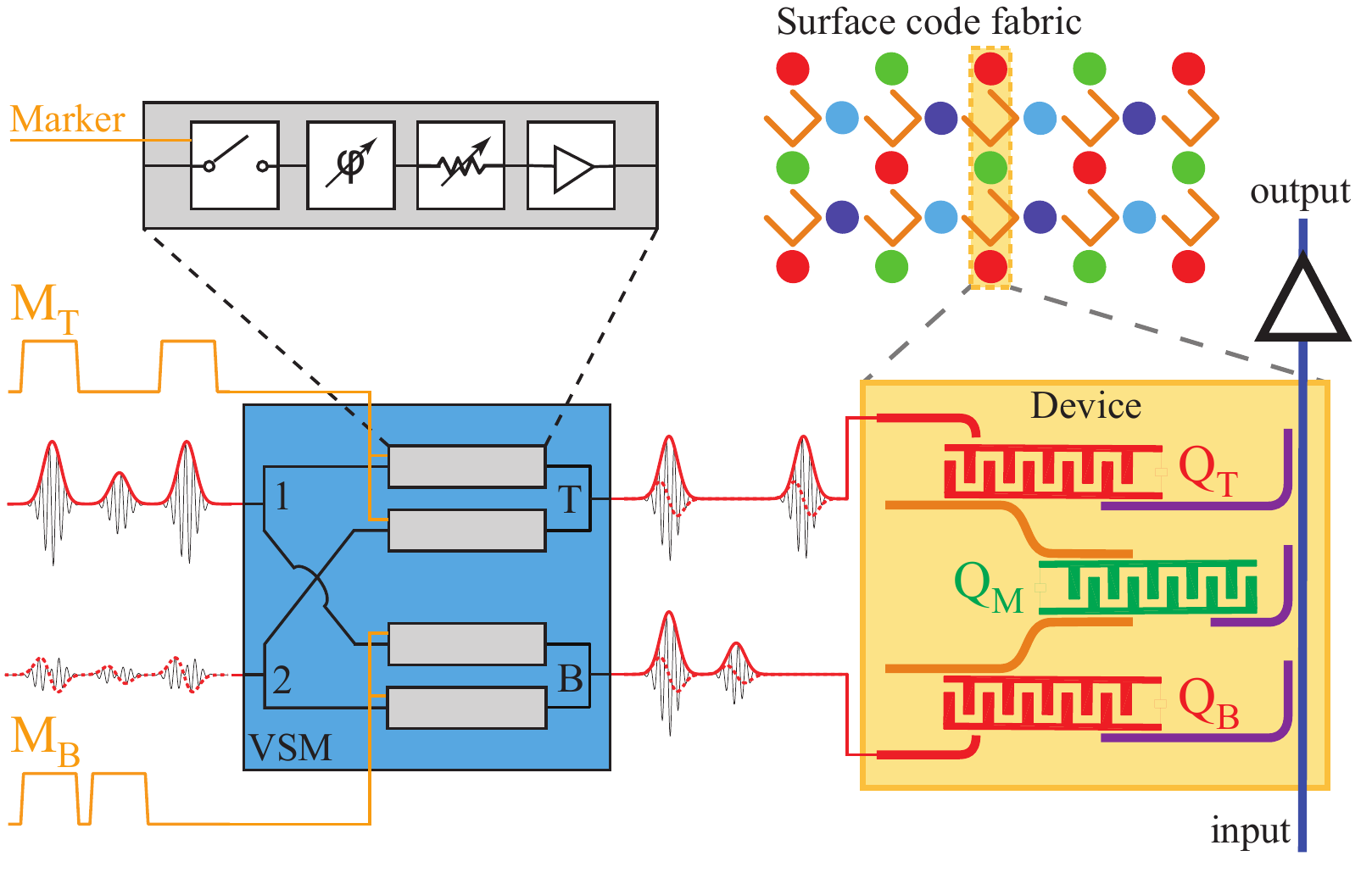}
  \caption{(color online).
    Schematic of independent control of same-frequency qubits using a vector switch matrix (VSM).
    The device (gold-colored box on right) connects two transmons with matched frequency ($\QT$ and $\QB$: $6.220~\GHz$) indirectly through coupling buses and a third non-matched transmon ($\QM$: $6.550~\GHz$).
    This provides the smallest relevant subunit of the four-frequency surface-code fabric illustrated above right.
    The VSM (blue box) allows independent, simultaneous transmon control with tailored DRAG pulsing by combining and directing Gaussian and derivative-of-Gaussian input pulses to separate outputs T and B connected to dedicated drive lines for each qubit.
    The link between inputs and outputs can be switched on nanosecond timescales using the digital marker inputs $\MT$ and $\MB$ (orange lines).
    DRAG pulses for the targeted qubit are independently tuned in both amplitude and phase for each input-output pair (top left).
     }
  \label{fig:VSM-qubits-schematic}
\end{figure}

Demonstrating frequency reuse in a context relevant for increasing system sizes requires two key elements: a method for distributing control pulses to multiple qubits using a single qubit-control source, and a multiqubit device containing same-frequency qubits with relevant connectivity.
Here, we focus on a particular implementation of the surface code, where the connectivity between nearest-neighbor qubits is achieved via bus resonators~\cite{Divincenzo09}.
Figure~\ref{fig:VSM-qubits-schematic} illustrates a conceptual design based on repeated tiling of a unit cell consisting of four qubits with unique frequencies that are coupled via bus resonators.

Our device contains a small block of this design, consisting of two same-frequency transmon qubits~($\QT$ and $\QB$), which are connected to a third qubit~($\QM$) via separate bus resonators (Fig.~\ref{fig:VSM-qubits-schematic}).
Each qubit has a capacitively-coupled drive line for individual qubit control~\cite{Fragner08}, a readout resonator coupled to a common feedline for frequency-division multiplexing readout~\cite{Groen13,Jerger12}, and a flux-bias line for individual frequency tuning~(Fig.~\ref{fig:VSM-qubits-schematic}).
While $\QT$ and $\QB$ were designed to be identical, fabrication uncertainties resulted in a sweet-spot (maximum) frequency of $\QT$ \SI{57}{\mega \hertz} higher than that of $\QB$.
With $\QB$ and $\QM$ kept at their respective sweet-spots (\SI{6.220}{\giga \hertz} and \SI{6.550}{\giga \hertz}, respectively), $\QT$ was then flux tuned to match $\QB$ with an accuracy of \SI{50}{\kilo \hertz}, determined using Ramsey measurements.
The coherence times at the operating frequency can be found in the Supplemental Material~\cite{SOMprappl}.
Because of the transmon's weak anharmonicity~\cite{Koch07}, high-fidelity fast single-qubit control is achieved using the method of derivative-removal-via-adiabatic-gate (DRAG) pulsing, where the in-phase Gaussian pulse is combined with an in-quadrature derivative-of-Gaussian pulse~\cite{Motzoi09,Chow10b}.
For each qubit, this requires independent amplitude control of the two constituent quadrature pulses.

The VSM was designed to accept multiple input pulses and selectively fan them out to multiple qubits with individual pulse tuning for each qubit (Fig.~\ref{fig:VSM-qubits-schematic}).
Our home-built room-temperature $4 \times 2$ (four input, two output) VSM allows independent control of amplitude and phase for each of its input-output combinations.
Fast marker-controlled digital switches enable routing of pulses to the qubits at nanosecond timescale, with approximately \SI{50}{\decibel} isolation in the frequency range from \SIrange[range-units = single]{4}{8}{\giga \hertz} (see~\cite{SOMprappl} for VSM specifications).
By directing the two consistuent pulses of DRAG control through separate inputs of the VSM, this allows independent, in-situ DRAG tuning for both same-frequency qubits using four AWG channels~\cite{SOMprappl}.

The first critical test of our control architecture is to assess the VSM's ability to implement high-precision control of one qubit while leaving the other qubit idle.
To do this, we use the standard technique of single-qubit RB based on Clifford gates~\cite{Knill08,Magesan11,Magesan12}, which allows us to characterize control performance independently of state preparation and measurement errors.
After initializing all qubits in the ground state by relaxation, we use the VSM to selectively apply random sequences of Cliffords gates to only one of the same-frequency qubits, with lengths $m$ ranging from 1 to 800 gates and the results averaged for 50 different random sequences.
We decompose each gate into the standard minimal sequence of $\pi$ and $\pm\pi/2$ pulses around the $x$ and $y$ axes (16~$\ns$ pulses separated by a 4~$\ns$ buffer; $t_p = 20~\ns$)~\cite{Epstein14}, requiring on average $\left< N_\mathrm{p} \right> = 1.875$ pulses per Clifford.
This is in contrast to atomic pulses, where the 24 single-qubit Cliffords can each be implemented with a single pulse~\cite{Johnson15}.
After applying a final Clifford that inverts the cumulative effect of all $m$ previous Cliffords, the driven qubit is ideally returned to the ground state, but as a result of imperfections such as gate errors and decoherence, the final ground-state population decays as a function of $m$.
The decay rate can be related to the average fidelity per Clifford $F_\mathrm{C}$~\cite{Knill08,Magesan11}.
In a strictly two-level system, the measured ground- and excited-state populations averaged over many sequences ($\langle P_0\rangle$ and $\langle P_1\rangle$) both converge to 0.5 for large $m$.
For weakly anharmonic transmon qubits,  leakage to the second-excited state can be an important additional source of gate error, which can lead to a shift of the asymptotic value away from 0.5.
We address this issue by performing the RB protocol both with and without an additional final $\pi$ pulse~\cite{Riste12b}, which allows us to explicitly estimate the populations of the first three transmon states (see~\cite{SOMprappl} for details).

\begin{figure}
  \begin{minipage}[t]{.5\linewidth}
    \flushright
    \includegraphics[width=.9\linewidth]{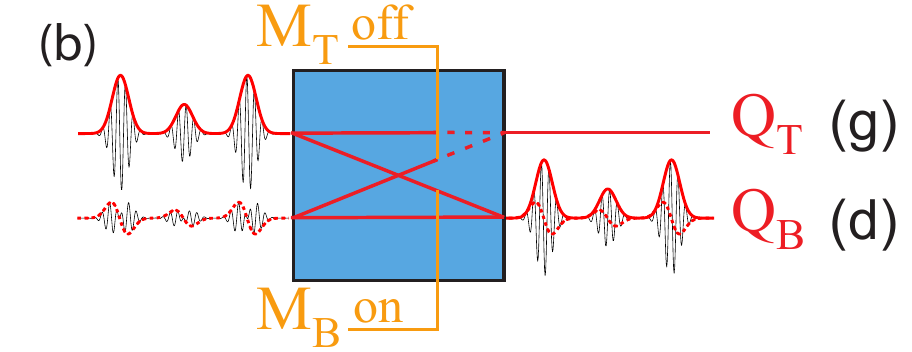}
  \end{minipage}
  \begin{minipage}[t]{.45\linewidth}
    \includegraphics[width=\linewidth]{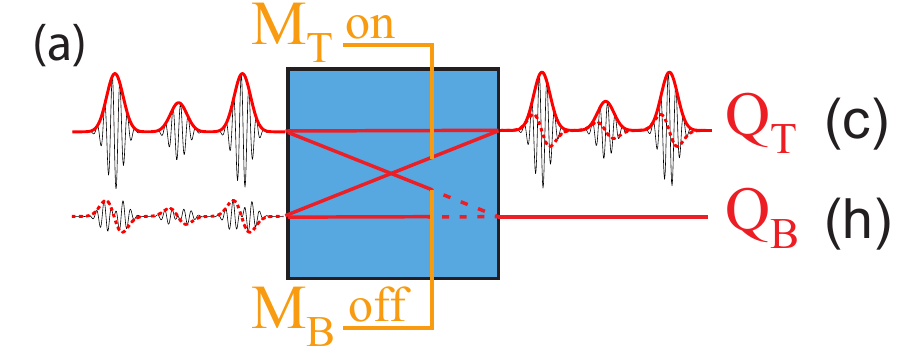}
  \end{minipage}
  \centering
  \includegraphics[width=\linewidth]{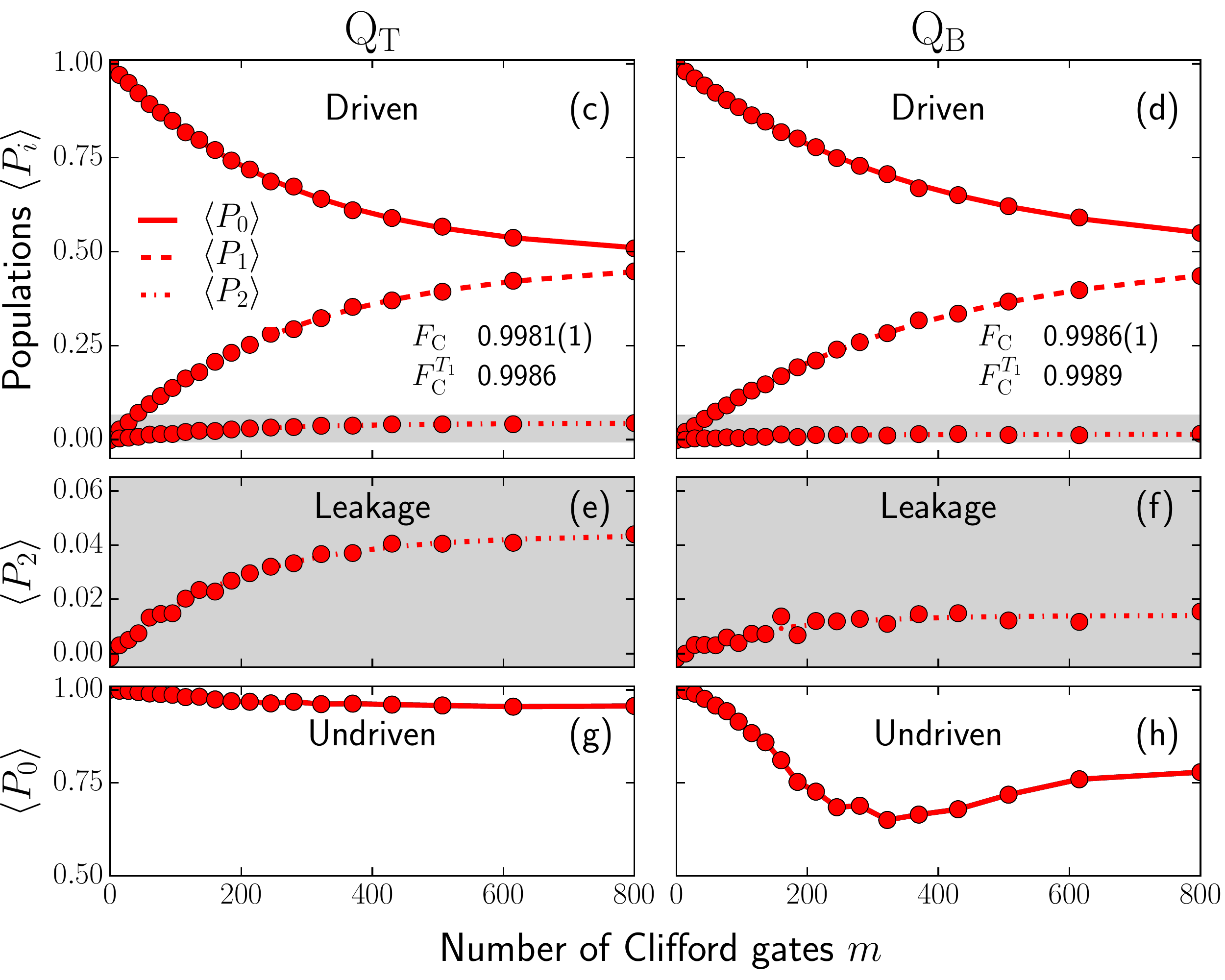}
  \caption{(color online).
  Single-qubit control of same-frequency qubits using the VSM.
  (a,b) Schematics showing DRAG pulses routed exclusively to either $\QT$ or $\QB$ using the corresponding markers (always on or off).
Details of the tune-up procedure for optimizing pulse amplitude and phase parameters for each qubit are provided in Ref.~\onlinecite{SOMprappl}.
  (c,d) Characterization of single-qubit control by randomized benchmarking (RB) of Clifford gates.
  Average populations of $\QT$ and $\QB$ in the ground, first- and second-excited states ($\langle P_0\rangle$, $\langle P_1\rangle$ and $\langle P_2\rangle$,  respectively) as a function of the number of Clifford gates applied.
  Curves are best fits of single exponentials with offsets.
  Single-qubit Clifford fidelities for each qubit are extracted from the decay of ground-state populations.
  For both qubits, the fidelity surpasses the surface-code fault-tolerance threshold and lies close to the theoretical value expected for $\Tone$-only relaxation-limited performance.
  (e,f) Expanded plots of second-excited state leakage during RB.
  Curves are best fits according to Eq.~(\ref{eq:second-excited state population equation}).
  (g,h) Cross-excitation of the undriven qubit resulting from control pulses applied to the driven qubit.
  Further measurements indicate that this effect can be attributed to microwave leakage.}
  \label{fig:RB 1Q populations}
\end{figure}

We implement the above characterization for one qubit at a time by switching off the marker for the undriven qubit at the VSM~[Fig.~\ref{fig:RB 1Q populations}(a,b)], in each case measuring the effect on both qubits simultaneously via multiplexed readout.
From the results in Fig.~\ref{fig:RB 1Q populations}(c,d), we calculate the average Clifford fidelities for the two individually-driven qubits to be $0.9982(2)$ ($\QT$) and $0.9986(2)$ ($\QB$), in both cases surpassing the best known surface-code fault-tolerance threshold for single-qubit gates of $\sim0.99$~\cite{Raussendorf07,Fowler09,Wang11}.
We compare these values with the expected average Clifford fidelities assuming only $\Tone$ decay~\cite{MagesanNote}:
\begin{equation}
  F_\mathrm{C}^{\Tone} \simeq \left[\frac{1}{6}\left(3 + 2 e^{-t_\mathrm{p}/2 \Tone} + e^{-t_\mathrm{p}/\Tone}\right)\right]^{\left< N_\mathrm{p} \right>}.
  \label{eq:RB T1 fidelity limit}
\end{equation}
This shows our results are predominately limited by relaxation effects, the difference in performance being consistent with the different $\Tone$ times for the two qubits.
Additional measurements in the Supplemental Material furthermore show that there is no difference in performance when both qubits are driven simultaneously by the same pulse sequence (both markers on).
From the measured leakage populations $\langle P_2 \rangle$, we also extract estimated per Clifford leakage rates $\kappa$ of $4.1(2) \times 10^{-6}$ ($\QT$) and $1.3(4) \times 10^{-6}$ ($\QB$) by fitting the following simple model to the data (see \cite{SOMprappl} for details):
\begin{equation}
  \langle P_2[m] \rangle \simeq \kappa \, T_{2\rightarrow1} \left( 1 - e^{-m \langle N_\mathrm{p} \rangle t_\mathrm{p} / T_{2\rightarrow1}} \right),
  \label{eq:second-excited state population equation}
\end{equation}
where $T_{2\rightarrow1}$ is the second- to first-excited-state relaxation time.
As these leakage rates are much smaller than the gate errors ($1 {-} F_\mathrm{C}$), it is reasonable to neglect them when estimating the Clifford fidelity.

We next explore the effect of the single-qubit control pulses on the undriven qubit~[Fig.~\ref{fig:RB 1Q populations}(g,h)], which should ideally remain in the ground state.
We fabricated the qubits as close to each other as possible in order to study the worst-case scenario for such cross-excitation effects.
While $\QT$ remains largely unaffected when driving $\QB$, a substantial deviation from the ground state is measured in $\QB$ when driving $\QT$.
There are several possible mechanisms for cross-excitation effects in our system: cross-coupling (higher-order quantum coupling mediated by the bus resonators and $\QM$) and cross-driving (spurious driving of the idle channel by microwave leakage resulting from imperfect isolation either in the VSM or on chip).
From single-excitation swap experiments (see Ref.~\onlinecite{SOMprappl}), we observe a residual exchange interaction~\cite{Blais04} between $\QT$ and $\QB$ with strength $J/2\pi \leq 36\pm1~\kHz$.
This symmetric swapping of excitation is therefore unlikely to explain the strong asymmetry in the amount of cross-excitation measured for the different qubits.
Furthermore, in RB experiments, where the state of the driven qubit is moved randomly around the Bloch sphere during each run of the experiment, we expect the effect of cross-coupling to be dramatically reduced, because the period $\pi/J$ is far longer than the average Clifford gate time.
Direct, independent measurements of microwave isolation in both the VSM and on chip indicate that the on-chip cross-driving significantly dominates~\cite{SOMprappl}.
This was also confirmed in-situ by performing the same RB measurements with the undriven qubit physically disconnected from the VSM.
This showed no significant reduction in cross-excitation.
Furthermore, numerical simulations show that the observed effects are consistent with cross-driving alone at levels similar to the measured values~\cite{SOMprappl}.
We note, however, that while the plots in Fig.~\ref{fig:RB 1Q populations}(g,h) are useful diagnostics for identifying the presence of the cross-driving effect, they should not be interpreted in the same way as the RB plots for the driven qubit.
A more comparable way to study this effect is to use interleaved RB~\cite{Magesan12b}, where pulses on the driven qubit (ideally identity operations on the undriven qubit) are interleaved with a random sequence of Cliffords applied to the undriven qubit~\cite{SOMprappl}.
From this, we estimate the average idling fidelity for $\QB$ to be 0.9986(5), which is consistent with the error due to $\Tone$ decay during idling.
This confirms that cross-excitation effects do not dominate the error per Clifford, as characterized by RB.

\begin{figure}
  \centering
  \includegraphics[width=\linewidth]{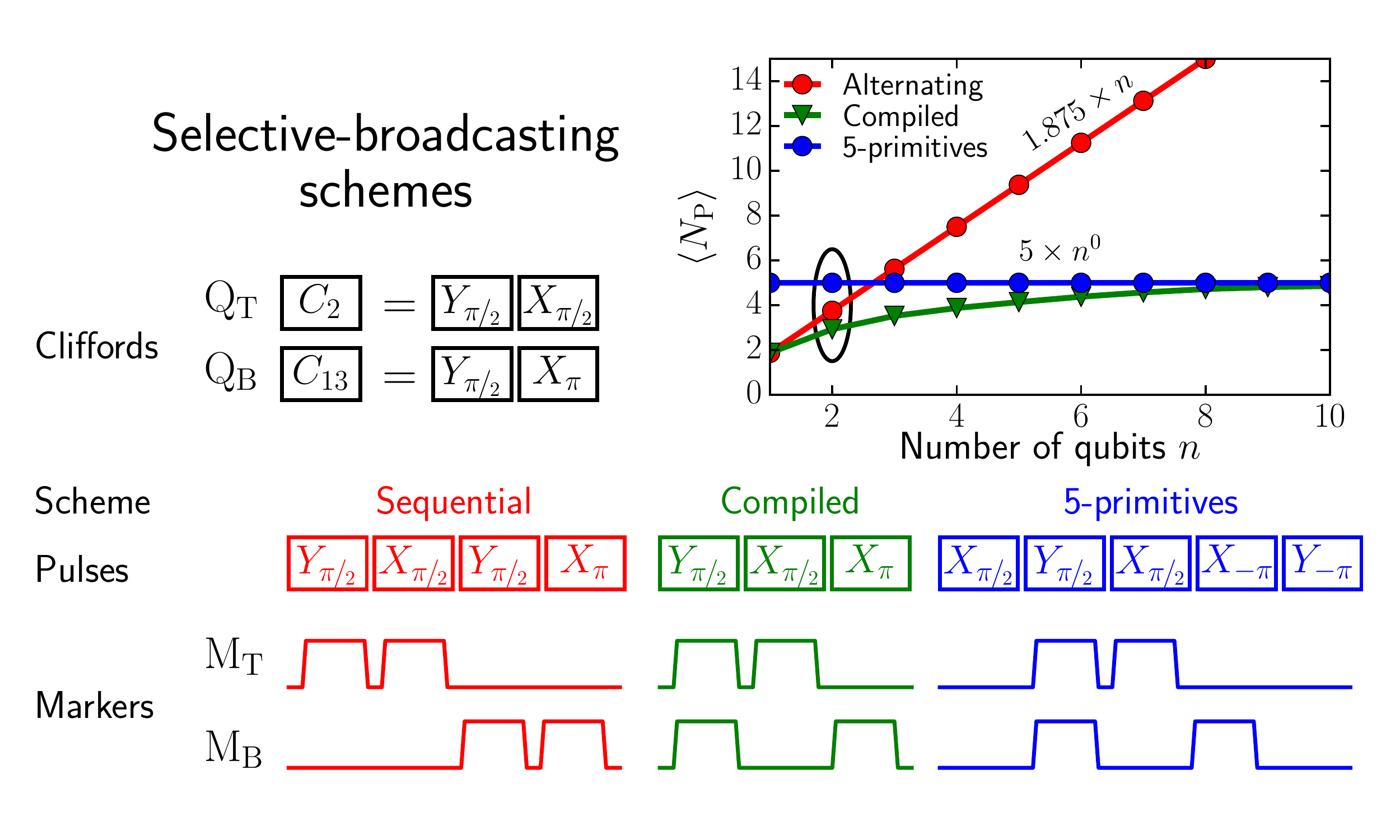}
  \caption{(color online).
   Selective broadcasting schemes for simultaneous single-qubit control of multiple qubits.
   Main figure: Example of a single Clifford round for $n{=}2$ qubits targeting $C_2$ ($C_{13}$) in $\QT$ ($\QB$) ($C_j$ defined in Ref.~\onlinecite{SOMprappl}).
  In the sequential scheme, the pulses implementing $C_2$ are directed to $\QT$, after which the pulses implementing $C_{13}$ are directed to $\QB$.
  In the compiled scheme, the two Cliffords are realized concurrently using a pre-determined pulse sequence, with appropriate markers, which minimizes the total number of pulses, $N_\mathrm{P}$ (see~\cite{SOMprappl} for the compilation algorithm).
  Finally, in the 5-primitives scheme, a fixed sequence of five pulses is repeated in each round ($\Np{=}5$).
  The targeted Cliffords are then applied simultaneously by selecting the appropriate subset of pulses for each qubit (see~\cite{SOMprappl} for the 5-primitives marker table).
  Top-right: Scaling of the average pulses per multiqubit combination of Cliffords, $\Npavg$, versus qubit number $n$.
  The constant scaling achieved by the 5-primitives scheme provides a dramatic improvement over the linear scaling of the sequential scheme.
  While $\Npavg$ is always lowest for the compiled scheme, pre-compiling the optimal pulse and marker combinations is impractical for $n \gtrsim 5$, and the improvement over the simpler 5-primitives scheme is negligible by $n \sim 10$.}
  \label{fig:RB schematic}
\end{figure}

The defining test of extensibility in our control architecture is to demonstrate the simultaneous, independent, single-qubit control over same-frequency qubits that is enabled by selective broadcasting using the VSM.
We explore three paradigmatic schemes for implementing selective broadcasting of Cliffords on an arbitrary number of qubits $n$ (Fig.~\ref{fig:RB schematic}).
In the most straightforward selective-broadcasting scheme, the individual qubits are driven sequentially, with each pulse being directed to one qubit at a time.
This results in a linear scaling of the average number of pulses per Clifford round $\left(\Npavg=1.875 \times n\right)$.
By contrast, the second paradigm takes best advantage of the VSM's capability to broadcast simultaneously to multiple qubits by compiling the constituent Clifford pulses to minimize $N_\mathrm{p}$ for each Clifford combination in the sequence.
The compilation is performed by searching all possible combinations of single-qubit Clifford decompositions and finding the one that minimizes $N_\mathrm{p}$ (see~\cite{SOMprappl} for further information).
In Fig.~\ref{fig:RB schematic}, we show exact values and estimates for $\Npavg$ for the compiled scheme up to $n=10$.
Unfortunately, the compilation run-time increases exponentially with the number of qubits.
This motivates our final broadcasting paradigm, where all Clifford gates can be implemented using the same fixed, ordered sequence of five pulse primitives~(Fig.~\ref{fig:RB schematic}).
Independent Cliffords can be applied to all qubits, irrespective of $n$, by selectively directing the appropriate subset of pulses to each qubit, achieving a constant overhead in time for single-qubit Clifford control.
While the compiled scheme by definition always provides the minimum sequence length, our estimates of $\Npavg$ for the compiled scheme suggest that it asymptotes to the same value of 5 achieved by the simple, prescriptive 5-primitives scheme.
We note that the specific choice of pulse primitives is not unique, but at least five primitives are required (four pulses allow a maximum of 16 unique gate decompositions, compared with the 24 single-qubit Cliffords).

\begin{figure}
  \centering
  \includegraphics[width=\linewidth]{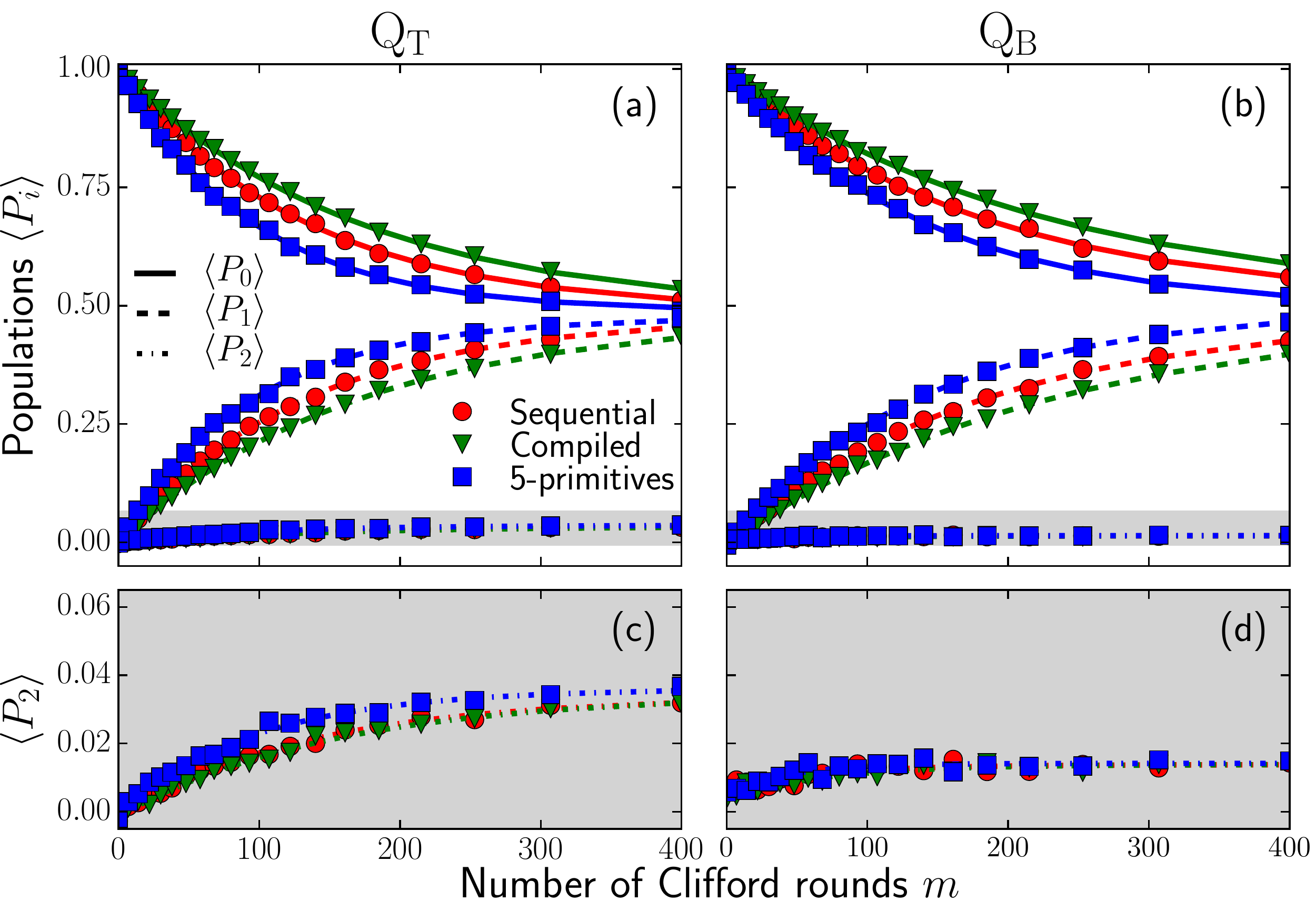}
    \vspace{-6mm}
        \flushleft
      {\usefont{T1}{phv}{m}{n}\scriptsize(e)}
      \vspace{1mm}
    \begin{ruledtabular}
      \begin{tabular}{l | c c | c c}
        \multirow{2}{*}{Scheme} & \multicolumn{2}{c|}{$\QT$} & \multicolumn{2}{c}{$\QB$} \\
        & $F_\mathrm{C}$ & $F_\mathrm{C}^{T_1}$ & $F_\mathrm{C}$ & $F_\mathrm{C}^{T_1}$ \\
        \hline
        Sequential  & 0.9962(4) & 0.9971 & 0.9972(5) & 0.9978\\
        Compiled     & 0.9971(4) & 0.9978 & 0.9978(5) & 0.9982\\
        5-primitives & 0.9947(5) & 0.9962 & 0.9964(5) & 0.9970\\
      \end{tabular}
    \end{ruledtabular}
  \caption{(color online).
   Characterization of sequential, compiled and 5-primitives selective-broadcasting schemes by RB.
  (a,b) Evolution of the average transmon populations for $\QT$ (a) and $\QB$ (b) as a function of the number of Clifford rounds.
  Curves are the best fits of single exponentials with offsets.
  (c,d) Expanded plots of second-excited-state leakage during RB.
  (e) Average single-qubit Clifford-gate fidelities for $\QT$ and $\QB$ in each scheme, extracted from the decay of the corresponding ground-state population.
  All fidelities surpass the surface-code fault-tolerance threshold and closely track those expected for $\Tone$-relaxation-limited performance.
  Compiled selective broadcasting performs best, as expected, having the lowest total number of pulses (see Fig.~\ref{fig:RB schematic}).}
  \label{fig:RB 2Q populations}
\end{figure}

To demonstrate the full functionality of this control architecture, we implement all three selective-broadcasting schemes and measure their performance using parallel single-qubit RB with independent Clifford sequences for each qubit.
Figure~\ref{fig:RB 2Q populations} shows that the compiled scheme performs best, followed by the sequential and then 5-primitives schemes, consistent with the average number of pulses required for each (Fig.~\ref{fig:RB schematic}).
In all cases, the average fidelity per Clifford still surpasses the surface-code fault-tolerance threshold, and the average error is again dominated by relaxation.
The results are completely consistent with the values obtained in the test for isolated single-qubit control, indicating no substantial decrease in gate performance using selective broadcasting schemes.

Our VSM allows efficient reuse of qubit control equipment on same-frequency qubits.
It enables high-precision single-qubit control of multiple qubits with a performance that is mainly limited by relaxation.
We have demonstrated three selective broadcasting schemes, all of which achieve a performance that surpasses the fault-tolerance threshold for the surface code.
In particular, the 5-primitives scheme implements arbitrary Clifford control with a fixed five-pulse sequence, where the target Clifford is selected by routing a subset of the pulses using digital markers.
By adding a sixth, non-Clifford gate to the five pulse primitives, this can be extended to achieve universal single-qubit control.
Combining the connectivity of our device, the VSM-based control, and the fixed pulse overhead of the 5-primitives broadcasting strategy, our experiment realizes the simplest element of an extensible qubit control architecture.
While we do not yet see explicit savings in control hardware for two qubits, this design can be expanded to more same-frequency qubits without any further increase in microwave sources or arbitrary waveform generators.
This experiment suggests that surface-code tiling with frequency reuse is a viable path towards large-scale quantum processors.

\begin{acknowledgments}
We thank R.~N.~Schouten, W.~Vlothuizen, and P.~Koobs de Hartog for experimental contributions and B.~Criger, T.~Chasseur, and D.~J.~Reilly for discussions.
We acknowledge funding by the Dutch Organization for Fundamental Research on Matter (FOM), the Netherlands Organisation for Scientific Research (NWO/OCW and Vidi scheme), the EU FP7 project ScaleQIT, and an ERC Synergy Grant.
\end{acknowledgments}

\putbib
\end{bibunit}

\clearpage
%\onecolumngrid

\begin{bibunit}
\renewcommand{\theequation}{S\arabic{equation}}
\renewcommand{\thefigure}{S\arabic{figure}}
\renewcommand{\thetable}{S\arabic{table}}
\setcounter{figure}{0}
\setcounter{equation}{0}
\setcounter{table}{0}

\section*{Supplemental Material}

 This supplement provides experimental details and additional data supporting the claims in the main text.
 The device and experimental setup, including images of the device and a full wiring diagram, are described in Section~\ref{sec:chip-setup}.
 The microwave performance of the vector switch matrix (VSM) and its use for independent control of two same-frequency qubits are demonstrated in Sec.~\ref{sec:Vector switch matrix isolation}.
 The techniques employed for tuning qubit pulses are discussed in Sec.~\ref{sec:Pulse-calibration routines}.
 Section~\ref{sec:global-broadcasting} contains the results of randomized benchmarking (RB) experiments in the two-qubit global broadcasting context.
 Our technique for assessing the effects of leakage is detailed in Sec.~\ref{sec:second-excited-state-leakage}.
 Cross-coupling and cross-driving effects are characterized in Sec.~\ref{sec:Cross-coupling and cross-driving effects}, along with numerical simulations of the effects of cross-excitations on RB.
 In Section~\ref{ssec:robust-pulse-sequences}, we introduce a method for generating selective-broadcasting pulse sequences that are robust to cross-excitation effects.
 The decompositions of the 24 single-qubit Cliffords into a minimal set of pulses and the 5-primitive pulses are given in Sec.~\ref{sec:Clifford pulse decomposition}.
 Finally, the algorithm used to compile optimal pulse sequences for implementing independent single-qubit Clifford gates on multiple qubits is explained in Sec.~\ref{sec:Compiled selective broadcasting algorithm}.

\section{Quantum chip and experimental setup}
\label{sec:chip-setup}

\subsection{Chip design and fabrication}

   \begin{figure}[t]
      \centering
      \includegraphics[width=\linewidth]{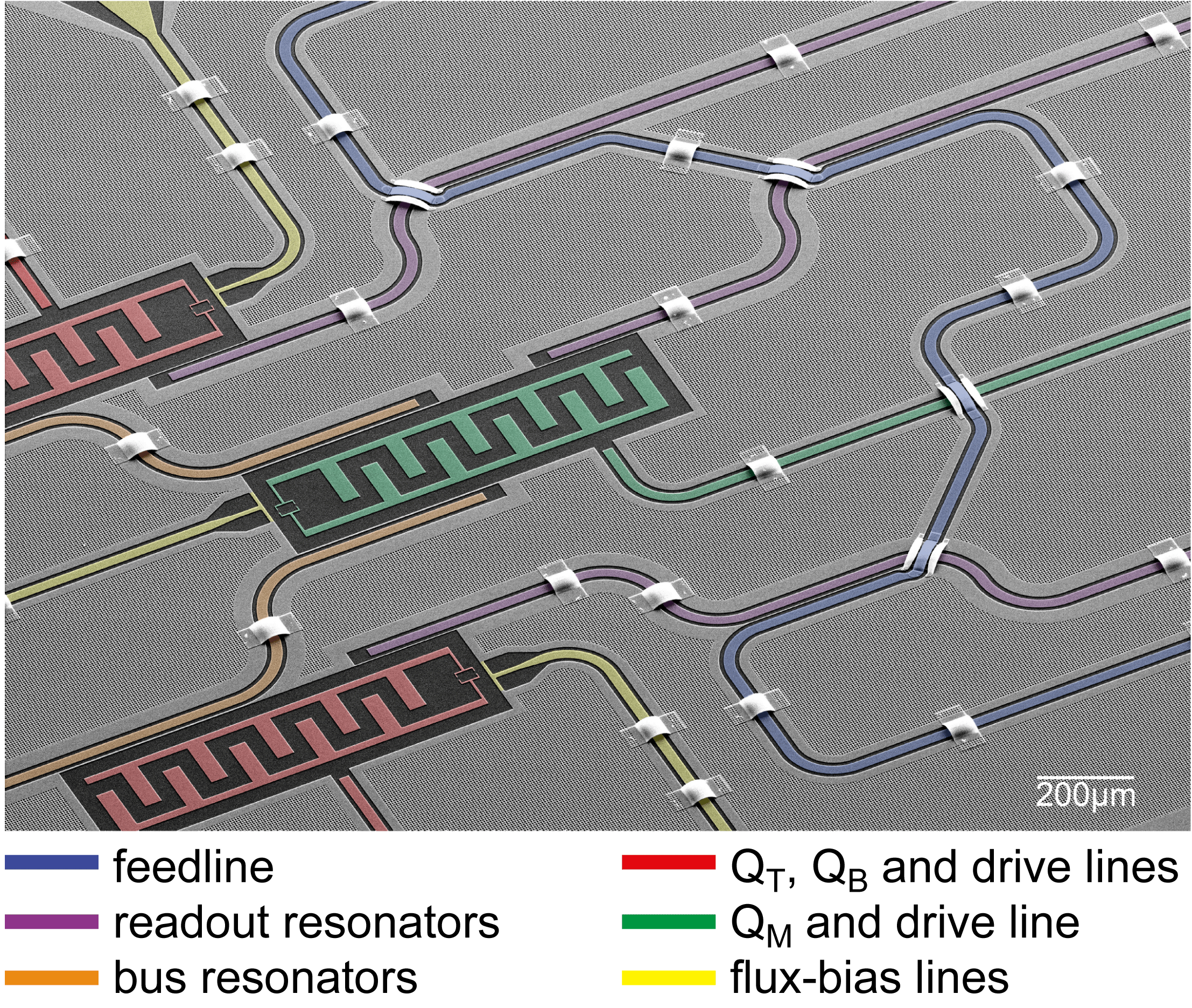}
      \caption{Scanning electron micrograph (with added false colour) of a device twin to the one used in the experiment and fabricated in the same batch. Notably, the feedline crosses the three readout resonators.}
      \label{fig:Muxmon SEM}
  \end{figure}
  \label{ssec:Device}

Our quantum chip consists of three transmons (top:~$\QT$, middle (ancilla): $\QM$, and bottom: $\QB$) with dedicated voltage drive lines ($\DT$, $\DM$ and $\DB$, respectively), flux-bias lines, and readout resonators. All readout resonators are capacitively coupled to one common feedline which crosses various on-chip components using airbridge crossovers. Qubits $\QT$ and $\QM$ are coupled by one bus resonator, and $\QM$ and $\QB$ by another (fundamental frequencies $4.9$ and $5.0~\GHz$, respectively). All resonators are open-ended on the coupling side, and short-circuited at the other.

The chip fabrication method is similar to that in Ref.~\onlinecite{Riste15}, but with some important differences which we now explain.
Rather than sapphire, we use a high-resistivity intrinsic silicon substrate prepared by HF dip and HMDS surface passivation before sputtering a 300-$\nm$-thick film of NbTiN, as introduced in Ref.~\onlinecite{Bruno15}.
This change aims to improve the substrate-metal interface and thereby intrinsic quality factors for both resonators and qubits.
After sputtering, the patterns are etched into the superconducting layer using reactive-ion etching with an SF$_6$/O$_2$ plasma.
In contrast with the Al transmon capacitor plates commonly used, in this experiment we make them also from NbTiN, with an aim to improve the substrate-metal interface and avoid large AlO$_x$ surfaces which may house unwanted two-level systems.
Only the Josephson junctions are made by the standard technique of Al-AlO$_x$-Al double-angle evaporation.
A further HF dip just prior to evaporation also helps to contact the junctions directly to the NbTiN capacitor plates.
In Ref.~\onlinecite{Riste15}, air bridges were already used to cross the feedline over flux-bias lines on chip.
Here, we extend this technique to allow the feedline to cross three readout resonators.

A key requirement for this experiment was the ability to match qubit frequencies without sacrificing coherence.
Flux-bias lines allow easy compensation for mismatch, but at the cost of reduced coherence in the qubit detuned from its maximal frequency (coherence sweet spot~\cite{Schreier08}).
We aimed for identical maximum frequencies of $\QT$ and $\QB$, as determined by capacitor and junction geometries.
The capacitors were easily matched in fabrication.
We then selected the chip with the closest matching room-temperature resistance values for the relevant qubit junctions.

\subsection{Experimental Setup}

 The chip is anchored to a copper cold-finger connected to the mixing chamber of a Leiden Cryogenics CS81 $\Hethree/\Hefour$ dilution refrigerator with $7~\mK$ base temperature.
  A copper can seals the sample space, with an inner surface that is coated with a mixture of Stycast 2850 and silicon carbide granules (15 to $1000~\nm$ diameter) used for infrared absorption~\cite{Barends11}. The copper can is in turn magnetically shielded by an aluminum enclosure and two outer Cryophy enclosures ($1~\mm$ thick)~\cite{Bruno15}.

\begin{figure}
    \includegraphics[width=\linewidth]{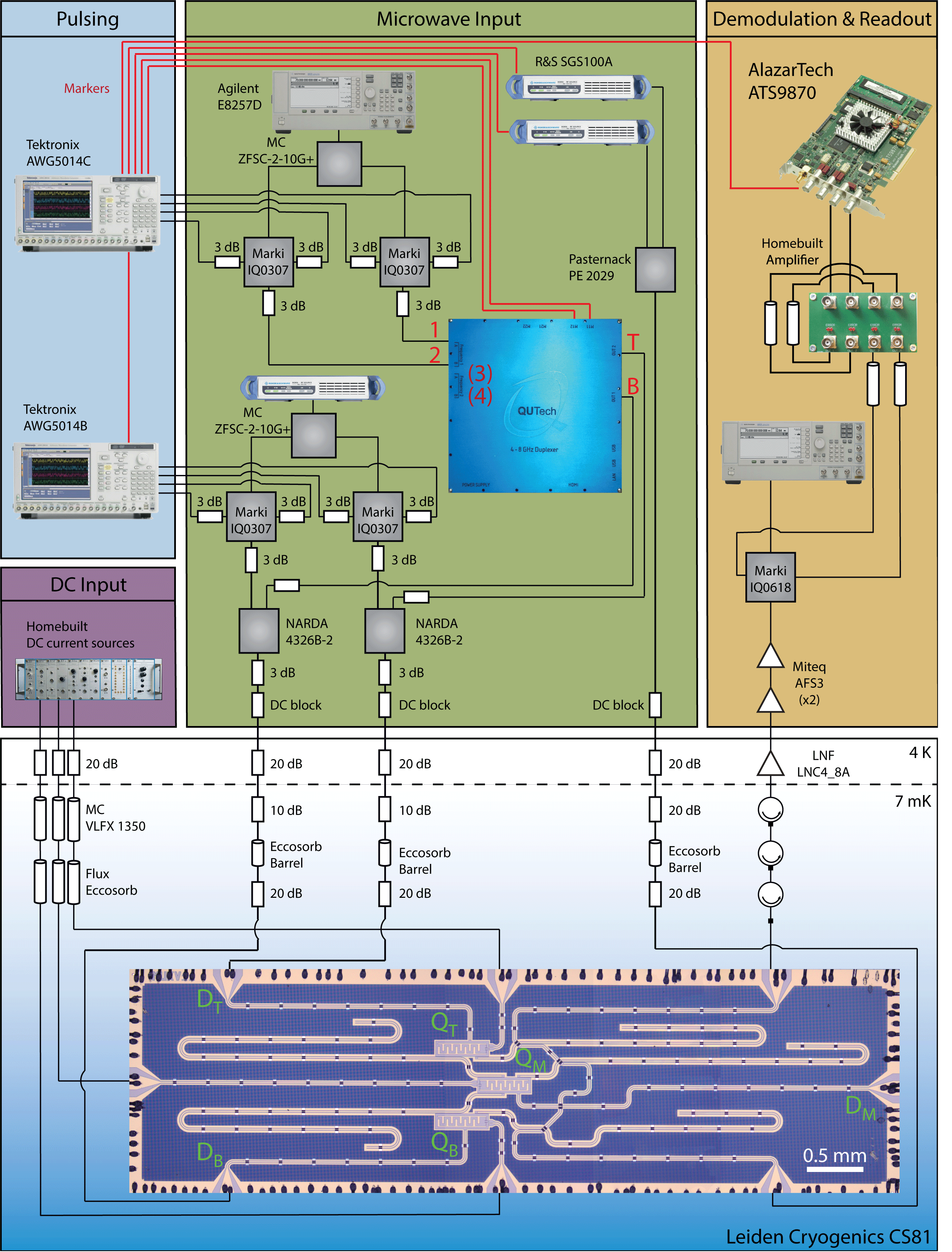}
    \caption{Detailed schematic of the experimental setup and optical image of the chip (installed in a Leiden Cryogenics CS81 dilution refrigerator).  Note that while our VSM has two pairs of analog inputs [(1,2) and (3,4)] and four gate inputs (activating links from each pair to each output),  the pair (3,4) and its associated gate inputs are not used throughout this experiment.}
    \label{fig:setup schematic}
  \end{figure}

  A complete wiring schematic showing all cryogenic and room-temperature components is shown in Fig.~\ref{fig:setup schematic}. The four analog channels of the Tektronix AWG5014C create the in-phase and in-quadrature pulses for $\QT$ and $\QB$ by single-sideband modulation of a common carrier.
  Because single-sideband modulation requires two AWG channels to modulate an IQ mixer, independent derivative-removal-via-adiabatic-gate (DRAG) tuning with the VSM therefore requires four AWG channels, irrespective of the number of qubits.
  The VSM can be scaled up to many output channels, and direct hardware savings can be realized as soon as three or more same-frequency qubits are driven by a single set of AWG inputs.
  These pulses are input at ports 1 and 2 of the VSM. The VSM combines these pulses with individually tuned insertion loss and phase to each of two outputs (labelled T and B).
  Input-output combinations can be switched on nanosecond timescales using the gate inputs of the VSM, provided by digital markers of the AWG5014C.
   A second AWG5014C with the appropriate carrier frequency is used to excite transmons $\QT$ and $\QB$ to the second-excited state~(Sec.~\ref{sec:second-excited-state-leakage}), to pulse measurement tones, and to trigger the AlazarTech ATS9870 acquisition card.

 \label{sec:Setup schematic}

\clearpage
  \subsection{Device frequencies}
  \label{ssec:Frequencies}

    \begin{table}[H]
      \begin{ruledtabular}
        \begin{tabular}{l c c c}
          Qubit & $f_\text{max}$ (GHz) & $f_\text{bias}$ (GHz) & $f_\text{res}$ (GHz)\\
          \hline
          $\QT$ & 6.277 & 6.220 & 6.700 \\
          $\QM$ & 6.551 & 6.551 & 6.733 \\
          $\QB$ & 6.220 & 6.220 & 6.800 \\
        \end{tabular}
      \end{ruledtabular}
      \caption{Table of sweet-spot frequencies $f_\text{max}$ and bias-point frequencies $f_\text{bias}$ of the three qubits, as well as the fundamental frequencies $f_\text{res}$ of their dedicated readout resonators at the bias point. The qubits are tuned into the bias point by a combination of spectroscopy and standard Ramsey experiments.}
      \label{tab:Muxmon0 qubit properties}
    \end{table}

  \subsection{Qubit coherence times}
    \label{ssec:Coherence times}
    
    \begin{figure}[H]
      \centering
      \includegraphics[width=\linewidth]{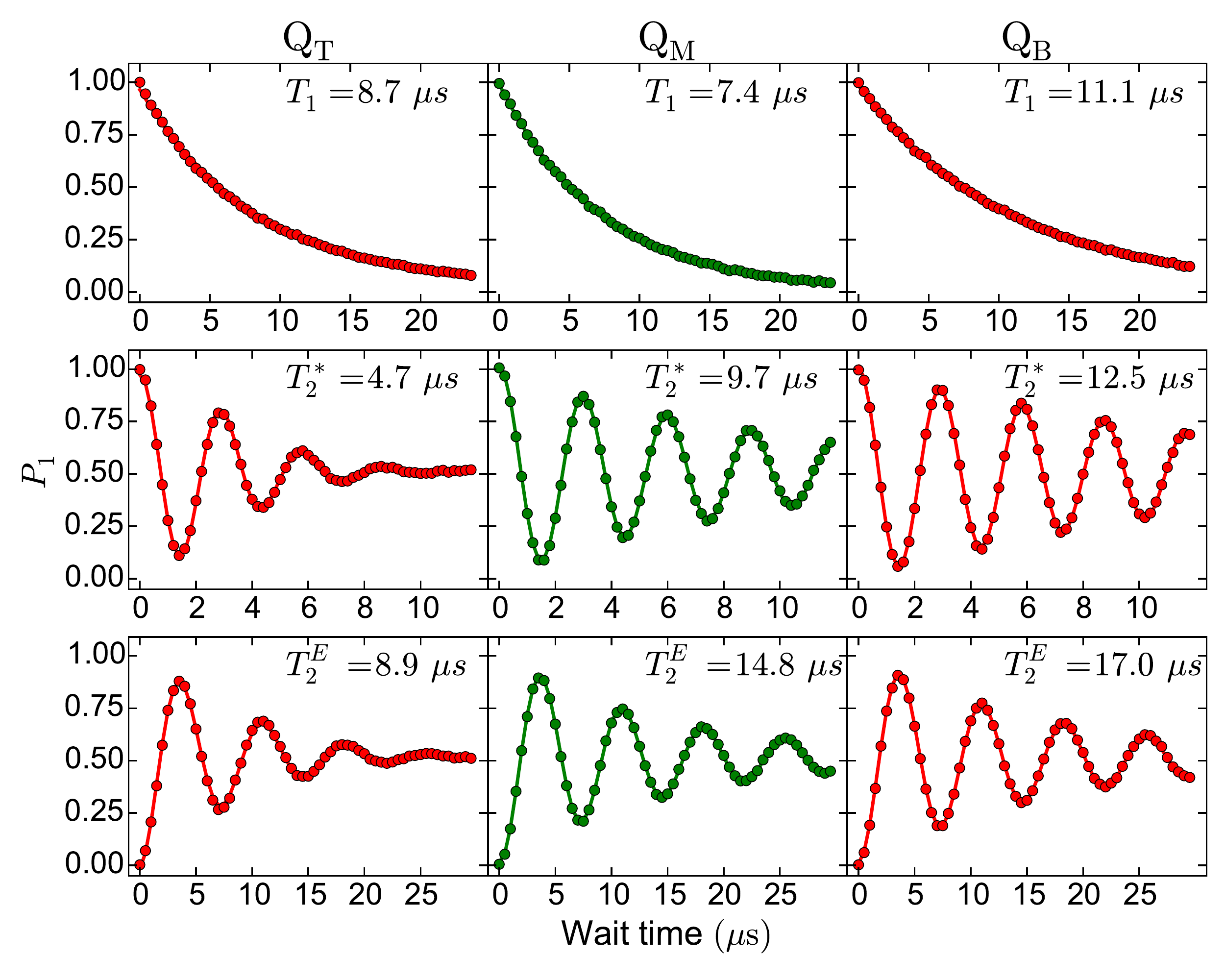}
      \caption{Measurements of relaxation ($\Tone$, top), Ramsey dephasing ($\Ttwostar$, middle) and echo dephasing ($\Ttwoecho$, bottom) times for the three qubits at the bias point. When measuring $\QT$ or $\QB$, the other qubit is detuned by $-50~\MHz$  to suppress cross-coupling effects. Ramsey fringes for $\QT$ (middle, left panel)  fit better to a  Gaussian (shown) than an exponential decay, reflecting the susceptibility of $\QT$ to low-frequency flux noise away from its sweet-spot. $P_1$ denotes excited-state population.}
      \label{fig:coherence times}
    \end{figure}

\section{Vector switch matrix}

  \subsection{Measured isolation}
  \label{sec:Vector switch matrix isolation}

  \begin{figure}
    \centering
    \includegraphics[width=\linewidth]{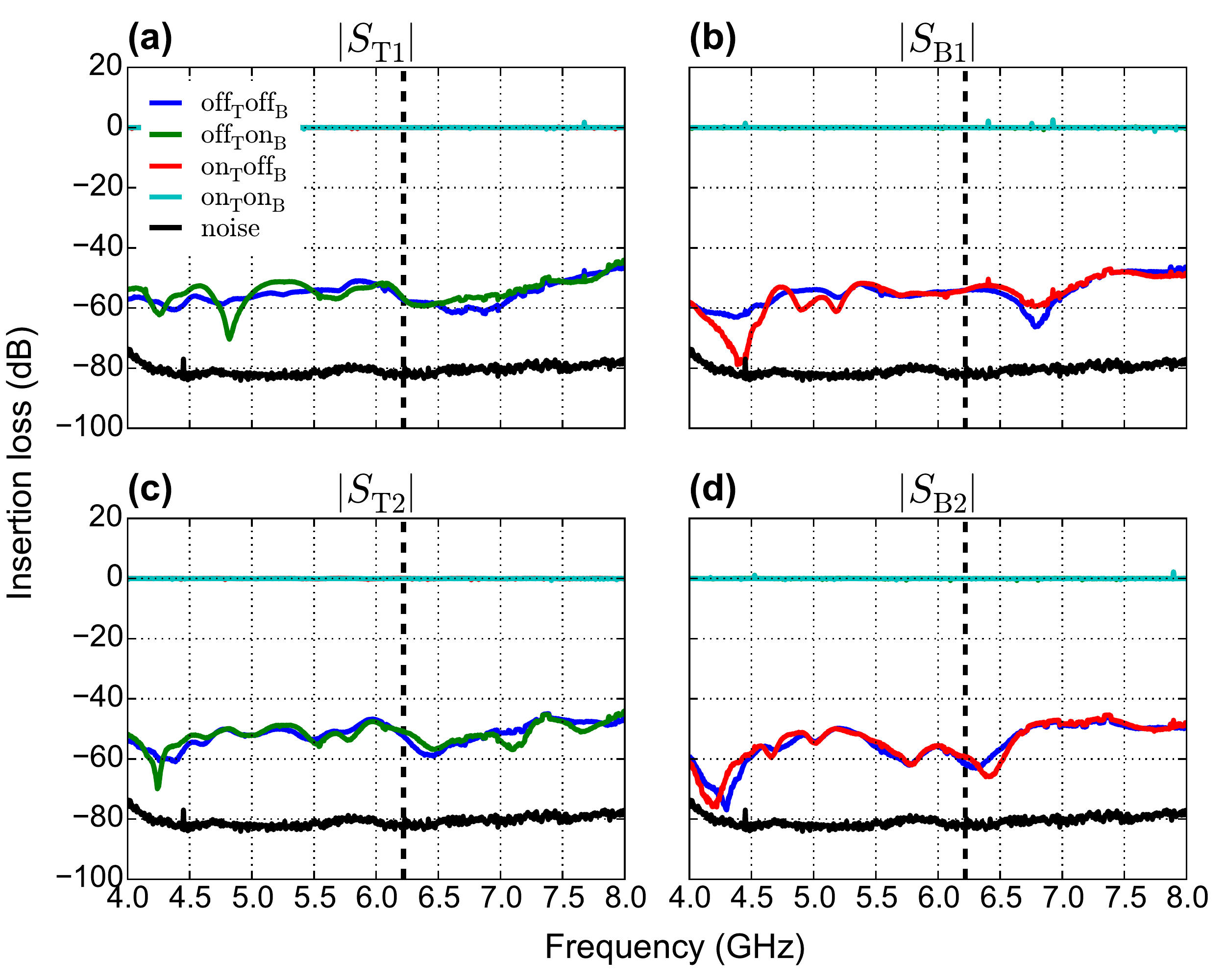}
    \caption{Insertion loss between inputs and outputs of the VSM for four static combinations of the gate inputs. The insertion loss is measured relative to the level with both gate inputs activated ($\mathrm{on}_\mathrm{T}\mathrm{on}_\mathrm{B}$). The black curves indicate the
    noise background in our scalar network analyzer measurement. The dashed vertical lines indicate the common frequency ($6.220~\GHz$) of $\QT$ and $\QB$ at the bias point.}
    \label{fig:VSM isolation}
  \end{figure}
  
To characterize the isolation of the VSM in the range $4$ to $8~\GHz$, we have measured the insertion loss between all input (1 and 2) and output ports (T and B) with static settings at the two gate inputs (Fig.~\ref{fig:VSM isolation}). Ideally, each gate activates (on state) and deactivates (off state) the link of both inputs 1 and 2 to one output, independent of the other gate. As shown in Fig.~\ref{fig:VSM isolation}, the typical relative isolation with the relevant gate in the off state is $\sim50~\dB$.

\subsection{Individual qubit tune-up}
\label{ssec:Individual qubit tuneup}
    
\begin{figure}
      \centering
       \includegraphics[width=1\linewidth]{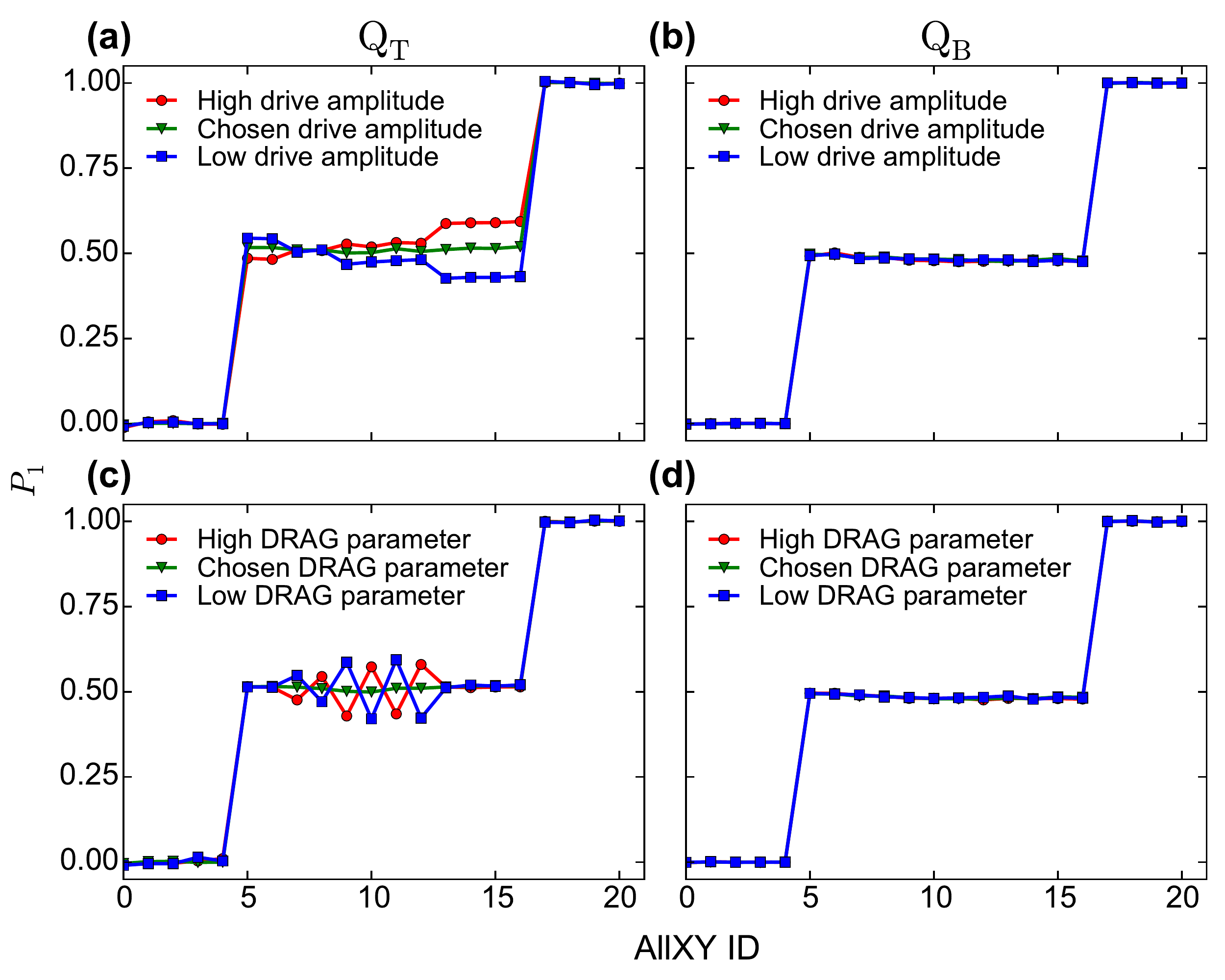}
       \caption{AllXY results for $\QT$ (left) and $\QB$ (right), as the in-phase amplitude (top) or in-quadrature amplitude (bottom) of $\QT$ pulses is varied.
       During each measurement, both qubits are driven simultaneously, but read out individually.
       The AllXY results of $\QB$ all overlap independent of the setting on $\QA$, showing independent control of pulse parameters for each qubit.}
       \label{fig:AllXY individual control}
    \end{figure}

The VSM enables independent control of the on/off state, insertion loss and phase for every input-output combination. We exploit this feature to perform DRAG-compensated pulses on $\QT$ and $\QB$, that are individually tailored for each qubit. Different types of gate errors, such as non-ideal in-phase and in-quadrature amplitudes, can be distinguished using an AllXY sequence~\cite{ReedPhD13}, consisting of 21 combinations of two pulses drawn from the set $\left\{I, X_{\pi}, Y_{\pi}, X_{\pi/2}, Y_{\pi/2}\right\}$  (Table~\ref{tab:AllXY sequence}). Figure~\ref{fig:AllXY individual control} shows AllXY sequence results for $\QT$ and $\QB$ as the amplitude of each quadrature on $\QT$ is varied independently. While the AllXY signature of $\QT$ reveals changing levels of amplitude and phase errors, there is no noticeable change in the AllXY signature of $\QB$. This demonstrates the use of the VSM for individual tune-up of pulses for same-frequency qubits.

    \begingroup
    \squeezetable
    \begin{table}
      \begin{ruledtabular}
        \centering
        \begin{tabular}{c c c c || c c c c}
          \multirow{2}{*}{ID} & \multirow{2}{*}{Ideal $P_1$} & \multicolumn{2}{c||}{Pulses} & \multirow{2}{*}{ID} & \multirow{2}{*}{Ideal $P_1$} & \multicolumn{2}{c}{Pulses} \\
          & & First & Second & & & First & Second\\
          \hline
          1  & 0  & $I$         & $I$         & 12 & 0.5  & $Y_\pi$     & $Y_{\pi/2}$ \\
          2  & 0  & $X_\pi$     & $X_\pi$     & 13 & 0.5  & $X_\pi$     & $X_{\pi/2}$ \\
          3  & 0  & $Y_\pi$     & $Y_\pi$     & 14 & 0.5  & $X_{\pi/2}$ & $X_\pi$     \\
          4  & 0  & $X_\pi$     & $Y_\pi$     & 15 & 0.5  & $X_\pi$     & $X_{\pi/2}$ \\
          5  & 0  & $Y_\pi$     & $X_\pi$     & 16 & 0.5  & $Y_{\pi/2}$ & $Y_\pi$     \\
          6  & 0.5  & $I$ & $X_{\pi/2}$         & 17 & 0.5  & $Y_\pi$     & $Y_{\pi/2}$ \\
          7  & 0.5  & $I$ & $Y_{\pi/2}$         & 18 & 1 & $I$         & $X_\pi$     \\
          8  & 0.5  & $X_{\pi/2}$ & $Y_{\pi/2}$ & 19 & 1 & $I$         & $Y_\pi$     \\
          9  & 0.5  & $Y_{\pi/2}$ & $X_{\pi/2}$ & 20 & 1 & $X_{\pi/2}$ & $X_{\pi/2}$ \\
          10 & 0.5  & $X_{\pi/2}$ & $Y_\pi$     & 21 & 1 & $Y_{\pi/2}$ & $Y_{\pi/2}$ \\
          11 & 0.5  & $Y_{\pi/2}$ & $X_\pi$     &         &             &             \\
        \end{tabular}
      \end{ruledtabular}
      \caption{The 21 two-pulse combinations comprising the AllXY pulse sequence~\cite{ReedPhD13}.}
      \label{tab:AllXY sequence}
    \end{table}
    \endgroup

\section{Pulse-calibration routines}
  \label{sec:Pulse-calibration routines}
   We tune up qubit pulses by alternating the calibration of in-phase and in-quadrature pulse amplitudes until a simultaneous optimum is found. The two calibration routines are discussed below.

  \subsection{Accurate in-phase pulse amplitude calibration}
    \label{ssec:Accurate drive-amplitude calibration}

\begin{figure}
      \centering
      \includegraphics[width=.7\linewidth]{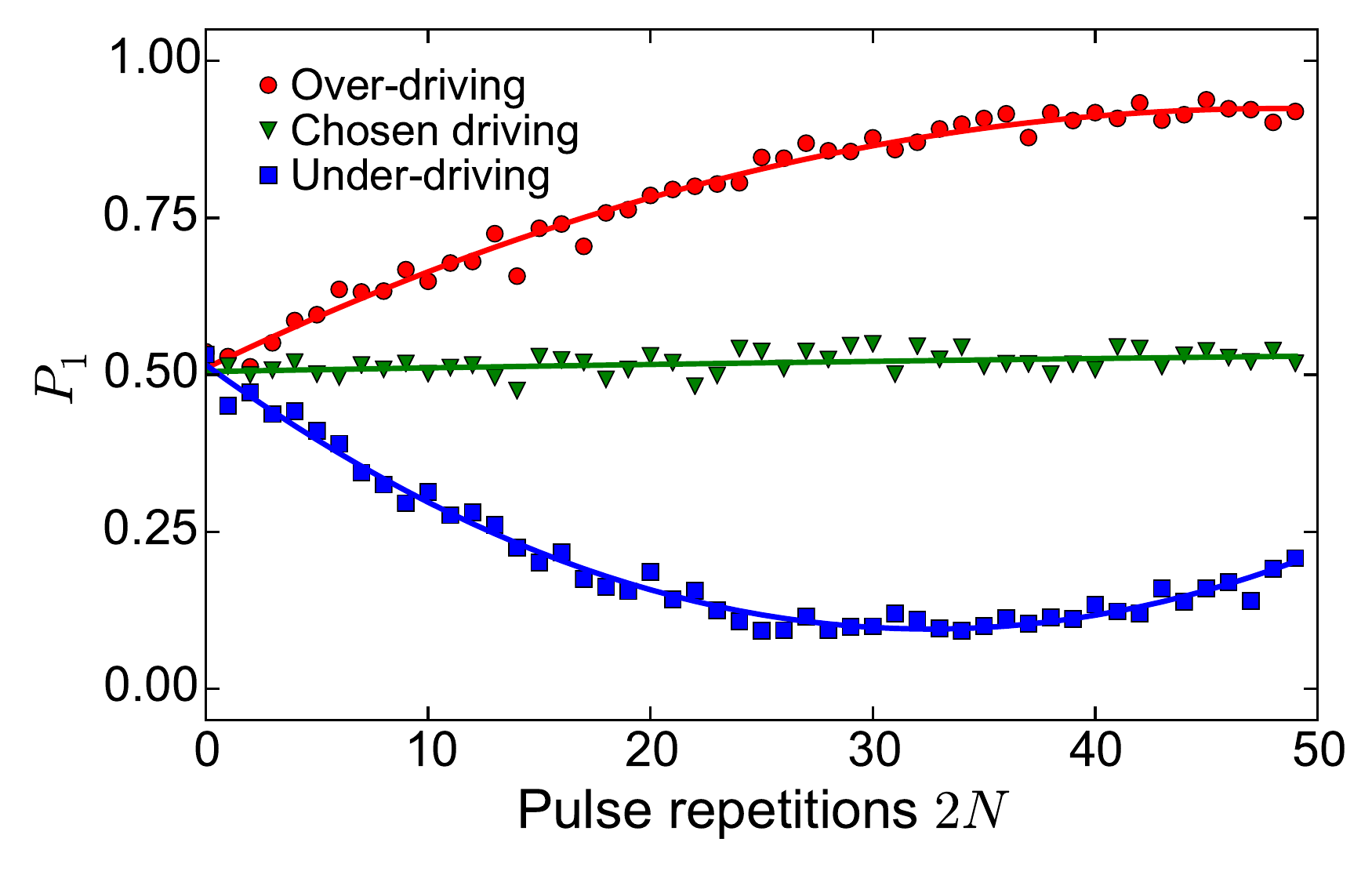}
      \caption{Fine calibration of pulse amplitude by initial $\pi/2$ pulse, followed by $2N$ repeated $\pi$ pulses. The initial slope determines if the qubit is over- (positive slope) or under-driven (negative slope).}
      \label{fig:PiX360}
    \end{figure}

    The in-phase quadrature amplitude is calibrated by first applying a $\pi/2$ pulse to the qubit, followed by a train of $\pi$ pulses. The pulse sequence is $\left( X_{\pi} \right)^{2N} X_{\pi/2} \ket{g}$, where $\ket{g}$ is the qubit ground state and $N\in \left[0,49\right]$. In the absence of gate errors and decoherence, the driven qubit would end on the equator of its Bloch sphere for all $N$. However, over- or under-driving produces a positive or negative initial slope on $P_1$ versus $N$, respectively (Fig.~\ref{fig:PiX360}). We choose the in-quadrature amplitude that minimizes the absolute slope.

  \subsection{DRAG-parameter calibration}
    \label{ssec:DRAG-parameter calibration}

    To minimize phase errors resulting from the presence of the second- and higher-excited states, we optimize the scaling of the in-quadrature pulse. As in conventional DRAG~\cite{Motzoi09,Chow10b}, we choose as the envelope of the in-quadrature pulse the derivative of the Gaussian envelope on the in-phase pulse. The DRAG scaling parameter is calibrated using the method detailed in Ref.~\cite{ReedPhD13}. Specifically, we measure the difference in excited-state population produced by the $Y_{\pi} X_{\pi/2}$ and $X_{\pi} Y_{\pi/2}$  pulse combinations (AllXY ID 10 and 11). Ideally, for both, the final qubit state would lie on the equator. However, any phase error shifts the final excited-state population in opposite directions in these cases. We choose the DRAG scaling parameter minimizing this shift.

\section{Global broadcasting}
\label{sec:global-broadcasting}

\begin{figure}
  \begin{minipage}[t]{.4\linewidth}
    {\normalsize Global broadcasting}
  \end{minipage}
  \begin{minipage}{.5\linewidth}
    \includegraphics[width=\linewidth]{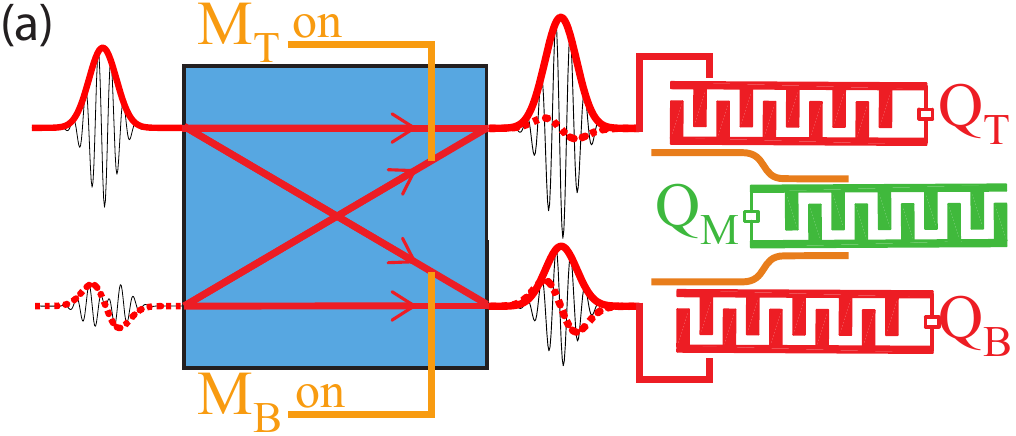}
  \end{minipage}
  \centering
  \includegraphics[width=\linewidth]{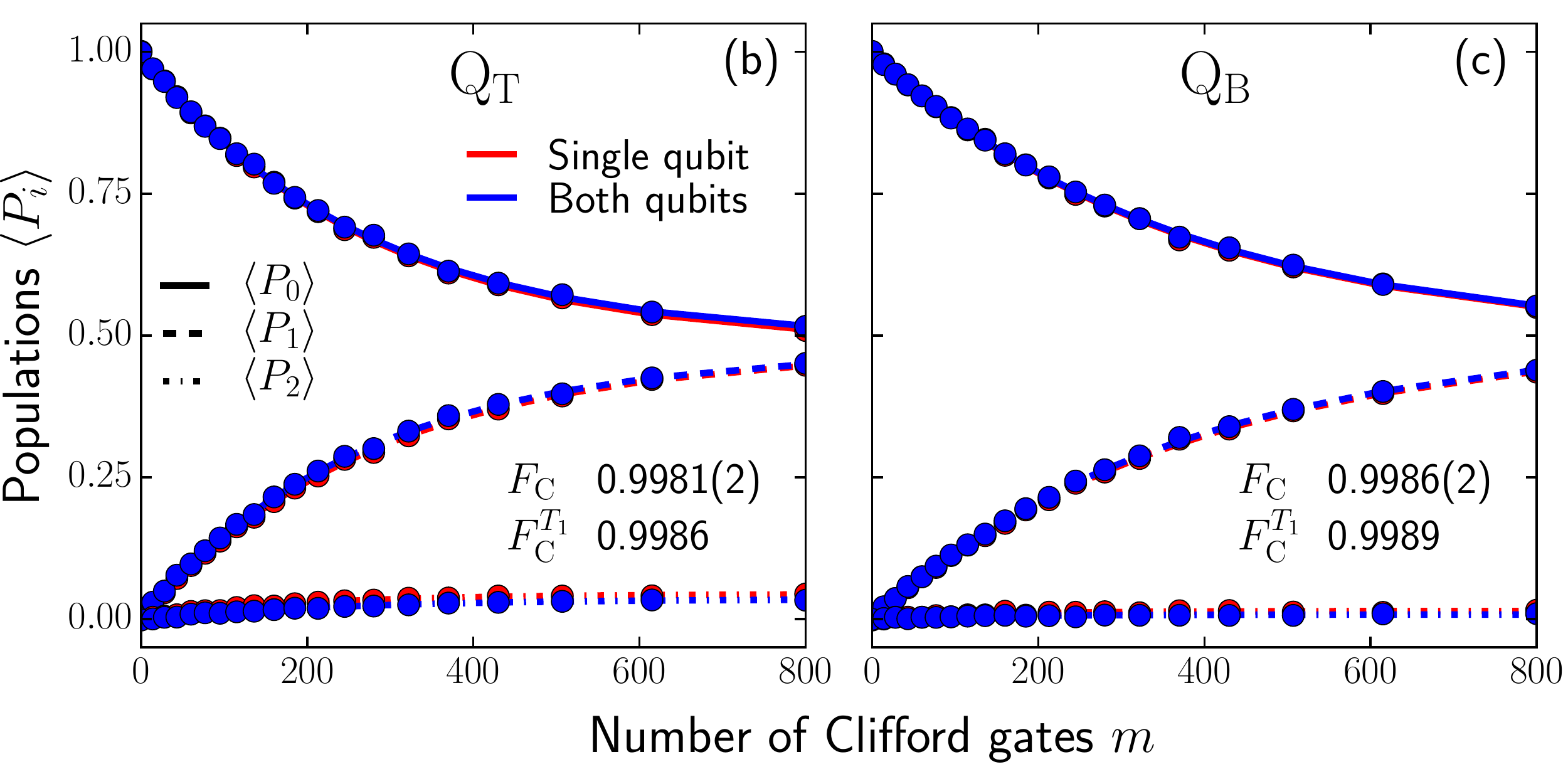}
  \caption{
  Global broadcasting of DRAG pulses to same-frequency qubits.
  (a) Illustration of global broadcasting.
  Two simultaneous pulses, one with Gaussian envelope at input 1, and another with derivative-of-Gaussian envelope at input 2, are simultaneously directed to $\QT$ and $\QB$ (both markers always on).
  The insertion loss and phase shift of each pulse is separately optimized for each output to produce precision DRAG pulses for each qubit.
  (b,c) Comparison of single-qubit driving versus driving both qubits (broadcasting) by RB of Clifford gates composed from $\pi/2$ and $\pi$ pulses~\cite{Epstein14}.
  Average population of $\QT$ and $\QB$ in the ground, first- and second-excited state ($\langle P_0\rangle$, $\langle P_1\rangle$ and $\langle P_2\rangle$, resp.) as a function of the number of Clifford gates applied.
  Curves are the best fits of single exponentials with offsets to the populations.
  The single-qubit Clifford-gate fidelity for each qubit is extracted from the decay of the corresponding ground-state population when using global broadcasting.
  }
  \label{fig:Broadcasting}
\end{figure}

Aside from single-qubit control and selective broadcasting, the VSM also allows global broadcasting of pulses to all qubits simultaneously by keeping the markers for both qubits on (Fig.~\ref{fig:Broadcasting}).
While this does provide simultaneous control of $\QT$ and $\QB$, marker control is needed to achieve independent control.
Using RB, we measure the performance of both qubits when broadcasting pulses to both qubits, and compare the results with those obtained from single-qubit control (Fig.~\ref{fig:Broadcasting}).
The global broadcasting RB measurements were alternated with the single-qubit RB measurements, and aside from marker settings all other settings were identical.
Comparison of the results in Fig.~\ref{fig:Broadcasting} show that the qubit gate performance does not depend on whether a single qubit is controlled, or both are controlled simultaneously through global broadcasting.

\section{Leakage to second excited state}
\label{sec:second-excited-state-leakage}

Leakage is fundamentally different from unitary qubit errors.
To quantify leakage, we monitor the populations $P_i$ of the three lowest energy states ($i \in \{0,1,2\}$) during RB and calculate the average values $\langle P_i\rangle$ over all seeds.
To do this, we calibrate the average signal levels $V_i$ for the transmons in level $i$, and perform each RB measurement twice, the second time with an added final $\pi$ pulse on the 0--1 transition.
This final $\pi$ pulse swaps $P_0$ and $P_1$, leaving $P_2$ unaffected.
Under the assumption that higher levels are unpopulated ($ P_0 + P_1 + P_2 = 1$),
\begin{align}
  \begin{bmatrix}
    V_0 - V_2 & V_1 - V_2 \\
    V_1 - V_2 & V_0 - V_2
  \end{bmatrix}
  \begin{bmatrix}
    P_0 \\
    P_1
  \end{bmatrix}
  =
  \begin{bmatrix}
    S - V_2 \\
    S^\prime - V_2
  \end{bmatrix},
\end{align}
where $S$ ($S^\prime$) is the measured signal level without (with) final $\pi$ pulse.  The populations are extracted by matrix inversion.

Measuring $\langle P_2\rangle$ as a function of the number of Clifford gates allows us to estimate an average leakage per Clifford, $\kappa$.
Because the populations are ensemble averages over different random seeds, we assume that leakage of the average qubit-space populations to $\langle P_2\rangle$ is incoherent, and, provided $\langle P_2\rangle$ remains small ($\kappa$ small), we also assume that leakage is irreversible.
We therefore model leakage using the following difference equation for $\langle P_2\rangle$:
\begin{equation}
\langle P_2[m{+}1]\rangle - \langle P_2[m]\rangle \simeq t_\mathrm{p} \langle N_\mathrm{P}\rangle \, \kappa -  \frac{t_\mathrm{p} \langle N_\mathrm{P}\rangle}{T_{2\rightarrow1}} \langle P_2[m] \rangle,
  \label{eq:second-excited state rate equation}
\end{equation}
where $T_{2\rightarrow1}$ is the second- to first-excited-state relaxation time.
Assuming no initial population in the second-excited state, the solution is Eq.~(2), which shows good agreement with measured data.
We extract $\kappa$ by fitting Eq.~(2) to $\langle P_2\rangle$ data.  $T_{2\rightarrow1}$ is obtained from the best-fit decay constant (not directly measured) and $\kappa$ from the best-fit prefactor.

\section{Cross-coupling and cross-driving effects}
  \label{sec:Cross-coupling and cross-driving effects}
  There are several sources of spurious cross-qubit interactions [see Fig.~2(g,h) of the main text] which play a role in our experiments.
  We divide these into two main classes: cross-coupling, where the qubits themselves are coupled via a direct or indirect quantum interaction, and cross-driving, where input microwaves directly drive the untargeted qubit with a reduced amplitude as a result of imperfect isolation either on or off chip.
  These cross-excitation effects depend strongly on specific chip design, in our case chosen according to our primary aim to study the potential of frequency reuse in a circuit fully compatible with a larger surface-code lattice.
  This governed both how qubits and resonators were connected, as well as the selection and arrangement of frequencies.
  In addition, in order to fully explore the limitations of such techniques, the qubits were positioned on the chip as close to each other as possible to provide a worst-case scenario for cross-excitation effects.

  \subsection{Cross-coupling}
    \label{ssec:Cross coupling}

\begin{figure}
      \centering
      \includegraphics[width=\linewidth]{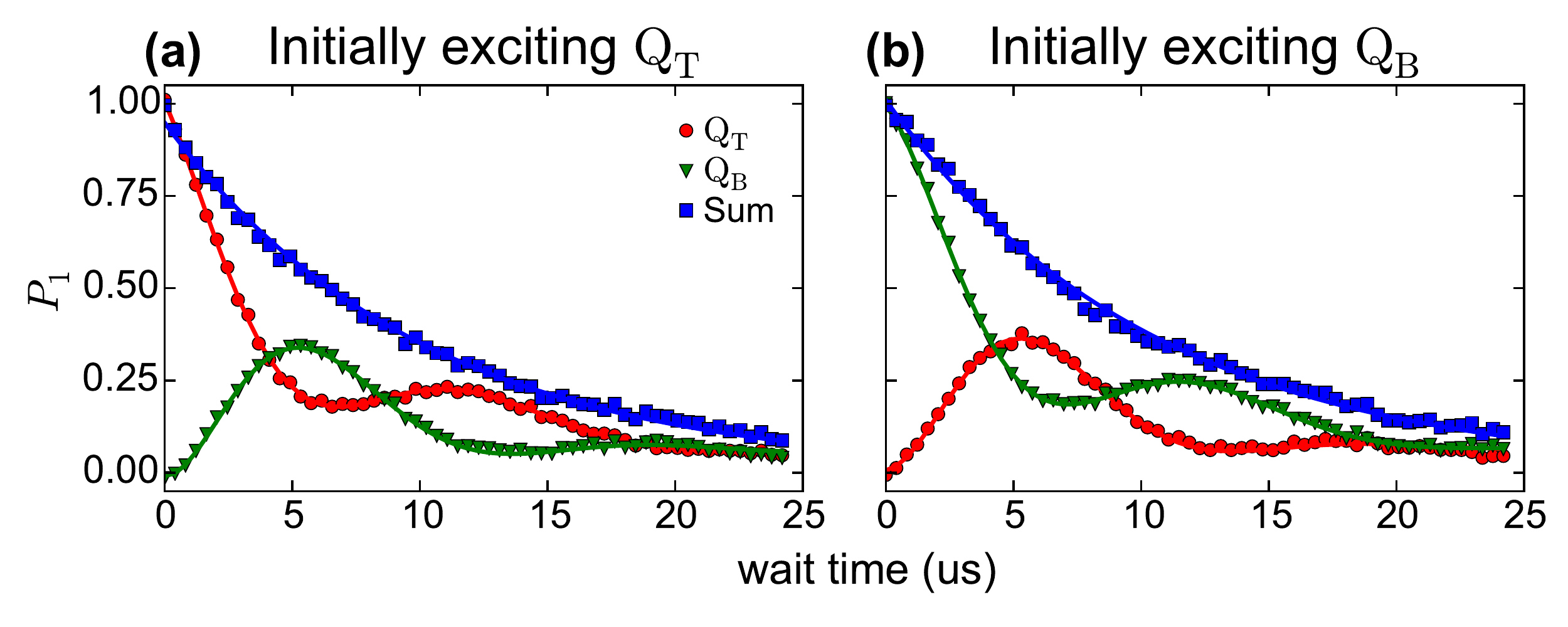}
      \caption{Temporal evolution of a single excitation after initially exciting $\QT$ \figlabel{a} or $\QB$ \figlabel{b}. The oscillations of population in both qubits are out of phase and have a common frequency. The sum of both populations shows approximately exponential decay, with time constant $10.2(2)~\us$, intermediate between the relaxation times of $\QT$ and $\QB$.}
      \label{fig:excitation swap}
    \end{figure}

    In our device, the same-frequency qubits $\QT$ and $\QB$ are connected through a linear chain of coupled quantum elements: the two bus resonators and intermediate ancilla qubit $\QM$.
    They are also coupled capacitively through the ground plane.
    When the intermediate elements are detuned in frequency away from the near-resonant qubits, such geometries typically give rise to higher-order exchange-type interactions between the qubits of the form $J (\sigma_\mathrm{T}^+ \sigma_\mathrm{B}^- + \sigma_\mathrm{T}^- \sigma_\mathrm{B}^+)$~\cite{Blais07, Majer07}.
    Consistent with this picture, we observe coherent swapping of excitations between $\QT$ and $\QB$ at a rate strongly dependent on the qubit detuning (Fig.~\ref{fig:excitation swap}).

    In order to characterise the cross-coupling between $\QT$ and $\QB$, we measure the evolution of excited-state populations after a single excitation is injected at one of the qubits with a $\pi$ pulse.
    To place a tight upper bound on the interaction strength $J$, the qubit frequencies must be matched as closely as possible.
    We achieve an accuracy of around 50~$\kHz$ using Ramsey experiments, limited by a combination of factors: the resolution of the flux tuning, the fitting resolution limit imposed by qubit $T_2$ dephasing times, and also the frequency shifting induced by the qubit-qubit exchange interaction itself.
    As shown in Fig.~\ref{fig:excitation swap}, the total excitation number exhibits a typical exponential $\Tone$ decay, while the individual populations show a symmetric oscillation of the excitation between the two qubits.
    The measured common frequency of decaying oscillations for both qubits sets an upper bound on the coupling strength of $J/2\pi\leq36\pm1~\kHz$.

  \subsection{Cross-driving}
    \label{ssec:Cross-driving}
    
\begin{figure}
      \centering
      \includegraphics[width=\linewidth]{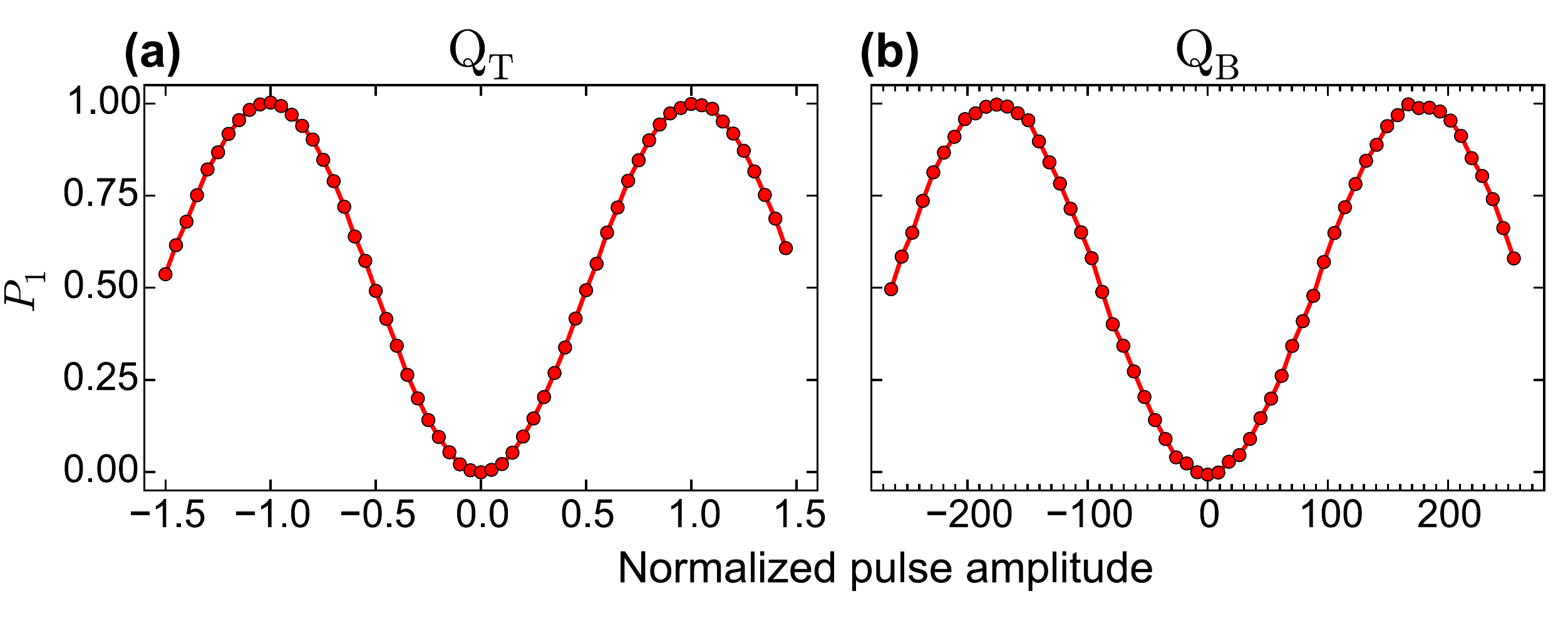}
    \begin{minipage}[t]{.54\linewidth}
      \vspace{-.28cm}
        \includegraphics[width=\linewidth]{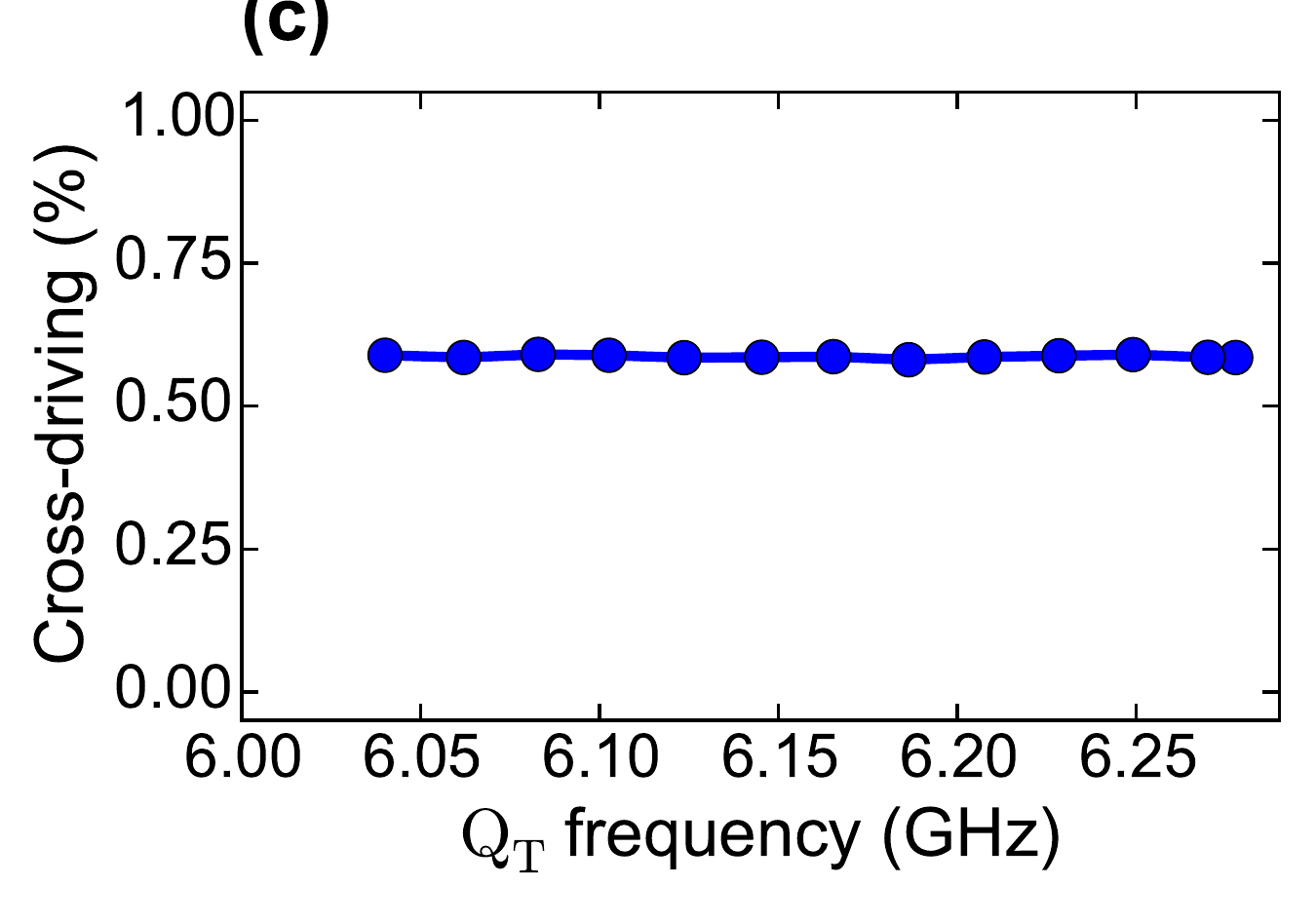}
    \end{minipage}
    \begin{minipage}[t]{0.43\linewidth}
      \vspace{-3mm}
      \flushleft
       {\usefont{T1}{phv}{b}{n}\scriptsize(d)}
      \vspace{.7mm}
      \begin{ruledtabular}
      \begin{tabular}{l c c}
        \multirow{2}{*}{Qubit} & \multicolumn{2}{c}{Driving strength} \\
             & $\DT$ & $\DB$ \\
        \hline
        \multirow{2}{*}{$\QT$}  & \multirow{2}{*}{$1.0000$}    & $0.0023$ \\
        && (-53 $\dB$) \\
        \multirow{2}{*}{$\QB$}  & $0.0057$ & \multirow{2}{*}{$1.0000$}    \\
        & (-45 $\dB$) & \\
      \end{tabular}
      \end{ruledtabular}
    \end{minipage}

      \caption{Rabi oscillations of $\QT$ \figlabel{a} and $\QB$ \figlabel{b} induced by pulses on $\DT$ at the bias point. The pulse amplitude is normalized to the $\pi$-pulse amplitude for $\QT$. \figlabel{c} Resonant cross-driving of $\QB$ via $\DT$ for a range of $\QT$ frequencies. The weak dependence of the cross-driving strength shows that this effect is most likely due to direct coupling of $\DT$ to $\QB$. \figlabel{d} Summary of cross-driving for $\DT$ and $\DB$ at the bias point.}
      \label{fig:cross-driving}
    \end{figure}

    When driving one qubit, the signal applied to its dedicated voltage line residually drives the distant qubits on the chip.
    This results from microwave leakage in the VSM, near the sample, or from direct on-chip coupling of the drive line to the distant qubits.
    In Section~\ref{sec:Vector switch matrix isolation}, we fully characterise the isolation in the VSM.
    This leads to leakage of around $-57~\dB$ and $-54~\dB$ on $\QT$ and $\QB$, respectively, at the qubit operating frequency for the conditions used in the main experiments.
    To characterize the residual cross-driving at the device, we disconnect the VSM and compare the amplitude required for pulses applied on one drive line (either $\DT$ or $\DB$) to perform a $\pi$ rotation on both $\QT$ and $\QB$.
    We define the cross-driving strength of each drive line as the ratio $r_\mathrm{c}$ of the $\pi$-pulse amplitude for the directly-driven to that for the cross-driven qubit.
    For this test, pulses are first amplified and then attenuated using a step attenuator to allow the large amplitude range required.
    Results shown in Fig.~\ref{fig:cross-driving} demonstrate on-chip cross-driving strengths below $1\%$ ($-53~\dB$ and $-45~\dB$ on $\QT$ and $\QB$, respectively), but still somewhat larger than cross-driving effects resulting from finite isolation in the VSM.
    During the main experiments, both VSM and on-chip leakage are able to contribute to cross-driving effects.

    In the excitation-swap experiments of the previous section, it is clear the effects result from cross-coupling rather than cross-driving, because the exchange dynamics occur while the drive pulses are off.
    Similarly, we can rule out these residual driving effects being related to cross-coupling, because the 16-$\ns$ Rabi pulses take only a fraction of the time required for a cross-coupling exchange oscillation.
    As a further check that the effects observed are indeed due to cross-driving, we measure the cross-driving on $\QB$ through $\DT$ as $\QT$ is tuned through resonance with $\QB$ [Fig.~\ref{fig:cross-driving}(c)].
    The cross-driving ratio is essentially constant, showing no significant dependence on the detuning between $\QT$ and $\QB$.

  \subsection{Cross-excitation in randomized benchmarking}
    \label{ssec:cross-excitation-in-RB}

    The isolated single-qubit control experiments in the main text [Fig.~2(g,h)] show that significant spurious excitations can build up in the idling qubit over the course of the long gate sequences tested in RB (particularly in the case of idling $\QB$ while driving $\QT$).
    It may therefore initially be somewhat surprising that virtually the same individual qubit-control performance is achieved in both selective-broadcasting (Fig.~4, main text) and global-broadcasting (Fig.~\ref{fig:Broadcasting}) scenarios.
    As discussed in the main text, the observed cross-excitation is unlikely to result from cross-coupling, primarily because a symmetrical quantum coupling should not result in strongly asymmetric effects on the different qubits.
    We now show the results are, however, consistent with the effects of cross-driving by numerically simulating RB with cross-driving under experimentally realistic conditions (using independently measured qubit and cross-driving parameters).
    Simulations are performed using QuTiP~\cite{Johansson12,Johansson13}.

    We model our system as two uncoupled qubits, $Q_\mathrm{T}$ and $Q_\mathrm{B}$, subject to $\Tone$ relaxation (with corresponding relaxation times) and cross-driving.
  We approximate the system dynamics using instantaneous unitary pulse operators from the standard Pauli set $\{ X_\pi, Y_\pi, X_{\pm\pi/2}, Y_{\pm\pi/2}\}$ with 20 $\ns$ delays of $\Tone$-only qubit relaxation between pulses implemented using a master equation.
  When applying a pulse to one qubit, cross-driving of the other qubit is implemented by applying a pulse with the same rotation axis, but with the original rotation angle multiplied by the relevant cross-driving ratio.
  We note that also trying to model the effects of qubit dephasing using a simple master equation does not produce RB data consistent with the experimental observations (e.g., Figs~2 and~4 of the main text).
  This reflects the non-uniform phase noise spectrum which affects the transmon qubit.
  The long RB pulse sequences consisting of $\pi$ and $\pi/2$ pulses around the $X$ and $Y$ axes seem to provide some form of dynamical decoupling which makes the RB measurements robust to qubit dephasing.
  As will be seen, the experimental results can be well modelled using only $\Tone$-type noise processes.

  \begin{figure}
    \centering
    \includegraphics[width=\linewidth]{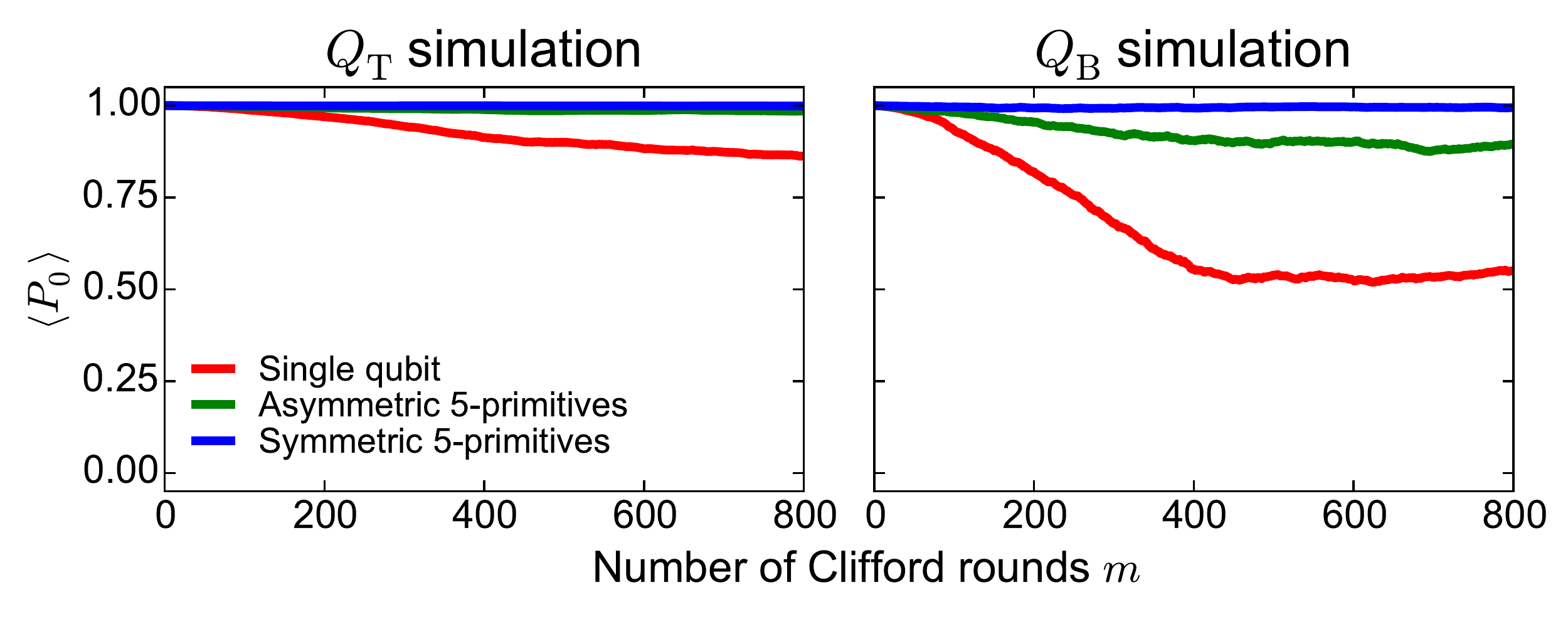}
    \caption{Simulation of cross-driving effects during RB. The results shown are averaged over ten runs, using the single-qubit minimal-set decomposition (red), and using the selective-broadcasting asymmetric and symmetric 5-primitives schemes (green and blue, respectively). Cross-driving effects are largely suppressed  in the 5-primitives schemes by choosing the five pulse primitives such that constituent pulses largely cancel out. The symmetric 5-primitives scheme further reduces cross-driving effects by alternating between the five pulse primitives and the inverse pulses.}
    \label{fig:simulations cross-driving}
  \end{figure}

  Randomized benchmarking is implemented by generating independent Clifford sequences for each qubit.
  We decompose Clifford gates using either the minimal set decomposition or one of the selective-broadcasting schemes.
  Figure~\ref{fig:simulations cross-driving} shows simulated results of cross-driving for the isolated single-qubit control scenario reported in Fig.~2 of the main text.
  In this section, we are only concerned with the red curves, which correspond to implementing single-qubit RB with the standard set of pulse decompositions~\cite{Epstein14}.
  These simulations can be compared directly with the curves in Fig.~2(g,h).
  While the maximum excitation population observed in the simulations is larger than the value observed in the experiments, the simulations for both qubits show the same qualitative behaviour as the measured data.
  The quantitative difference may be explained by the fact that the direct measurements of cross-driving were made at a different time from the main measurement run and we observed some small fluctuations in cross-driving levels over time.

    \begin{figure}
    \centering
    \includegraphics[width=\linewidth]{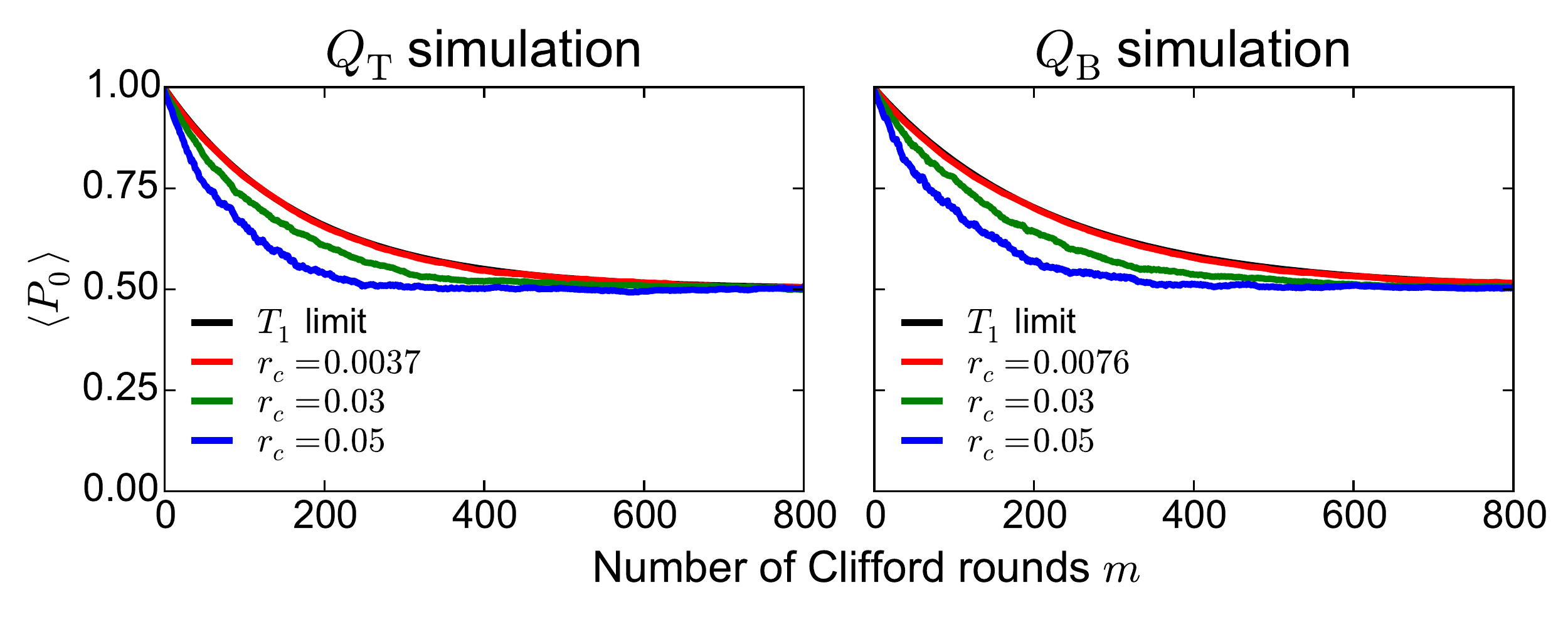}
    \caption{Simulation of sequential (interleaved) RB for several cross-driving ratios. For each cross-driving ratio, 50 simulation runs were performed, using sequential RB up to $m=800$ Clifford rounds.  Under the cross-driving levels pertaining in our experiments (0.0037 and 0.0076 for $Q_\mathrm{T}$ and $Q_\mathrm{B}$, respectively), the error per Clifford for the idling operation is dominated by the effects of $T_1$ relaxation as calculated from Eq.~(1).}
    \label{fig:simulations RB vs cross ratio}
  \end{figure}

\begin{figure}
    \centering
    \includegraphics[width=\linewidth]{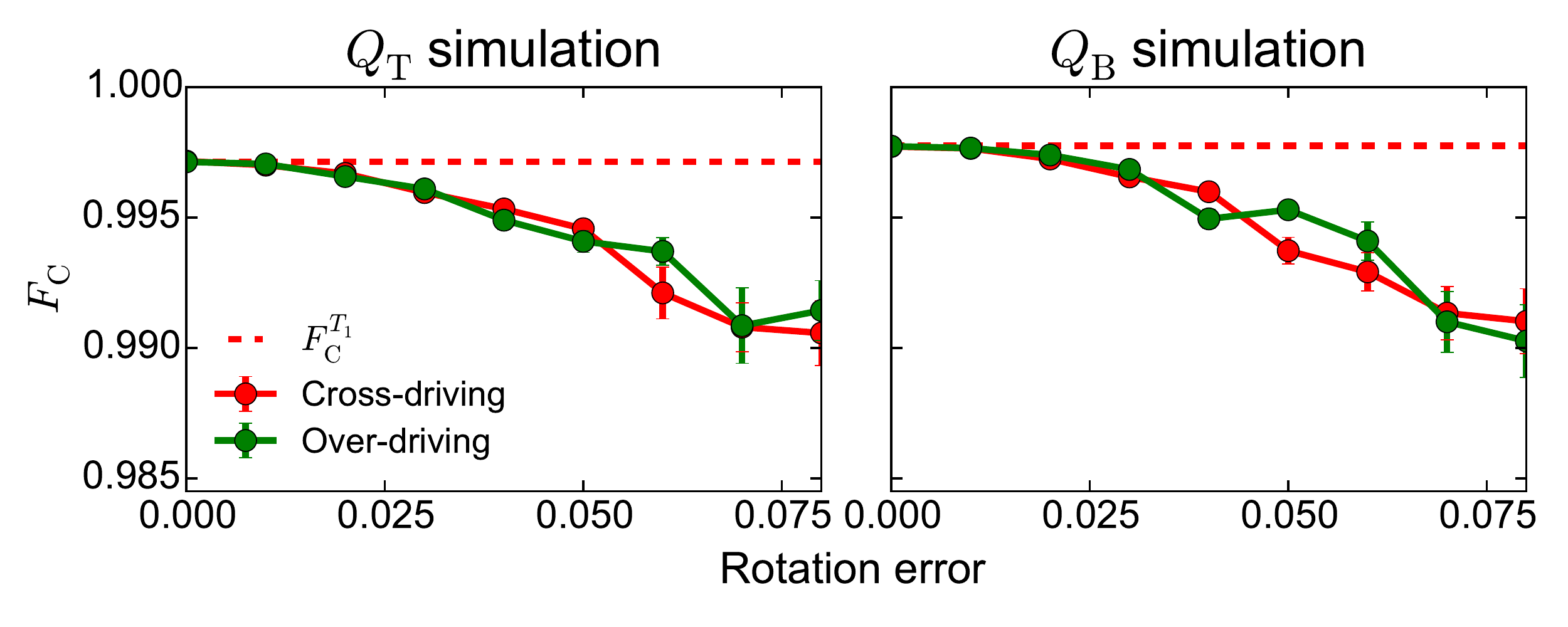}
    \caption{Simulations of Clifford fidelity $F_C$ as a function of the cross-driving ratio ($r_\mathrm{c}$: red) and the relative over-driving rotation error ($r_\mathrm{o}{-}1$: green). For each error type, 50 simulation runs were performed, using sequential RB on both qubits simultaneously up to $m=800$ Clifford rounds, after which $F_C$ was extracted from an exponential fit to averaged data. The $T_1$ limit [Eq.~(1)] is given by a horizontal dashed line. The data shows that $F_C$ is first-order insensitive to both cross-driving and to over-rotations. }
    \label{fig:simulations FC vs ratio}
  \end{figure}

    As discussed in the main text, while the plots of cross-excitation during RB are useful diagnostics of the presence of a spurious cross-driving effect, they may give a misleading impression when presented in parallel with RB results.
    Although the decay curves look superficially similar, they should not be interpreted in the same way.
    By contrast, the technique of interleaved RB (IRB), which was introduced to enable rigorous quantification of the performance of individual gates, allows us to calculate a meaningful error per Clifford for the idling operation~\cite{Magesan12b}.
    In IRB, the usual random sequence of Cliffords is alternated with identical repetitions of an individual gate.
    By comparing the interleaved decay rate with the decay rate for a standard RB measurement, it is possible to calculate a robust error per gate for the individual gate in question.
    In this context, the target gate is the nominal identity operation on one qubit which results from a random Clifford being applied to the other qubit.
    The IRB pulse sequence is therefore identical to the sequence implemented in the sequential selective-broadcasting scheme (see Fig.~4 of the main text).
    In the main text, we use the formulas in Ref.~\cite{Magesan12b} to calculate the idling error per Clifford, but for these simulations, the performance of sequential selective-broadcasting already provides a simple way to assess the performance of cross-excitation during idling.
    Figure~\ref{fig:simulations RB vs cross ratio} shows that idling performance as quantified by RB is limited mainly by $\Tone$ relaxation.
    Finally, when identical gate sequences are being applied to both qubits, cross-driving will result in a small amount of over-driving on each qubit (overdriving ratio $r_\mathrm{o}$), which would also look like an error in pulse rotation angle.
    Figure~\ref{fig:simulations FC vs ratio} shows that the Clifford error is insensitive to both cross-driving and over-driving to first order.

  \subsection{Making pulse sequences robust to cross-driving}
  \label{ssec:robust-pulse-sequences}

    We have already shown that cross-excitation does not have a dominant effect on single-qubit control in both global and selective broadcasting.
    We show here that any residual effect can be largely eliminated also while a qubit is idling by choosing robust pulse sequences for decomposing the Clifford gates.

\begin{figure}
      \centering
      \includegraphics[width=\linewidth]{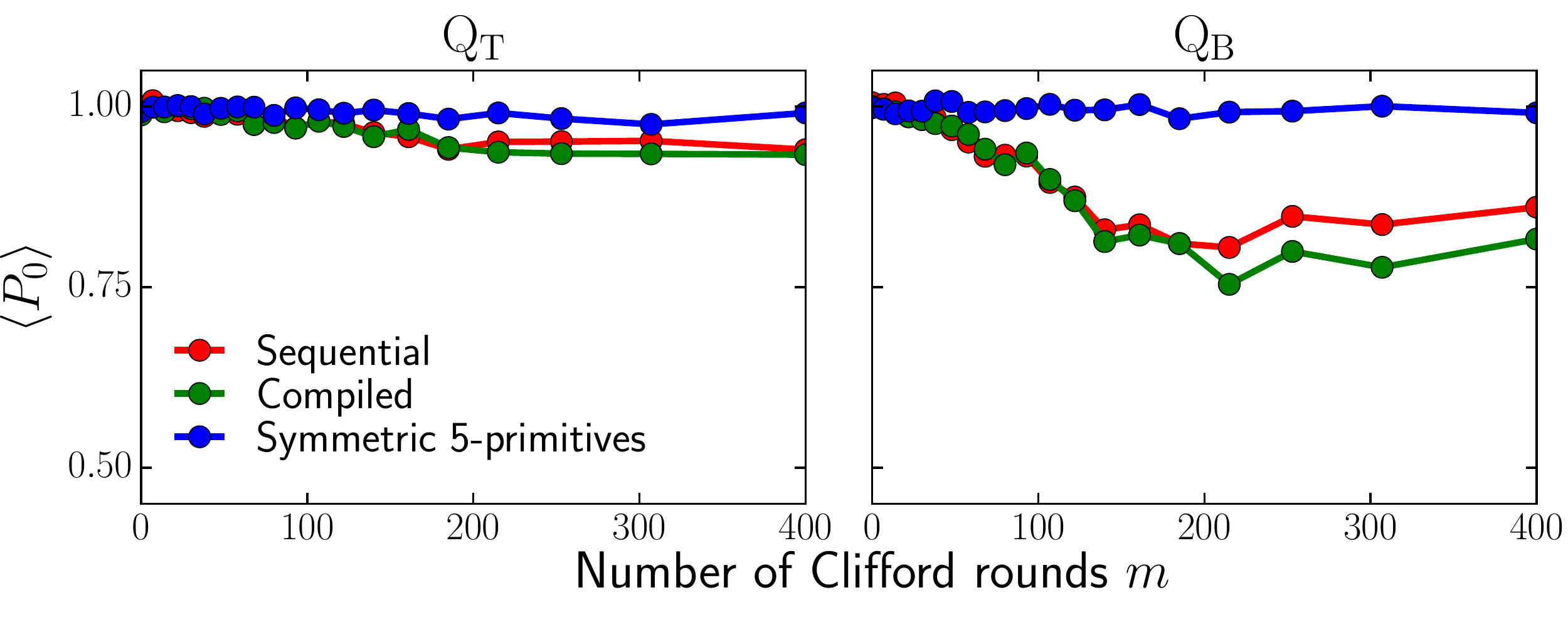}
      \caption{Measured cross-driving results when performing selective broadcasting.
      Even though the selective broadcasting schemes are meant for multiple qubits, the markers of the measured qubit are turned off to measure cross-driving effects.
      The cross-driving effects of $\QB$ are stronger than for $\QT$, in agreement with single-qubit results.
      With pulse decompositions whose cumulative effect largely cancels out, cross-driving is strongly reduced in the 5-primitives method.}
      \label{fig:cross-driving selective broadcasting}
    \end{figure}

    If a qubit is idle, every pulse that is applied to the driven qubit rotates the idle qubit by an amount depending on the cross-driving ratio.
    The random application of successive pulses to the driven qubit can therefore be viewed as a random walk for the idle qubit.
    As we will discuss in more detail in Sec.~\ref{sec:Compiled selective broadcasting algorithm}, there are many ways to compose a given Clifford gate from a small set of standard rotations.
    By choosing the constituent pulses in such a way that their combined application largely cancels out, cross-driving effects can be greatly reduced.
    In the standard set of Clifford decompositions~\cite{Epstein14}, the decompositions involve a majority of pulses rotating in the positive direction, biasing the random walk and producing a pronounced net cross-driving effect.
    This effect can be countered by choosing decompositions which minimize the bias.
    We have implemented this in the 5-primitives scheme, by choosing the first three pulses, $\{X_{\pi/2}, Y_{\pi/2}, X_{\pi/2}\}$, to be positive rotations, and the last two, $\{X_{-\pi}, Y_{-\pi}\}$, to move in the negative direction.
    Even though the pulse subset that is applied depends on the Clifford chosen, the pulses still largely cancel out after applying many Cliffords.
    Furthermore, as the single-qubit Clifford operations form a group, the inverse of all Cliffords also form the Clifford group.
    The complete inverse of the five pulse primitives, $\{X_{\pi}, Y_{\pi}, X_{-\pi/2}, Y_{-\pi/2}, X_{-\pi/2}\}$, can therefore also generate each of the 24 Cliffords using an appropriate subset of the pulses.
    By alternating between the normal five-pulse primitives and the inverted five pulses, cross-driving effects can be further reduced.
    (In fact, we note that this exactly eliminates all cross-driving that occurs via leakage in the VSM, because all pulses are always present at that distribution stage.)
    We refer to this as the symmetric 5-primitives technique and this is the technique we implement in the main experiments described in Fig.~4 of the main text.
    Our simulations in Fig.~\ref{fig:simulations cross-driving} show that the asymmetric 5-primitives technique already dramatically reduces the effect of cross-driving, and in the case of isolated single-qubit control, cross-driving is effectively eliminated completely using the symmetric 5-primitives scheme.
    This is also confirmed by measurements of cross-driving for the three selective-broadcasting schemes (Fig.~\ref{fig:cross-driving selective broadcasting}), where we implement the symmetric 5-primitives technique.
    While we have only demonstrated this technique for the 5-primitives scheme, it could also be relatively straightforwardly applied to the sequential scheme, but the far better scaling of the 5-primitives scheme make it more interesting for scaling up to larger system sizes.

\newpage
\section{Clifford pulse decomposition}
  \label{sec:Clifford pulse decomposition}

   The decompositions of the 24 single-qubit Clifford gates into a minimal set of $\pi/2$ and $\pi$ pulses and into the 5-primitives scheme are shown in Table~\ref{tab:Clifford pulse decomposition}.

  \begingroup
  \squeezetable
  \begin{table}[h!]
    \begin{ruledtabular}
    \begin{tabular}{c l l l c c c c c }
      Clifford ID     & \multicolumn{3}{c}{Minimal set decomposition} & \multicolumn{5}{c}{5-primitives decomposition} \\
                      & First & Second & Third      & $X_{\pi/2}$ & $Y_{\pi/2}$ & $X_{\pi/2}$ & $X_{-\pi}$ & $Y_{-\pi}$ \\
      \hline
      1 & $ I$            &                 &                & 0  &  0  &  0  &  0  &  0 \\
      2 & $ Y_{\pi/2}$    &     $X_{\pi/2}$ &                & 0  &  1  &  1  &  0  &  0 \\
      3 & $ X_{-\pi/2}$   &   $Y_{-\pi/2}$  &                & 1  &  1  &  0  &  1  &  0 \\
      4 & $ X_\pi$        &                 &                & 0  &  0  &  0  &  1  &  0 \\
      5 & $ Y_{-\pi/2}$   &    $X_{-\pi/2}$ &                & 0  &  1  &  1  &  0  &  1 \\
      6 & $ X_{\pi/2}$    &   $Y_{-\pi/2}$  &                & 1  &  1  &  0  &  0  &  1 \\
      7 & $ Y_\pi$        &                 &                & 0  &  0  &  0  &  0  &  1 \\
      8 & $ Y_{-\pi/2}$   &    $X_{\pi/2}$  &                & 0  &  1  &  1  &  1  &  1 \\
      9 & $ X_{\pi/2}$    &   $Y_{\pi/2}$   &                & 1  &  1  &  0  &  0  &  0 \\
      10 & $ X_\pi$       &   $Y_\pi$       &                & 0  &  0  &  0  &  1  &  1 \\
      11 & $ Y_{\pi/2}$   &    $X_{-\pi/2}$ &                & 0  &  1  &  1  &  1  &  0 \\
      12 & $ X_{-\pi/2}$  &   $Y_{\pi/2}$   &                & 1  &  1  &  0  &  1  &  1 \\
      13 & $ Y_{\pi/2}$   &    $X_\pi$      &                & 0  &  1  &  0  &  1  &  0 \\
      14 & $ X_{-\pi/2}$  &                 &                & 0  &  0  &  1  &  1  &  0 \\
      15 & $ X_{\pi/2}$   &   $Y_{-\pi/2}$  &  $X_{-\pi/2}$  & 1  &  1  &  1  &  0  &  1 \\
      16 & $ Y_{-\pi/2}$  &                 &                & 0  &  1  &  0  &  0  &  1 \\
      17 & $ X_{\pi/2}$   &                 &                & 0  &  0  &  1  &  0  &  0 \\
      18 & $ X_{\pi/2}$   &   $Y_{\pi/2}$   &  $X_{\pi/2}$   & 1  &  1  &  1  &  0  &  0 \\
      19 & $ Y_{-\pi/2}$  &    $X_\pi$      &                & 0  &  1  &  0  &  1  &  1 \\
      20 & $ X_{\pi/2}$   &   $Y_\pi$       &                & 1  &  0  &  0  &  0  &  1 \\
      21 & $ X_{\pi/2}$   &   $Y_{-\pi/2}$  &  $X_{\pi/2}$   & 1  &  1  &  1  &  1  &  1 \\
      22 & $ Y_{\pi/2}$   &                 &                & 0  &  1  &  0  &  0  &  0 \\
      23 & $ X_{-\pi/2}$  &   $Y_\pi$       &                & 1  &  0  &  0  &  1  &  1 \\
      24 & $ X_{\pi/2}$   &   $Y_{\pi/2}$   &  $X_{-\pi/2}$  & 1  &  1  &  1  &  1  &  0 \\

    \end{tabular}
    \end{ruledtabular}
    \caption{Two decompositions of the 24 single-qubit Clifford gates. The first, taken from Ref.~\cite{Epstein14}, minimizes the number of
  $\pi/2$ and $\pi$ pulses around the $\pm x$  and $\pm y$ axes. The second is our decomposition into 5 primitives. Pulses are applied from left to right.}
    \label{tab:Clifford pulse decomposition}
  \end{table}
  \endgroup

\section{Compiled selective broadcasting algorithm}
  \label{sec:Compiled selective broadcasting algorithm}

  \subsection{Finding the optimal pulse sequence}
    \label{ssec:Finding the optimal pulse sequence}
    When using a selective broadcasting architecture to send pulse sequences to multiple qubits, pulses can be directed to any subset of the qubits, but distinct pulses may not be applied simultaneously.
    In compiled selective broadcasting, the total number of pulses required to implement single-qubit gates on all qubits of a multi-qubit system is minimized by searching all possible combinations of single-qubit Clifford decompositions and grouping together like pulses where possible.
    In this section, we introduce an algorithm for determining the shortest compiled pulse sequence implementing independent single-qubit Clifford gates on $n$ qubits.

    On average, there are approximately 38 distinct decompositions for each single-qubit Clifford gate, given the basis set of $X$ and $Y$ pulses: $\{I, X_\pi, Y_\pi, X_{\pm\pi/2}, Y_{\pm\pi/2}\}$, resulting in approximately $38^n$ different decompositions for a given $n$-qubit combination of Cliffords.
    Here, we only consider sequences of up to four pulses, because the 5-primitives decomposition identified in Table~\ref{tab:Clifford pulse decomposition} already provides a recipe for decomposing an arbitrary $n$-qubit Clifford combination into five pulses.
    We do not include trivial decompositions where sequential pulses cancel out.

    Given a particular choice of $n$ Cliffords $\left(C_{\alpha_1}, \dots, C_{\alpha_n}\right)$, where $\alpha_i$ is the Clifford ID for qubit $i$, we write a specific decomposition as $\left(\left( P_1^1, ..., P_{m_1}^1 \right) , ..., \left(P_1^n, ..., P_{m_n}^n\right)\right)$, where $P_j^i$ is the $j$th of $m_i$ pulses which implement $C_{\alpha_i}$.
    While this already fixes the order in which pulses must be applied to individual qubits, we still have the freedom to choose in which order the distinct pulses are applied to different qubits.
    For each possible decomposition, we use the following recursive algorithm to search and minimize over all possible pulse orderings.

    We first define an empty broadcasting sequence $\bm{P}_\text{seq}$ to store the compiled multi-qubit pulse sequence.
    In order to convert from parallel single-qubit pulse sequences to the single broadcasting sequence $\bm{P}_\text{seq}$, we define a vector of indices $\bm{\beta}=\left(\beta_1, \dots, \beta_n\right)$ to store the current position in each single-qubit sequence.
    Initially, $\bm{\beta} = \left(1, ..., 1\right)$.
    At each instant, $\bm{P_\bm{\beta}}=( P_{\beta_1}^1, \dots, P_{\beta_n}^n )$ contains the next pulses to be applied to each qubit.
    When $\beta_i=m_i + 1$, the pulse sequence for that qubit is completed, and so there is no $P_{\beta_i}$ to be added to $\bm{P_\bm{\beta}}$.
    The recursive part of the algorithm then proceeds as follows:
    \begin{enumerate}
      \item Define $\bm{P}^\prime_{\bm{\beta}}$ to be the set of distinct pulses in $\bm{P_\bm{\beta}}$.
      \begin{description}
        \item[If:] $\bm{P}^\prime_{\bm{\beta}}$ is empty, store the number of pulses in $\bm{P}_\text{seq}$ and abort this recursion branch ($\bm{P}_\text{seq}$ is a completed pulse sequence such that all Cliffords are applied to the corresponding qubits).
        \item[Else:] Continue.
      \end{description}
      \item For each pulse $P$ in $\bm{P}^\prime_{\bm{\beta}}$, perform the following steps:
      \begin{enumerate}[(i)]
        \item Append $P$ to $\bm{P}_\text{seq}$;
        \item Copy indices $\bm{\beta}$ to $\bm{\beta}^\text{new}=\left(\beta_1^\text{new}, ..., \beta_n^\text{new}\right)$;
        \item For all indices $i$ for which $P_{\beta_i}^i=P$, increase the index $\beta_i^\text{new} = \beta_i+1$;
        \item Recursively loop to step 1 using the new indices $\bm{\beta}^\text{new}$ for $\bm{\beta}$.
      \end{enumerate}
    \end{enumerate}
    After considering all possible pulse sequences and looping over all possible decompositions, choose the sequence with the minimum number of pulses $N_\mathrm{P}$ in $\bm{P}_\text{seq}$.

    This algorithm determines the minimum number of pulses $N_\mathrm{P}$ required to implement a given $n$-qubit combination of single-qubit Clifford gates.
    However, this becomes prohibitively resource intensive as the number of qubits increases.
    It is therefore important to assess how the performance of the optimal compiled decomposition compares with the 5-primitives decomposition, which can be applied to any number of qubits without any extra overhead in resources (in neither calculation time nor sequence length).
    To do this, we use the average number of pulses $\left<N_\mathrm{P}\right>$ required per $n$-qubit combination of Cliffords (per Clifford).

    In the case of compiled selective broadcasting, finding $\left<N_\mathrm{P}\right>$ requires minimizing the sequence length for all $24^n$ possible Clifford combinations.
    This problem again scales exponentially with $n$.
    For example, for $n=5$ qubits, this requires $24^5\cdot38^5 \approx 6.3\cdot10^{14}$ repetitions of the complete recursive search described above.
    Nevertheless, by employing a number of optimizations desribed in the next section, we have exactly calculated $\left<N_\mathrm{P}\right>$ for $1\leq n \leq 5$ qubits in under 2 hours.
    Using random sampling (finding the shortest pulse sequence for a random sample of Clifford combinations), we also approximated $\left<N_\mathrm{P}\right>$ for $5\leq n \leq 10$.
    Exact and approximate results for $n=5$ agree.
    As shown in Table~\ref{tab:pulses per Clifford}, the improvement offered by compiled selective broadcasting over the 5-primitives method is already less than one pulse per Clifford and continues to decrease rapidly.
    Considering how badly the resource overhead scales for increasing numbers of qubits for finding a compiled sequence, it is questionable whether compiling offers any significant benefit over using the prescriptive 5-primitives approach when scaling up to larger system sizes.

    \begin{table}
      \begin{ruledtabular}
      \begin{tabular}{c c || c c}
        \multicolumn{2}{l||}{Exact calculation} & \multicolumn{2}{l}{Random sampling}\\
        $n$  & $\left<N_\mathrm{P}\right>$ & $n$  & $\left<N_\mathrm{P}\right>$ \\
        \hline
        1 & 1.875 & 5  & 4.139 (2) \\
        2 & 2.925 & 6  & 4.380 (12)\\
        3 & 3.521 & 7  & 4.570 (15)\\
        4 & 3.874 & 8  & 4.721 (10)\\
        5 & 4.137 & 9  & 4.808 (14)\\
        &         & 10 & 4.857 (24)\\
      \end{tabular}
      \end{ruledtabular}
      \caption{The average number of pulses $\left<N_\mathrm{P}\right>$ required to perform one Clifford gate on each of $n$ qubits in compiled selective broadcasting. Exact values were obtained for $1\leq n \leq5$  in under two hours, using the outlined optimizations. Approximate values were obtained for $5\leq n\leq 10$ using random sampling. These results are plotted in Fig.~4 of main text.}
      \label{tab:pulses per Clifford}
    \end{table}

\newpage
  \subsection{Optimizations for the Clifford compilation algorithm}
    \label{ssec:Optimizations for the Clifford compilation algorithm}

    In the first optimization, we place an upper bound $N^\text{ub}$ on the pulse sequence length. The upper bound $N^\text{ub}$ is given by the minimum number of pulses found so far that can compile a given Clifford combination. At each stage, the algorithm checks if the sum of the pulses in $P_\text{seq}$ and all distinct pulses left is equal to or greater than $N^\text{ub}$. If this is the case, a shorter combination of pulses using $P_\text{seq}$ is not possible, so we stop considering this sequence and proceed to the next one. Initially $N^\text{ub}=5$, as the $5$-primitives method proves that there is always a decomposition of an arbitrary number of Cliffords into $5$ pulses. Note that, as the limit $N^\text{ub}$ moves down, the frequency at which the algorithm stops considering sequences increases.

    The second optimization relies on decompositions with fewer pulses being more likely to result in an optimal Clifford compilation. The decompositions of every Clifford are therefore arranged in ascending number of pulses. The first decompositions compared are then those with the minimum number of pulses; these have the highest probability of finding an optimal Clifford compilation. Even if an optimal Clifford compilation is not found, it is more likely that $N^\text{ub}$ will be low. This optimization is especially effective in combination with the first optimization.

    The third optimization places a lower bound $N^\text{lb}$ on the number of pulses. For a given Clifford combination $\left(C_{\alpha_1}, \dots, C_{\alpha_n}\right)$, $N^\text{lb}$ is found by looking at the minimum number of pulses $N_\mathrm{P}$ previously found for all $n-1$ Clifford subsets. Since $N_\mathrm{P}$ for the $n$ Cliffords can never be less than $N_\mathrm{P}$ for any of the $n-1$ Clifford subsets, the maximum length of the $n-1$ Clifford subsets therefore places a lower bound $N^\text{lb}$ on $N_\mathrm{P}$ for the $n$ Cliffords. This means that if a pulse sequence is found whose length is equal to $N^\text{lb}$, it is an optimal Clifford compilation, and all further search is aborted. This is in contrast to the first optimization, where only the particular sequence of pulses is aborted upon reaching $N^\text{ub}$. Furthermore, as $n$ increases, it becomes increasingly likely that the lower bound is 5. In this case, $N^\text{lb}=N^\text{ub}$, and so the $5$-primitives method is an optimal Clifford compilation. This optimization results in the largest gain in computation time, by several orders of magnitude.

    In the fourth and most complicated optimization, all decompositions composed of three pulses or less are separated from those composed of four pulses. First, all combinations of Clifford decompositions composed of three pulses or less are compared. This reduces the average number of decompositions per Clifford, from $38$ to $7$, resulting in an exponentially reduced number of total decomposition combinations. It is, however, not always the case that the optimal Clifford compilation is found using only up to three pulses per decomposition; sometimes optimal Clifford compilations requires that one of the decompositions is composed of four pulses. However, after comparing decompositions of three pulses or less, these four-pulse decompositions only need to be considered when $N^\text{lb} \leq 4$ and $N^\text{ub} =5$. If there is a sequence containing a four-pulse decomposition that outperforms any found using up to three-pulse decompositions and the $5$-primitives method, the sequence must consist of four pulses. Only one Clifford then has a four-pulse decomposition, while all other Cliffords are subsets of these four pulses. We therefore loop, for every Clifford, over each of the four-pulse decompositions, and test whether every other Cliffords can be decomposed into a subset of these four pulses. This changes the comparison of four-pulse decompositions from scaling exponentially with $n$ to scaling linearly.

    The fifth and final optimization is only of use when all different Clifford combinations need to be considered to determine $\left<N_\mathrm{P}\right>$. It stems from the observation that an optimal compilation for a certain Clifford combination $\left(C_{\alpha_1}, \dots, C_{\alpha_n}\right)$ is the same as for any permutation of those Cliffords. We therefore only determine an optimal Clifford compilation when $\beta_1 \leq \dots \leq \beta_n$. This  reduces the number of calculations exponentially ($81$ times fewer computations for $n=5$).

\putbib
\end{bibunit}


\begin{thebibliography}{39}%
\makeatletter
\providecommand \@ifxundefined [1]{%
 \@ifx{#1\undefined}
}%
\providecommand \@ifnum [1]{%
 \ifnum #1\expandafter \@firstoftwo
 \else \expandafter \@secondoftwo
 \fi
}%
\providecommand \@ifx [1]{%
 \ifx #1\expandafter \@firstoftwo
 \else \expandafter \@secondoftwo
 \fi
}%
\providecommand \natexlab [1]{#1}%
\providecommand \enquote  [1]{``#1''}%
\providecommand \bibnamefont  [1]{#1}%
\providecommand \bibfnamefont [1]{#1}%
\providecommand \citenamefont [1]{#1}%
\providecommand \href@noop [0]{\@secondoftwo}%
\providecommand \href [0]{\begingroup \@sanitize@url \@href}%
\providecommand \@href[1]{\@@startlink{#1}\@@href}%
\providecommand \@@href[1]{\endgroup#1\@@endlink}%
\providecommand \@sanitize@url [0]{\catcode `\\12\catcode `\$12\catcode
  `\&12\catcode `\#12\catcode `\^12\catcode `\_12\catcode `\%12\relax}%
\providecommand \@@startlink[1]{}%
\providecommand \@@endlink[0]{}%
\providecommand \url  [0]{\begingroup\@sanitize@url \@url }%
\providecommand \@url [1]{\endgroup\@href {#1}{\urlprefix }}%
\providecommand \urlprefix  [0]{URL }%
\providecommand \Eprint [0]{\href }%
\providecommand \doibase [0]{http://dx.doi.org/}%
\providecommand \selectlanguage [0]{\@gobble}%
\providecommand \bibinfo  [0]{\@secondoftwo}%
\providecommand \bibfield  [0]{\@secondoftwo}%
\providecommand \translation [1]{[#1]}%
\providecommand \BibitemOpen [0]{}%
\providecommand \bibitemStop [0]{}%
\providecommand \bibitemNoStop [0]{.\EOS\space}%
\providecommand \EOS [0]{\spacefactor3000\relax}%
\providecommand \BibitemShut  [1]{\csname bibitem#1\endcsname}%
\let\auto@bib@innerbib\@empty
%</preamble>
\bibitem [{\citenamefont {Monroe}\ and\ \citenamefont {Kim}(2013)}]{Monroe13}%
  \BibitemOpen
  \bibfield  {author} {\bibinfo {author} {\bibfnamefont {C.}~\bibnamefont
  {Monroe}}\ and\ \bibinfo {author} {\bibfnamefont {J.}~\bibnamefont {Kim}},\
  }\href@noop {} {\bibfield  {journal} {\bibinfo  {journal} {Science}\ }\textbf
  {\bibinfo {volume} {339}},\ \bibinfo {pages} {1164} (\bibinfo {year}
  {2013})}\BibitemShut {NoStop}%
\bibitem [{\citenamefont {Awschalom}\ \emph {et~al.}(2013)\citenamefont
  {Awschalom}, \citenamefont {Bassett}, \citenamefont {Dzurak}, \citenamefont
  {Hu},\ and\ \citenamefont {Petta}}]{Awschalom13}%
  \BibitemOpen
  \bibfield  {author} {\bibinfo {author} {\bibfnamefont {D.~D.}\ \bibnamefont
  {Awschalom}}, \bibinfo {author} {\bibfnamefont {L.~C.}\ \bibnamefont
  {Bassett}}, \bibinfo {author} {\bibfnamefont {A.~S.}\ \bibnamefont {Dzurak}},
  \bibinfo {author} {\bibfnamefont {E.~L.}\ \bibnamefont {Hu}}, \ and\ \bibinfo
  {author} {\bibfnamefont {J.~R.}\ \bibnamefont {Petta}},\ }\href@noop {}
  {\bibfield  {journal} {\bibinfo  {journal} {Science}\ }\textbf {\bibinfo
  {volume} {339}},\ \bibinfo {pages} {1174} (\bibinfo {year}
  {2013})}\BibitemShut {NoStop}%
\bibitem [{\citenamefont {Devoret}\ and\ \citenamefont
  {Schoelkopf}(2013)}]{Devoret13}%
  \BibitemOpen
  \bibfield  {author} {\bibinfo {author} {\bibfnamefont {M.~H.}\ \bibnamefont
  {Devoret}}\ and\ \bibinfo {author} {\bibfnamefont {R.~J.}\ \bibnamefont
  {Schoelkopf}},\ }\href@noop {} {\bibfield  {journal} {\bibinfo  {journal}
  {Science}\ }\textbf {\bibinfo {volume} {339}},\ \bibinfo {pages} {1169}
  (\bibinfo {year} {2013})}\BibitemShut {NoStop}%
\bibitem [{\citenamefont {Kelly}\ \emph {et~al.}(2015)\citenamefont {Kelly},
  \citenamefont {Barends}, \citenamefont {Fowler}, \citenamefont {Megrant},
  \citenamefont {Jeffrey}, \citenamefont {White}, \citenamefont {Sank},
  \citenamefont {Mutus}, \citenamefont {Campbell}, \citenamefont {Chen} \emph
  {et~al.}}]{Kelly15}%
  \BibitemOpen
  \bibfield  {author} {\bibinfo {author} {\bibfnamefont {J.}~\bibnamefont
  {Kelly}}, \bibinfo {author} {\bibfnamefont {R.}~\bibnamefont {Barends}},
  \bibinfo {author} {\bibfnamefont {A.}~\bibnamefont {Fowler}}, \bibinfo
  {author} {\bibfnamefont {A.}~\bibnamefont {Megrant}}, \bibinfo {author}
  {\bibfnamefont {E.}~\bibnamefont {Jeffrey}}, \bibinfo {author} {\bibfnamefont
  {T.}~\bibnamefont {White}}, \bibinfo {author} {\bibfnamefont
  {D.}~\bibnamefont {Sank}}, \bibinfo {author} {\bibfnamefont {J.}~\bibnamefont
  {Mutus}}, \bibinfo {author} {\bibfnamefont {B.}~\bibnamefont {Campbell}},
  \bibinfo {author} {\bibfnamefont {Y.}~\bibnamefont {Chen}},  \emph {et~al.},\
  }\href@noop {} {\bibfield  {journal} {\bibinfo  {journal} {Nature}\ }\textbf
  {\bibinfo {volume} {519}},\ \bibinfo {pages} {66} (\bibinfo {year}
  {2015})}\BibitemShut {NoStop}%
\bibitem [{\citenamefont {Corcoles}\ \emph {et~al.}(2015)\citenamefont
  {Corcoles}, \citenamefont {Magesan}, \citenamefont {Srinivasan},
  \citenamefont {Cross}, \citenamefont {Steffen}, \citenamefont {Gambetta},\
  and\ \citenamefont {Chow}}]{Corcoles15}%
  \BibitemOpen
  \bibfield  {author} {\bibinfo {author} {\bibfnamefont {A.~D.}\ \bibnamefont
  {Corcoles}}, \bibinfo {author} {\bibfnamefont {E.}~\bibnamefont {Magesan}},
  \bibinfo {author} {\bibfnamefont {S.~J.}\ \bibnamefont {Srinivasan}},
  \bibinfo {author} {\bibfnamefont {A.~W.}\ \bibnamefont {Cross}}, \bibinfo
  {author} {\bibfnamefont {M.}~\bibnamefont {Steffen}}, \bibinfo {author}
  {\bibfnamefont {J.~M.}\ \bibnamefont {Gambetta}}, \ and\ \bibinfo {author}
  {\bibfnamefont {J.~M.}\ \bibnamefont {Chow}},\ }\href@noop {} {\bibfield
  {journal} {\bibinfo  {journal} {Nat.\ Commun.}\ }\textbf {\bibinfo {volume}
  {{6}}} (\bibinfo {year} {{2015}})}\BibitemShut {NoStop}%
\bibitem [{\citenamefont {Rist\`{e}}\ \emph {et~al.}(2015)\citenamefont
  {Rist\`{e}}, \citenamefont {Poletto}, \citenamefont {Huang}, \citenamefont
  {Bruno}, \citenamefont {Vesterinen}, \citenamefont {Saira},\ and\
  \citenamefont {DiCarlo}}]{Riste15}%
  \BibitemOpen
  \bibfield  {author} {\bibinfo {author} {\bibfnamefont {D.}~\bibnamefont
  {Rist\`{e}}}, \bibinfo {author} {\bibfnamefont {S.}~\bibnamefont {Poletto}},
  \bibinfo {author} {\bibfnamefont {M.~Z.}\ \bibnamefont {Huang}}, \bibinfo
  {author} {\bibfnamefont {A.}~\bibnamefont {Bruno}}, \bibinfo {author}
  {\bibfnamefont {V.}~\bibnamefont {Vesterinen}}, \bibinfo {author}
  {\bibfnamefont {O.~P.}\ \bibnamefont {Saira}}, \ and\ \bibinfo {author}
  {\bibfnamefont {L.}~\bibnamefont {DiCarlo}},\ }\href@noop {} {\bibfield
  {journal} {\bibinfo  {journal} {Nat.\ Commun.}\ }\textbf {\bibinfo {volume}
  {{6}}} (\bibinfo {year} {{2015}})}\BibitemShut {NoStop}%
\bibitem [{\citenamefont {Hornibrook}\ \emph {et~al.}(2015)\citenamefont
  {Hornibrook}, \citenamefont {Colless}, \citenamefont {Conway~Lamb},
  \citenamefont {Pauka}, \citenamefont {Lu}, \citenamefont {Gossard},
  \citenamefont {Watson}, \citenamefont {Gardner}, \citenamefont {Fallahi},
  \citenamefont {Manfra},\ and\ \citenamefont {Reilly}}]{Hornibrook15}%
  \BibitemOpen
  \bibfield  {author} {\bibinfo {author} {\bibfnamefont {J.~M.}\ \bibnamefont
  {Hornibrook}}, \bibinfo {author} {\bibfnamefont {J.~I.}\ \bibnamefont
  {Colless}}, \bibinfo {author} {\bibfnamefont {I.~D.}\ \bibnamefont
  {Conway~Lamb}}, \bibinfo {author} {\bibfnamefont {S.~J.}\ \bibnamefont
  {Pauka}}, \bibinfo {author} {\bibfnamefont {H.}~\bibnamefont {Lu}}, \bibinfo
  {author} {\bibfnamefont {A.~C.}\ \bibnamefont {Gossard}}, \bibinfo {author}
  {\bibfnamefont {J.~D.}\ \bibnamefont {Watson}}, \bibinfo {author}
  {\bibfnamefont {G.~C.}\ \bibnamefont {Gardner}}, \bibinfo {author}
  {\bibfnamefont {S.}~\bibnamefont {Fallahi}}, \bibinfo {author} {\bibfnamefont
  {M.~J.}\ \bibnamefont {Manfra}}, \ and\ \bibinfo {author} {\bibfnamefont
  {D.~J.}\ \bibnamefont {Reilly}},\ }\href@noop {} {\bibfield  {journal}
  {\bibinfo  {journal} {Phys. Rev. Appl.}\ }\textbf {\bibinfo {volume} {3}},\
  \bibinfo {pages} {024010} (\bibinfo {year} {2015})}\BibitemShut {NoStop}%
\bibitem [{\citenamefont {Knoernschild}\ \emph {et~al.}(2010)\citenamefont
  {Knoernschild}, \citenamefont {Zhang}, \citenamefont {Isenhower},
  \citenamefont {Gill}, \citenamefont {Lu}, \citenamefont {Saffman},\ and\
  \citenamefont {Kim}}]{Knoernschild10}%
  \BibitemOpen
  \bibfield  {author} {\bibinfo {author} {\bibfnamefont {C.}~\bibnamefont
  {Knoernschild}}, \bibinfo {author} {\bibfnamefont {X.~L.}\ \bibnamefont
  {Zhang}}, \bibinfo {author} {\bibfnamefont {L.}~\bibnamefont {Isenhower}},
  \bibinfo {author} {\bibfnamefont {A.~T.}\ \bibnamefont {Gill}}, \bibinfo
  {author} {\bibfnamefont {F.~P.}\ \bibnamefont {Lu}}, \bibinfo {author}
  {\bibfnamefont {M.}~\bibnamefont {Saffman}}, \ and\ \bibinfo {author}
  {\bibfnamefont {J.}~\bibnamefont {Kim}},\ }\href@noop {} {\bibfield
  {journal} {\bibinfo  {journal} {Appl. Phys. Lett.}\ }\textbf {\bibinfo
  {volume} {97}},\ \bibinfo {eid} {134101} (\bibinfo {year}
  {2010})}\BibitemShut {NoStop}%
\bibitem [{\citenamefont {Weitenberg}\ \emph {et~al.}(2011)\citenamefont
  {Weitenberg}, \citenamefont {Endres}, \citenamefont {Sherson}, \citenamefont
  {Cheneau}, \citenamefont {Schauss}, \citenamefont {Fukuhara}, \citenamefont
  {Bloch},\ and\ \citenamefont {Kuhr}}]{Weitenberg11}%
  \BibitemOpen
  \bibfield  {author} {\bibinfo {author} {\bibfnamefont {C.}~\bibnamefont
  {Weitenberg}}, \bibinfo {author} {\bibfnamefont {M.}~\bibnamefont {Endres}},
  \bibinfo {author} {\bibfnamefont {J.~F.}\ \bibnamefont {Sherson}}, \bibinfo
  {author} {\bibfnamefont {M.}~\bibnamefont {Cheneau}}, \bibinfo {author}
  {\bibfnamefont {P.}~\bibnamefont {Schauss}}, \bibinfo {author} {\bibfnamefont
  {T.}~\bibnamefont {Fukuhara}}, \bibinfo {author} {\bibfnamefont
  {I.}~\bibnamefont {Bloch}}, \ and\ \bibinfo {author} {\bibfnamefont
  {S.}~\bibnamefont {Kuhr}},\ }\href@noop {} {\bibfield  {journal} {\bibinfo
  {journal} {Nature}\ }\textbf {\bibinfo {volume} {471}},\ \bibinfo {pages}
  {319} (\bibinfo {year} {2011})}\BibitemShut {NoStop}%
\bibitem [{\citenamefont {Crain}\ \emph {et~al.}(2014)\citenamefont {Crain},
  \citenamefont {Mount}, \citenamefont {Baek},\ and\ \citenamefont
  {Kim}}]{Crain14}%
  \BibitemOpen
  \bibfield  {author} {\bibinfo {author} {\bibfnamefont {S.}~\bibnamefont
  {Crain}}, \bibinfo {author} {\bibfnamefont {E.}~\bibnamefont {Mount}},
  \bibinfo {author} {\bibfnamefont {S.}~\bibnamefont {Baek}}, \ and\ \bibinfo
  {author} {\bibfnamefont {J.}~\bibnamefont {Kim}},\ }\href@noop {} {\bibfield
  {journal} {\bibinfo  {journal} {Appl. Phys. Lett.}\ }\textbf {\bibinfo
  {volume} {105}},\ \bibinfo {eid} {181115} (\bibinfo {year}
  {2014})}\BibitemShut {NoStop}%
\bibitem [{\citenamefont {Xia}\ \emph {et~al.}(2015)\citenamefont {Xia},
  \citenamefont {Lichtman}, \citenamefont {Maller}, \citenamefont {Carr},
  \citenamefont {Piotrowicz}, \citenamefont {Isenhower},\ and\ \citenamefont
  {Saffman}}]{Xia15}%
  \BibitemOpen
  \bibfield  {author} {\bibinfo {author} {\bibfnamefont {T.}~\bibnamefont
  {Xia}}, \bibinfo {author} {\bibfnamefont {M.}~\bibnamefont {Lichtman}},
  \bibinfo {author} {\bibfnamefont {K.}~\bibnamefont {Maller}}, \bibinfo
  {author} {\bibfnamefont {A.~W.}\ \bibnamefont {Carr}}, \bibinfo {author}
  {\bibfnamefont {M.~J.}\ \bibnamefont {Piotrowicz}}, \bibinfo {author}
  {\bibfnamefont {L.}~\bibnamefont {Isenhower}}, \ and\ \bibinfo {author}
  {\bibfnamefont {M.}~\bibnamefont {Saffman}},\ }\href@noop {} {\bibfield
  {journal} {\bibinfo  {journal} {Phys. Rev. Lett.}\ }\textbf {\bibinfo
  {volume} {114}},\ \bibinfo {pages} {100503} (\bibinfo {year}
  {2015})}\BibitemShut {NoStop}%
\bibitem [{\citenamefont {Blais}\ \emph {et~al.}(2004)\citenamefont {Blais},
  \citenamefont {Huang}, \citenamefont {Wallraff}, \citenamefont {Girvin},\
  and\ \citenamefont {Schoelkopf}}]{Blais04}%
  \BibitemOpen
  \bibfield  {author} {\bibinfo {author} {\bibfnamefont {A.}~\bibnamefont
  {Blais}}, \bibinfo {author} {\bibfnamefont {R.-S.}\ \bibnamefont {Huang}},
  \bibinfo {author} {\bibfnamefont {A.}~\bibnamefont {Wallraff}}, \bibinfo
  {author} {\bibfnamefont {S.~M.}\ \bibnamefont {Girvin}}, \ and\ \bibinfo
  {author} {\bibfnamefont {R.~J.}\ \bibnamefont {Schoelkopf}},\ }\href@noop {}
  {\bibfield  {journal} {\bibinfo  {journal} {Phys. Rev. A}\ }\textbf {\bibinfo
  {volume} {69}},\ \bibinfo {pages} {062320} (\bibinfo {year}
  {2004})}\BibitemShut {NoStop}%
\bibitem [{\citenamefont {DiVincenzo}(2009)}]{Divincenzo09}%
  \BibitemOpen
  \bibfield  {author} {\bibinfo {author} {\bibfnamefont {D.~P.}\ \bibnamefont
  {DiVincenzo}},\ }\href@noop {} {\bibfield  {journal} {\bibinfo  {journal}
  {Physica Scripta}\ }\textbf {\bibinfo {volume} {2009}},\ \bibinfo {pages}
  {014020} (\bibinfo {year} {2009})}\BibitemShut {NoStop}%
\bibitem [{\citenamefont {Helmer}\ \emph {et~al.}(2009)\citenamefont {Helmer},
  \citenamefont {Mariantoni}, \citenamefont {Fowler}, \citenamefont
  {Von~Delft}, \citenamefont {Solano},\ and\ \citenamefont
  {Marquardt}}]{Helmer09cavity}%
  \BibitemOpen
  \bibfield  {author} {\bibinfo {author} {\bibfnamefont {F.}~\bibnamefont
  {Helmer}}, \bibinfo {author} {\bibfnamefont {M.}~\bibnamefont {Mariantoni}},
  \bibinfo {author} {\bibfnamefont {A.}~\bibnamefont {Fowler}}, \bibinfo
  {author} {\bibfnamefont {J.}~\bibnamefont {Von~Delft}}, \bibinfo {author}
  {\bibfnamefont {E.}~\bibnamefont {Solano}}, \ and\ \bibinfo {author}
  {\bibfnamefont {F.}~\bibnamefont {Marquardt}},\ }\href@noop {} {\bibfield
  {journal} {\bibinfo  {journal} {Europhys. Lett.}\ }\textbf {\bibinfo {volume}
  {85}},\ \bibinfo {pages} {50007} (\bibinfo {year} {2009})}\BibitemShut
  {NoStop}%
\bibitem [{\citenamefont {Ghosh}\ \emph {et~al.}(2012)\citenamefont {Ghosh},
  \citenamefont {Fowler},\ and\ \citenamefont {Geller}}]{Ghosh12}%
  \BibitemOpen
  \bibfield  {author} {\bibinfo {author} {\bibfnamefont {J.}~\bibnamefont
  {Ghosh}}, \bibinfo {author} {\bibfnamefont {A.~G.}\ \bibnamefont {Fowler}}, \
  and\ \bibinfo {author} {\bibfnamefont {M.~R.}\ \bibnamefont {Geller}},\
  }\href@noop {} {\bibfield  {journal} {\bibinfo  {journal} {Phys. Rev. A}\
  }\textbf {\bibinfo {volume} {86}},\ \bibinfo {pages} {062318} (\bibinfo
  {year} {2012})}\BibitemShut {NoStop}%
\bibitem [{\citenamefont {Gambetta}\ and\ \citenamefont
  {Smolin}(2014)}]{Gambetta2014frequency}%
  \BibitemOpen
  \bibfield  {author} {\bibinfo {author} {\bibfnamefont {J.}~\bibnamefont
  {Gambetta}}\ and\ \bibinfo {author} {\bibfnamefont {J.}~\bibnamefont
  {Smolin}},\ }\href@noop {} {\enquote {\bibinfo {title} {Frequency arrangement
  for surface code on a superconducting lattice},}\ } (\bibinfo {year}
  {2014}),\ \bibinfo {note} {{US} Patent App. 13/827,326}\BibitemShut {NoStop}%
\bibitem [{\citenamefont {Schutjens}\ \emph {et~al.}(2013)\citenamefont
  {Schutjens}, \citenamefont {Dagga}, \citenamefont {Egger},\ and\
  \citenamefont {Wilhelm}}]{Schutjens13}%
  \BibitemOpen
  \bibfield  {author} {\bibinfo {author} {\bibfnamefont {R.}~\bibnamefont
  {Schutjens}}, \bibinfo {author} {\bibfnamefont {F.~A.}\ \bibnamefont
  {Dagga}}, \bibinfo {author} {\bibfnamefont {D.~J.}\ \bibnamefont {Egger}}, \
  and\ \bibinfo {author} {\bibfnamefont {F.~K.}\ \bibnamefont {Wilhelm}},\
  }\href@noop {} {\bibfield  {journal} {\bibinfo  {journal} {Phys. Rev. A}\
  }\textbf {\bibinfo {volume} {88}},\ \bibinfo {pages} {052330} (\bibinfo
  {year} {2013})}\BibitemShut {NoStop}%
\bibitem [{\citenamefont {Vesterinen}\ \emph {et~al.}(2014)\citenamefont
  {Vesterinen}, \citenamefont {Saira}, \citenamefont {Bruno},\ and\
  \citenamefont {DiCarlo}}]{Vesterinen14}%
  \BibitemOpen
  \bibfield  {author} {\bibinfo {author} {\bibfnamefont {V.}~\bibnamefont
  {Vesterinen}}, \bibinfo {author} {\bibfnamefont {O.-P.}\ \bibnamefont
  {Saira}}, \bibinfo {author} {\bibfnamefont {A.}~\bibnamefont {Bruno}}, \ and\
  \bibinfo {author} {\bibfnamefont {L.}~\bibnamefont {DiCarlo}},\ }\href@noop
  {} {\bibfield  {journal} {\bibinfo  {journal} {arXiv:cond-mat/1405.0450}\ }
  (\bibinfo {year} {2014})}\BibitemShut {NoStop}%
\bibitem [{\citenamefont {Bravyi}\ and\ \citenamefont
  {Kitaev}(1998)}]{Bravyi98}%
  \BibitemOpen
  \bibfield  {author} {\bibinfo {author} {\bibfnamefont {S.~B.}\ \bibnamefont
  {Bravyi}}\ and\ \bibinfo {author} {\bibfnamefont {A.~Y.}\ \bibnamefont
  {Kitaev}},\ }\href@noop {} {\bibfield  {journal} {\bibinfo  {journal}
  {arXiv:quant-ph/9811052}\ } (\bibinfo {year} {1998})}\BibitemShut {NoStop}%
\bibitem [{\citenamefont {Fowler}\ \emph {et~al.}(2012)\citenamefont {Fowler},
  \citenamefont {Mariantoni}, \citenamefont {Martinis},\ and\ \citenamefont
  {Cleland}}]{Fowler12}%
  \BibitemOpen
  \bibfield  {author} {\bibinfo {author} {\bibfnamefont {A.~G.}\ \bibnamefont
  {Fowler}}, \bibinfo {author} {\bibfnamefont {M.}~\bibnamefont {Mariantoni}},
  \bibinfo {author} {\bibfnamefont {J.~M.}\ \bibnamefont {Martinis}}, \ and\
  \bibinfo {author} {\bibfnamefont {A.~N.}\ \bibnamefont {Cleland}},\
  }\href@noop {} {\bibfield  {journal} {\bibinfo  {journal} {Phys. Rev. A}\
  }\textbf {\bibinfo {volume} {86}},\ \bibinfo {pages} {032324} (\bibinfo
  {year} {2012})}\BibitemShut {NoStop}%
\bibitem [{\citenamefont {Chasseur}\ and\ \citenamefont
  {Wilhelm}(2015)}]{chasseur15}%
  \BibitemOpen
  \bibfield  {author} {\bibinfo {author} {\bibfnamefont {T.}~\bibnamefont
  {Chasseur}}\ and\ \bibinfo {author} {\bibfnamefont {F.}~\bibnamefont
  {Wilhelm}},\ }\href@noop {} {\bibfield  {journal} {\bibinfo  {journal}
  {arXiv:1505.00580}\ } (\bibinfo {year} {2015})}\BibitemShut {NoStop}%
\bibitem [{\citenamefont {Epstein}\ \emph {et~al.}(2014)\citenamefont
  {Epstein}, \citenamefont {Cross}, \citenamefont {Magesan},\ and\
  \citenamefont {Gambetta}}]{Epstein14}%
  \BibitemOpen
  \bibfield  {author} {\bibinfo {author} {\bibfnamefont {J.~M.}\ \bibnamefont
  {Epstein}}, \bibinfo {author} {\bibfnamefont {A.~W.}\ \bibnamefont {Cross}},
  \bibinfo {author} {\bibfnamefont {E.}~\bibnamefont {Magesan}}, \ and\
  \bibinfo {author} {\bibfnamefont {J.~M.}\ \bibnamefont {Gambetta}},\
  }\href@noop {} {\bibfield  {journal} {\bibinfo  {journal} {Phys. Rev. A}\
  }\textbf {\bibinfo {volume} {89}},\ \bibinfo {pages} {062321} (\bibinfo
  {year} {2014})}\BibitemShut {NoStop}%
\bibitem [{\citenamefont {Fragner}\ \emph {et~al.}(2008)\citenamefont
  {Fragner}, \citenamefont {G\"{o}ppl}, \citenamefont {Fink}, \citenamefont
  {Baur}, \citenamefont {Bianchetti}, \citenamefont {Leek}, \citenamefont
  {Blais},\ and\ \citenamefont {Wallraff}}]{Fragner08}%
  \BibitemOpen
  \bibfield  {author} {\bibinfo {author} {\bibfnamefont {A.}~\bibnamefont
  {Fragner}}, \bibinfo {author} {\bibfnamefont {M.}~\bibnamefont {G\"{o}ppl}},
  \bibinfo {author} {\bibfnamefont {J.~M.}\ \bibnamefont {Fink}}, \bibinfo
  {author} {\bibfnamefont {M.}~\bibnamefont {Baur}}, \bibinfo {author}
  {\bibfnamefont {R.}~\bibnamefont {Bianchetti}}, \bibinfo {author}
  {\bibfnamefont {P.~J.}\ \bibnamefont {Leek}}, \bibinfo {author}
  {\bibfnamefont {A.}~\bibnamefont {Blais}}, \ and\ \bibinfo {author}
  {\bibfnamefont {A.}~\bibnamefont {Wallraff}},\ }\href@noop {} {\bibfield
  {journal} {\bibinfo  {journal} {Science}\ }\textbf {\bibinfo {volume}
  {322}},\ \bibinfo {pages} {1357} (\bibinfo {year} {2008})}\BibitemShut
  {NoStop}%
\bibitem [{\citenamefont {Groen}\ \emph {et~al.}(2013)\citenamefont {Groen},
  \citenamefont {Rist\`e}, \citenamefont {Tornberg}, \citenamefont {Cramer},
  \citenamefont {de~Groot}, \citenamefont {Picot}, \citenamefont {Johansson},\
  and\ \citenamefont {DiCarlo}}]{Groen13}%
  \BibitemOpen
  \bibfield  {author} {\bibinfo {author} {\bibfnamefont {J.~P.}\ \bibnamefont
  {Groen}}, \bibinfo {author} {\bibfnamefont {D.}~\bibnamefont {Rist\`e}},
  \bibinfo {author} {\bibfnamefont {L.}~\bibnamefont {Tornberg}}, \bibinfo
  {author} {\bibfnamefont {J.}~\bibnamefont {Cramer}}, \bibinfo {author}
  {\bibfnamefont {P.~C.}\ \bibnamefont {de~Groot}}, \bibinfo {author}
  {\bibfnamefont {T.}~\bibnamefont {Picot}}, \bibinfo {author} {\bibfnamefont
  {G.}~\bibnamefont {Johansson}}, \ and\ \bibinfo {author} {\bibfnamefont
  {L.}~\bibnamefont {DiCarlo}},\ }\href@noop {} {\bibfield  {journal} {\bibinfo
   {journal} {Phys. Rev. Lett.}\ }\textbf {\bibinfo {volume} {111}},\ \bibinfo
  {pages} {090506} (\bibinfo {year} {2013})}\BibitemShut {NoStop}%
\bibitem [{\citenamefont {Jerger}\ \emph {et~al.}(2012)\citenamefont {Jerger},
  \citenamefont {Poletto}, \citenamefont {Macha}, \citenamefont {H{\"u}bner},
  \citenamefont {Il'ichev},\ and\ \citenamefont {Ustinov}}]{Jerger12}%
  \BibitemOpen
  \bibfield  {author} {\bibinfo {author} {\bibfnamefont {M.}~\bibnamefont
  {Jerger}}, \bibinfo {author} {\bibfnamefont {S.}~\bibnamefont {Poletto}},
  \bibinfo {author} {\bibfnamefont {P.}~\bibnamefont {Macha}}, \bibinfo
  {author} {\bibfnamefont {U.}~\bibnamefont {H{\"u}bner}}, \bibinfo {author}
  {\bibfnamefont {E.}~\bibnamefont {Il'ichev}}, \ and\ \bibinfo {author}
  {\bibfnamefont {A.~V.}\ \bibnamefont {Ustinov}},\ }\href@noop {} {\bibfield
  {journal} {\bibinfo  {journal} {Appl. Phys. Lett.}\ }\textbf {\bibinfo
  {volume} {101}},\ \bibinfo {pages} {042604} (\bibinfo {year}
  {2012})}\BibitemShut {NoStop}%
\bibitem [{SOM()}]{SOMprappl}%
  \BibitemOpen
  \href@noop {} {}\bibinfo {howpublished} {See supplemental material below for
  additional data.}\BibitemShut {Stop}%
\bibitem [{\citenamefont {Koch}\ \emph {et~al.}(2007)\citenamefont {Koch},
  \citenamefont {Yu}, \citenamefont {Gambetta}, \citenamefont {Houck},
  \citenamefont {Schuster}, \citenamefont {Majer}, \citenamefont {Blais},
  \citenamefont {Devoret}, \citenamefont {Girvin},\ and\ \citenamefont
  {Schoelkopf}}]{Koch07}%
  \BibitemOpen
  \bibfield  {author} {\bibinfo {author} {\bibfnamefont {J.}~\bibnamefont
  {Koch}}, \bibinfo {author} {\bibfnamefont {T.~M.}\ \bibnamefont {Yu}},
  \bibinfo {author} {\bibfnamefont {J.}~\bibnamefont {Gambetta}}, \bibinfo
  {author} {\bibfnamefont {A.~A.}\ \bibnamefont {Houck}}, \bibinfo {author}
  {\bibfnamefont {D.~I.}\ \bibnamefont {Schuster}}, \bibinfo {author}
  {\bibfnamefont {J.}~\bibnamefont {Majer}}, \bibinfo {author} {\bibfnamefont
  {A.}~\bibnamefont {Blais}}, \bibinfo {author} {\bibfnamefont {M.~H.}\
  \bibnamefont {Devoret}}, \bibinfo {author} {\bibfnamefont {S.~M.}\
  \bibnamefont {Girvin}}, \ and\ \bibinfo {author} {\bibfnamefont {R.~J.}\
  \bibnamefont {Schoelkopf}},\ }\href@noop {} {\bibfield  {journal} {\bibinfo
  {journal} {Phys. Rev. A}\ }\textbf {\bibinfo {volume} {76}},\ \bibinfo
  {pages} {042319} (\bibinfo {year} {2007})}\BibitemShut {NoStop}%
\bibitem [{\citenamefont {Motzoi}\ \emph {et~al.}(2009)\citenamefont {Motzoi},
  \citenamefont {Gambetta}, \citenamefont {Rebentrost},\ and\ \citenamefont
  {Wilhelm}}]{Motzoi09}%
  \BibitemOpen
  \bibfield  {author} {\bibinfo {author} {\bibfnamefont {F.}~\bibnamefont
  {Motzoi}}, \bibinfo {author} {\bibfnamefont {J.~M.}\ \bibnamefont
  {Gambetta}}, \bibinfo {author} {\bibfnamefont {P.}~\bibnamefont
  {Rebentrost}}, \ and\ \bibinfo {author} {\bibfnamefont {F.~K.}\ \bibnamefont
  {Wilhelm}},\ }\href@noop {} {\bibfield  {journal} {\bibinfo  {journal} {Phys.
  Rev. Lett.}\ }\textbf {\bibinfo {volume} {103}},\ \bibinfo {pages} {110501}
  (\bibinfo {year} {2009})}\BibitemShut {NoStop}%
\bibitem [{\citenamefont {Chow}\ \emph {et~al.}(2010)\citenamefont {Chow},
  \citenamefont {DiCarlo}, \citenamefont {Gambetta}, \citenamefont {Motzoi},
  \citenamefont {Frunzio}, \citenamefont {Girvin},\ and\ \citenamefont
  {Schoelkopf}}]{Chow10b}%
  \BibitemOpen
  \bibfield  {author} {\bibinfo {author} {\bibfnamefont {J.~M.}\ \bibnamefont
  {Chow}}, \bibinfo {author} {\bibfnamefont {L.}~\bibnamefont {DiCarlo}},
  \bibinfo {author} {\bibfnamefont {J.~M.}\ \bibnamefont {Gambetta}}, \bibinfo
  {author} {\bibfnamefont {F.}~\bibnamefont {Motzoi}}, \bibinfo {author}
  {\bibfnamefont {L.}~\bibnamefont {Frunzio}}, \bibinfo {author} {\bibfnamefont
  {S.~M.}\ \bibnamefont {Girvin}}, \ and\ \bibinfo {author} {\bibfnamefont
  {R.~J.}\ \bibnamefont {Schoelkopf}},\ }\href@noop {} {\bibfield  {journal}
  {\bibinfo  {journal} {Phys. Rev. A}\ }\textbf {\bibinfo {volume} {82}},\
  \bibinfo {pages} {040305} (\bibinfo {year} {2010})}\BibitemShut {NoStop}%
\bibitem [{\citenamefont {Knill}\ \emph {et~al.}(2008)\citenamefont {Knill},
  \citenamefont {Leibfried}, \citenamefont {Reichle}, \citenamefont {Britton},
  \citenamefont {Blakestad}, \citenamefont {Jost}, \citenamefont {Langer},
  \citenamefont {Ozeri}, \citenamefont {Seidelin},\ and\ \citenamefont
  {Wineland}}]{Knill08}%
  \BibitemOpen
  \bibfield  {author} {\bibinfo {author} {\bibfnamefont {E.}~\bibnamefont
  {Knill}}, \bibinfo {author} {\bibfnamefont {D.}~\bibnamefont {Leibfried}},
  \bibinfo {author} {\bibfnamefont {R.}~\bibnamefont {Reichle}}, \bibinfo
  {author} {\bibfnamefont {J.}~\bibnamefont {Britton}}, \bibinfo {author}
  {\bibfnamefont {R.~B.}\ \bibnamefont {Blakestad}}, \bibinfo {author}
  {\bibfnamefont {J.~D.}\ \bibnamefont {Jost}}, \bibinfo {author}
  {\bibfnamefont {C.}~\bibnamefont {Langer}}, \bibinfo {author} {\bibfnamefont
  {R.}~\bibnamefont {Ozeri}}, \bibinfo {author} {\bibfnamefont
  {S.}~\bibnamefont {Seidelin}}, \ and\ \bibinfo {author} {\bibfnamefont
  {D.~J.}\ \bibnamefont {Wineland}},\ }\href@noop {} {\bibfield  {journal}
  {\bibinfo  {journal} {Phys. Rev. A}\ }\textbf {\bibinfo {volume} {77}},\
  \bibinfo {pages} {012307} (\bibinfo {year} {2008})}\BibitemShut {NoStop}%
\bibitem [{\citenamefont {Magesan}\ \emph {et~al.}(2011)\citenamefont
  {Magesan}, \citenamefont {Gambetta},\ and\ \citenamefont
  {Emerson}}]{Magesan11}%
  \BibitemOpen
  \bibfield  {author} {\bibinfo {author} {\bibfnamefont {E.}~\bibnamefont
  {Magesan}}, \bibinfo {author} {\bibfnamefont {J.~M.}\ \bibnamefont
  {Gambetta}}, \ and\ \bibinfo {author} {\bibfnamefont {J.}~\bibnamefont
  {Emerson}},\ }\href@noop {} {\bibfield  {journal} {\bibinfo  {journal} {Phys.
  Rev. Lett.}\ }\textbf {\bibinfo {volume} {106}},\ \bibinfo {pages} {180504}
  (\bibinfo {year} {2011})}\BibitemShut {NoStop}%
\bibitem [{\citenamefont {Magesan}\ \emph
  {et~al.}(2012{\natexlab{a}})\citenamefont {Magesan}, \citenamefont
  {Gambetta},\ and\ \citenamefont {Emerson}}]{Magesan12}%
  \BibitemOpen
  \bibfield  {author} {\bibinfo {author} {\bibfnamefont {E.}~\bibnamefont
  {Magesan}}, \bibinfo {author} {\bibfnamefont {J.~M.}\ \bibnamefont
  {Gambetta}}, \ and\ \bibinfo {author} {\bibfnamefont {J.}~\bibnamefont
  {Emerson}},\ }\href@noop {} {\bibfield  {journal} {\bibinfo  {journal} {Phys.
  Rev. A}\ }\textbf {\bibinfo {volume} {85}},\ \bibinfo {pages} {042311}
  (\bibinfo {year} {2012}{\natexlab{a}})}\BibitemShut {NoStop}%
\bibitem [{\citenamefont {Johnson}\ \emph {et~al.}(2015)\citenamefont
  {Johnson}, \citenamefont {da~Silva}, \citenamefont {Ryan}, \citenamefont
  {Kimmel}, \citenamefont {Chow},\ and\ \citenamefont {Ohki}}]{Johnson15}%
  \BibitemOpen
  \bibfield  {author} {\bibinfo {author} {\bibfnamefont {B.~R.}\ \bibnamefont
  {Johnson}}, \bibinfo {author} {\bibfnamefont {M.~P.}\ \bibnamefont
  {da~Silva}}, \bibinfo {author} {\bibfnamefont {C.~A.}\ \bibnamefont {Ryan}},
  \bibinfo {author} {\bibfnamefont {S.}~\bibnamefont {Kimmel}}, \bibinfo
  {author} {\bibfnamefont {J.~M.}\ \bibnamefont {Chow}}, \ and\ \bibinfo
  {author} {\bibfnamefont {T.~A.}\ \bibnamefont {Ohki}},\ }\href@noop {}
  {\bibfield  {journal} {\bibinfo  {journal} {arXiv:1505.06686}\ } (\bibinfo
  {year} {2015})}\BibitemShut {NoStop}%
\bibitem [{\citenamefont {Rist\`e}\ \emph {et~al.}(2012)\citenamefont
  {Rist\`e}, \citenamefont {Bultink}, \citenamefont {Lehnert},\ and\
  \citenamefont {DiCarlo}}]{Riste12b}%
  \BibitemOpen
  \bibfield  {author} {\bibinfo {author} {\bibfnamefont {D.}~\bibnamefont
  {Rist\`e}}, \bibinfo {author} {\bibfnamefont {C.~C.}\ \bibnamefont
  {Bultink}}, \bibinfo {author} {\bibfnamefont {K.~W.}\ \bibnamefont
  {Lehnert}}, \ and\ \bibinfo {author} {\bibfnamefont {L.}~\bibnamefont
  {DiCarlo}},\ }\href@noop {} {\bibfield  {journal} {\bibinfo  {journal} {Phys.
  Rev. Lett.}\ }\textbf {\bibinfo {volume} {109}},\ \bibinfo {pages} {240502}
  (\bibinfo {year} {2012})}\BibitemShut {NoStop}%
\bibitem [{\citenamefont {Raussendorf}\ and\ \citenamefont
  {Harrington}(2007)}]{Raussendorf07}%
  \BibitemOpen
  \bibfield  {author} {\bibinfo {author} {\bibfnamefont {R.}~\bibnamefont
  {Raussendorf}}\ and\ \bibinfo {author} {\bibfnamefont {J.}~\bibnamefont
  {Harrington}},\ }\href@noop {} {\bibfield  {journal} {\bibinfo  {journal}
  {Phys. Rev. Lett.}\ }\textbf {\bibinfo {volume} {98}},\ \bibinfo {pages}
  {190504} (\bibinfo {year} {2007})}\BibitemShut {NoStop}%
\bibitem [{\citenamefont {Fowler}\ \emph {et~al.}(2009)\citenamefont {Fowler},
  \citenamefont {Stephens},\ and\ \citenamefont {Groszkowski}}]{Fowler09}%
  \BibitemOpen
  \bibfield  {author} {\bibinfo {author} {\bibfnamefont {A.~G.}\ \bibnamefont
  {Fowler}}, \bibinfo {author} {\bibfnamefont {A.~M.}\ \bibnamefont
  {Stephens}}, \ and\ \bibinfo {author} {\bibfnamefont {P.}~\bibnamefont
  {Groszkowski}},\ }\href@noop {} {\bibfield  {journal} {\bibinfo  {journal}
  {Phys. Rev. A}\ }\textbf {\bibinfo {volume} {80}},\ \bibinfo {pages} {052312}
  (\bibinfo {year} {2009})}\BibitemShut {NoStop}%
\bibitem [{\citenamefont {Wang}\ \emph {et~al.}(2011)\citenamefont {Wang},
  \citenamefont {Fowler},\ and\ \citenamefont {Hollenberg}}]{Wang11}%
  \BibitemOpen
  \bibfield  {author} {\bibinfo {author} {\bibfnamefont {D.~S.}\ \bibnamefont
  {Wang}}, \bibinfo {author} {\bibfnamefont {A.~G.}\ \bibnamefont {Fowler}}, \
  and\ \bibinfo {author} {\bibfnamefont {L.~C.~L.}\ \bibnamefont
  {Hollenberg}},\ }\href@noop {} {\bibfield  {journal} {\bibinfo  {journal}
  {Phys. Rev. A}\ }\textbf {\bibinfo {volume} {83}},\ \bibinfo {pages} {020302}
  (\bibinfo {year} {2011})}\BibitemShut {NoStop}%
\bibitem [{Mag()}]{MagesanNote}%
  \BibitemOpen
  \href@noop {} {}\bibinfo {howpublished} {E.~Magesan, private
  communication}\BibitemShut {NoStop}%
\bibitem [{\citenamefont {Magesan}\ \emph
  {et~al.}(2012{\natexlab{b}})\citenamefont {Magesan}, \citenamefont
  {Gambetta}, \citenamefont {Johnson}, \citenamefont {Ryan}, \citenamefont
  {Chow}, \citenamefont {Merkel}, \citenamefont {da~Silva}, \citenamefont
  {Keefe}, \citenamefont {Rothwell}, \citenamefont {Ohki}, \citenamefont
  {Ketchen},\ and\ \citenamefont {Steffen}}]{Magesan12b}%
  \BibitemOpen
  \bibfield  {author} {\bibinfo {author} {\bibfnamefont {E.}~\bibnamefont
  {Magesan}}, \bibinfo {author} {\bibfnamefont {J.~M.}\ \bibnamefont
  {Gambetta}}, \bibinfo {author} {\bibfnamefont {B.~R.}\ \bibnamefont
  {Johnson}}, \bibinfo {author} {\bibfnamefont {C.~A.}\ \bibnamefont {Ryan}},
  \bibinfo {author} {\bibfnamefont {J.~M.}\ \bibnamefont {Chow}}, \bibinfo
  {author} {\bibfnamefont {S.~T.}\ \bibnamefont {Merkel}}, \bibinfo {author}
  {\bibfnamefont {M.~P.}\ \bibnamefont {da~Silva}}, \bibinfo {author}
  {\bibfnamefont {G.~A.}\ \bibnamefont {Keefe}}, \bibinfo {author}
  {\bibfnamefont {M.~B.}\ \bibnamefont {Rothwell}}, \bibinfo {author}
  {\bibfnamefont {T.~A.}\ \bibnamefont {Ohki}}, \bibinfo {author}
  {\bibfnamefont {M.~B.}\ \bibnamefont {Ketchen}}, \ and\ \bibinfo {author}
  {\bibfnamefont {M.}~\bibnamefont {Steffen}},\ }\href@noop {} {\bibfield
  {journal} {\bibinfo  {journal} {Phys. Rev. Lett.}\ }\textbf {\bibinfo
  {volume} {109}},\ \bibinfo {pages} {080505} (\bibinfo {year}
  {2012}{\natexlab{b}})}\BibitemShut {NoStop}%
\end{thebibliography}%


\begin{thebibliography}{13}%
\makeatletter
\providecommand \@ifxundefined [1]{%
 \@ifx{#1\undefined}
}%
\providecommand \@ifnum [1]{%
 \ifnum #1\expandafter \@firstoftwo
 \else \expandafter \@secondoftwo
 \fi
}%
\providecommand \@ifx [1]{%
 \ifx #1\expandafter \@firstoftwo
 \else \expandafter \@secondoftwo
 \fi
}%
\providecommand \natexlab [1]{#1}%
\providecommand \enquote  [1]{``#1''}%
\providecommand \bibnamefont  [1]{#1}%
\providecommand \bibfnamefont [1]{#1}%
\providecommand \citenamefont [1]{#1}%
\providecommand \href@noop [0]{\@secondoftwo}%
\providecommand \href [0]{\begingroup \@sanitize@url \@href}%
\providecommand \@href[1]{\@@startlink{#1}\@@href}%
\providecommand \@@href[1]{\endgroup#1\@@endlink}%
\providecommand \@sanitize@url [0]{\catcode `\\12\catcode `\$12\catcode
  `\&12\catcode `\#12\catcode `\^12\catcode `\_12\catcode `\%12\relax}%
\providecommand \@@startlink[1]{}%
\providecommand \@@endlink[0]{}%
\providecommand \url  [0]{\begingroup\@sanitize@url \@url }%
\providecommand \@url [1]{\endgroup\@href {#1}{\urlprefix }}%
\providecommand \urlprefix  [0]{URL }%
\providecommand \Eprint [0]{\href }%
\providecommand \doibase [0]{http://dx.doi.org/}%
\providecommand \selectlanguage [0]{\@gobble}%
\providecommand \bibinfo  [0]{\@secondoftwo}%
\providecommand \bibfield  [0]{\@secondoftwo}%
\providecommand \translation [1]{[#1]}%
\providecommand \BibitemOpen [0]{}%
\providecommand \bibitemStop [0]{}%
\providecommand \bibitemNoStop [0]{.\EOS\space}%
\providecommand \EOS [0]{\spacefactor3000\relax}%
\providecommand \BibitemShut  [1]{\csname bibitem#1\endcsname}%
\let\auto@bib@innerbib\@empty
%</preamble>
\bibitem [{\citenamefont {Rist\`{e}}\ \emph {et~al.}(2015)\citenamefont
  {Rist\`{e}}, \citenamefont {Poletto}, \citenamefont {Huang}, \citenamefont
  {Bruno}, \citenamefont {Vesterinen}, \citenamefont {Saira},\ and\
  \citenamefont {DiCarlo}}]{Riste15}%
  \BibitemOpen
  \bibfield  {author} {\bibinfo {author} {\bibfnamefont {D.}~\bibnamefont
  {Rist\`{e}}}, \bibinfo {author} {\bibfnamefont {S.}~\bibnamefont {Poletto}},
  \bibinfo {author} {\bibfnamefont {M.~Z.}\ \bibnamefont {Huang}}, \bibinfo
  {author} {\bibfnamefont {A.}~\bibnamefont {Bruno}}, \bibinfo {author}
  {\bibfnamefont {V.}~\bibnamefont {Vesterinen}}, \bibinfo {author}
  {\bibfnamefont {O.~P.}\ \bibnamefont {Saira}}, \ and\ \bibinfo {author}
  {\bibfnamefont {L.}~\bibnamefont {DiCarlo}},\ }\href@noop {} {\bibfield
  {journal} {\bibinfo  {journal} {Nat.\ Commun.}\ }\textbf {\bibinfo {volume}
  {{6}}} (\bibinfo {year} {{2015}})}\BibitemShut {NoStop}%
\bibitem [{\citenamefont {Bruno}\ \emph {et~al.}(2015)\citenamefont {Bruno},
  \citenamefont {de~Lange}, \citenamefont {Asaad}, \citenamefont {van~der
  Enden}, \citenamefont {Langford},\ and\ \citenamefont {DiCarlo}}]{Bruno15}%
  \BibitemOpen
  \bibfield  {author} {\bibinfo {author} {\bibfnamefont {A.}~\bibnamefont
  {Bruno}}, \bibinfo {author} {\bibfnamefont {G.}~\bibnamefont {de~Lange}},
  \bibinfo {author} {\bibfnamefont {S.}~\bibnamefont {Asaad}}, \bibinfo
  {author} {\bibfnamefont {K.~L.}\ \bibnamefont {van~der Enden}}, \bibinfo
  {author} {\bibfnamefont {N.~K.}\ \bibnamefont {Langford}}, \ and\ \bibinfo
  {author} {\bibfnamefont {L.}~\bibnamefont {DiCarlo}},\ }\href@noop {}
  {\bibfield  {journal} {\bibinfo  {journal} {Appl. Phys. Lett.}\ }\textbf
  {\bibinfo {volume} {106}},\ \bibinfo {pages} {182601} (\bibinfo {year}
  {2015})}\BibitemShut {NoStop}%
\bibitem [{\citenamefont {Schreier}\ \emph {et~al.}(2008)\citenamefont
  {Schreier}, \citenamefont {Houck}, \citenamefont {Koch}, \citenamefont
  {Schuster}, \citenamefont {Johnson}, \citenamefont {Chow}, \citenamefont
  {Gambetta}, \citenamefont {Majer}, \citenamefont {Frunzio}, \citenamefont
  {Devoret}, \citenamefont {Girvin},\ and\ \citenamefont
  {Schoelkopf}}]{Schreier08}%
  \BibitemOpen
  \bibfield  {author} {\bibinfo {author} {\bibfnamefont {J.~A.}\ \bibnamefont
  {Schreier}}, \bibinfo {author} {\bibfnamefont {A.~A.}\ \bibnamefont {Houck}},
  \bibinfo {author} {\bibfnamefont {J.}~\bibnamefont {Koch}}, \bibinfo {author}
  {\bibfnamefont {D.~I.}\ \bibnamefont {Schuster}}, \bibinfo {author}
  {\bibfnamefont {B.~R.}\ \bibnamefont {Johnson}}, \bibinfo {author}
  {\bibfnamefont {J.~M.}\ \bibnamefont {Chow}}, \bibinfo {author}
  {\bibfnamefont {J.~M.}\ \bibnamefont {Gambetta}}, \bibinfo {author}
  {\bibfnamefont {J.}~\bibnamefont {Majer}}, \bibinfo {author} {\bibfnamefont
  {L.}~\bibnamefont {Frunzio}}, \bibinfo {author} {\bibfnamefont {M.~H.}\
  \bibnamefont {Devoret}}, \bibinfo {author} {\bibfnamefont {S.~M.}\
  \bibnamefont {Girvin}}, \ and\ \bibinfo {author} {\bibfnamefont {R.~J.}\
  \bibnamefont {Schoelkopf}},\ }\href@noop {} {\bibfield  {journal} {\bibinfo
  {journal} {Phys. Rev. B}\ }\textbf {\bibinfo {volume} {77}},\ \bibinfo
  {pages} {180502} (\bibinfo {year} {2008})}\BibitemShut {NoStop}%
\bibitem [{\citenamefont {Barends}\ \emph {et~al.}(2011)\citenamefont
  {Barends}, \citenamefont {Wenner}, \citenamefont {Lenander}, \citenamefont
  {Chen}, \citenamefont {Bialczak}, \citenamefont {Kelly}, \citenamefont
  {Lucero}, \citenamefont {O'Malley}, \citenamefont {Mariantoni}, \citenamefont
  {Sank}, \citenamefont {Wang}, \citenamefont {White}, \citenamefont {Yin},
  \citenamefont {Zhao}, \citenamefont {Cleland}, \citenamefont {Martinis},\
  and\ \citenamefont {Baselmans}}]{Barends11}%
  \BibitemOpen
  \bibfield  {author} {\bibinfo {author} {\bibfnamefont {R.}~\bibnamefont
  {Barends}}, \bibinfo {author} {\bibfnamefont {J.}~\bibnamefont {Wenner}},
  \bibinfo {author} {\bibfnamefont {M.}~\bibnamefont {Lenander}}, \bibinfo
  {author} {\bibfnamefont {Y.}~\bibnamefont {Chen}}, \bibinfo {author}
  {\bibfnamefont {R.~C.}\ \bibnamefont {Bialczak}}, \bibinfo {author}
  {\bibfnamefont {J.}~\bibnamefont {Kelly}}, \bibinfo {author} {\bibfnamefont
  {E.}~\bibnamefont {Lucero}}, \bibinfo {author} {\bibfnamefont
  {P.}~\bibnamefont {O'Malley}}, \bibinfo {author} {\bibfnamefont
  {M.}~\bibnamefont {Mariantoni}}, \bibinfo {author} {\bibfnamefont
  {D.}~\bibnamefont {Sank}}, \bibinfo {author} {\bibfnamefont {H.}~\bibnamefont
  {Wang}}, \bibinfo {author} {\bibfnamefont {T.~C.}\ \bibnamefont {White}},
  \bibinfo {author} {\bibfnamefont {Y.}~\bibnamefont {Yin}}, \bibinfo {author}
  {\bibfnamefont {J.}~\bibnamefont {Zhao}}, \bibinfo {author} {\bibfnamefont
  {A.~N.}\ \bibnamefont {Cleland}}, \bibinfo {author} {\bibfnamefont {J.~M.}\
  \bibnamefont {Martinis}}, \ and\ \bibinfo {author} {\bibfnamefont {J.~J.~A.}\
  \bibnamefont {Baselmans}},\ }\href@noop {} {\bibfield  {journal} {\bibinfo
  {journal} {Appl. Phys. Lett.}\ }\textbf {\bibinfo {volume} {99}},\ \bibinfo
  {pages} {113507} (\bibinfo {year} {2011})}\BibitemShut {NoStop}%
\bibitem [{\citenamefont {Reed}(2013)}]{ReedPhD13}%
  \BibitemOpen
  \bibfield  {author} {\bibinfo {author} {\bibfnamefont {M.}~\bibnamefont
  {Reed}},\ }\emph {\bibinfo {title} {Entanglement and quantum error correction
  with superconducting qubits}},\ \href@noop {} {\bibinfo {type} {Ph{D}
  {D}issertation}},\ \bibinfo  {school} {Yale University} (\bibinfo {year}
  {2013})\BibitemShut {NoStop}%
\bibitem [{\citenamefont {Motzoi}\ \emph {et~al.}(2009)\citenamefont {Motzoi},
  \citenamefont {Gambetta}, \citenamefont {Rebentrost},\ and\ \citenamefont
  {Wilhelm}}]{Motzoi09}%
  \BibitemOpen
  \bibfield  {author} {\bibinfo {author} {\bibfnamefont {F.}~\bibnamefont
  {Motzoi}}, \bibinfo {author} {\bibfnamefont {J.~M.}\ \bibnamefont
  {Gambetta}}, \bibinfo {author} {\bibfnamefont {P.}~\bibnamefont
  {Rebentrost}}, \ and\ \bibinfo {author} {\bibfnamefont {F.~K.}\ \bibnamefont
  {Wilhelm}},\ }\href@noop {} {\bibfield  {journal} {\bibinfo  {journal} {Phys.
  Rev. Lett.}\ }\textbf {\bibinfo {volume} {103}},\ \bibinfo {pages} {110501}
  (\bibinfo {year} {2009})}\BibitemShut {NoStop}%
\bibitem [{\citenamefont {Chow}\ \emph {et~al.}(2010)\citenamefont {Chow},
  \citenamefont {DiCarlo}, \citenamefont {Gambetta}, \citenamefont {Motzoi},
  \citenamefont {Frunzio}, \citenamefont {Girvin},\ and\ \citenamefont
  {Schoelkopf}}]{Chow10b}%
  \BibitemOpen
  \bibfield  {author} {\bibinfo {author} {\bibfnamefont {J.~M.}\ \bibnamefont
  {Chow}}, \bibinfo {author} {\bibfnamefont {L.}~\bibnamefont {DiCarlo}},
  \bibinfo {author} {\bibfnamefont {J.~M.}\ \bibnamefont {Gambetta}}, \bibinfo
  {author} {\bibfnamefont {F.}~\bibnamefont {Motzoi}}, \bibinfo {author}
  {\bibfnamefont {L.}~\bibnamefont {Frunzio}}, \bibinfo {author} {\bibfnamefont
  {S.~M.}\ \bibnamefont {Girvin}}, \ and\ \bibinfo {author} {\bibfnamefont
  {R.~J.}\ \bibnamefont {Schoelkopf}},\ }\href@noop {} {\bibfield  {journal}
  {\bibinfo  {journal} {Phys. Rev. A}\ }\textbf {\bibinfo {volume} {82}},\
  \bibinfo {pages} {040305} (\bibinfo {year} {2010})}\BibitemShut {NoStop}%
\bibitem [{\citenamefont {Epstein}\ \emph {et~al.}(2014)\citenamefont
  {Epstein}, \citenamefont {Cross}, \citenamefont {Magesan},\ and\
  \citenamefont {Gambetta}}]{Epstein14}%
  \BibitemOpen
  \bibfield  {author} {\bibinfo {author} {\bibfnamefont {J.~M.}\ \bibnamefont
  {Epstein}}, \bibinfo {author} {\bibfnamefont {A.~W.}\ \bibnamefont {Cross}},
  \bibinfo {author} {\bibfnamefont {E.}~\bibnamefont {Magesan}}, \ and\
  \bibinfo {author} {\bibfnamefont {J.~M.}\ \bibnamefont {Gambetta}},\
  }\href@noop {} {\bibfield  {journal} {\bibinfo  {journal} {Phys. Rev. A}\
  }\textbf {\bibinfo {volume} {89}},\ \bibinfo {pages} {062321} (\bibinfo
  {year} {2014})}\BibitemShut {NoStop}%
\bibitem [{\citenamefont {Blais}\ \emph {et~al.}(2007)\citenamefont {Blais},
  \citenamefont {Gambetta}, \citenamefont {Wallraff}, \citenamefont {Schuster},
  \citenamefont {Girvin}, \citenamefont {Devoret},\ and\ \citenamefont
  {Schoelkopf}}]{Blais07}%
  \BibitemOpen
  \bibfield  {author} {\bibinfo {author} {\bibfnamefont {A.}~\bibnamefont
  {Blais}}, \bibinfo {author} {\bibfnamefont {J.}~\bibnamefont {Gambetta}},
  \bibinfo {author} {\bibfnamefont {A.}~\bibnamefont {Wallraff}}, \bibinfo
  {author} {\bibfnamefont {D.~I.}\ \bibnamefont {Schuster}}, \bibinfo {author}
  {\bibfnamefont {S.~M.}\ \bibnamefont {Girvin}}, \bibinfo {author}
  {\bibfnamefont {M.~H.}\ \bibnamefont {Devoret}}, \ and\ \bibinfo {author}
  {\bibfnamefont {R.~J.}\ \bibnamefont {Schoelkopf}},\ }\href@noop {}
  {\bibfield  {journal} {\bibinfo  {journal} {Phys. Rev. A}\ }\textbf {\bibinfo
  {volume} {75}},\ \bibinfo {pages} {032329} (\bibinfo {year}
  {2007})}\BibitemShut {NoStop}%
\bibitem [{\citenamefont {Majer}\ \emph {et~al.}(2007)\citenamefont {Majer},
  \citenamefont {Chow}, \citenamefont {Gambetta}, \citenamefont {Johnson},
  \citenamefont {Schreier}, \citenamefont {Frunzio}, \citenamefont {Schuster},
  \citenamefont {Houck}, \citenamefont {Wallraff}, \citenamefont {Blais},
  \citenamefont {Devoret}, \citenamefont {Girvin},\ and\ \citenamefont
  {Schoelkopf}}]{Majer07}%
  \BibitemOpen
  \bibfield  {author} {\bibinfo {author} {\bibfnamefont {J.}~\bibnamefont
  {Majer}}, \bibinfo {author} {\bibfnamefont {J.~M.}\ \bibnamefont {Chow}},
  \bibinfo {author} {\bibfnamefont {J.~M.}\ \bibnamefont {Gambetta}}, \bibinfo
  {author} {\bibfnamefont {B.~R.}\ \bibnamefont {Johnson}}, \bibinfo {author}
  {\bibfnamefont {J.~A.}\ \bibnamefont {Schreier}}, \bibinfo {author}
  {\bibfnamefont {L.}~\bibnamefont {Frunzio}}, \bibinfo {author} {\bibfnamefont
  {D.~I.}\ \bibnamefont {Schuster}}, \bibinfo {author} {\bibfnamefont {A.~A.}\
  \bibnamefont {Houck}}, \bibinfo {author} {\bibfnamefont {A.}~\bibnamefont
  {Wallraff}}, \bibinfo {author} {\bibfnamefont {A.}~\bibnamefont {Blais}},
  \bibinfo {author} {\bibfnamefont {M.~H.}\ \bibnamefont {Devoret}}, \bibinfo
  {author} {\bibfnamefont {S.~M.}\ \bibnamefont {Girvin}}, \ and\ \bibinfo
  {author} {\bibfnamefont {R.~J.}\ \bibnamefont {Schoelkopf}},\ }\href@noop {}
  {\bibfield  {journal} {\bibinfo  {journal} {Nature}\ }\textbf {\bibinfo
  {volume} {449}},\ \bibinfo {pages} {443} (\bibinfo {year}
  {2007})}\BibitemShut {NoStop}%
\bibitem [{\citenamefont {Johansson}\ \emph {et~al.}(2012)\citenamefont
  {Johansson}, \citenamefont {Nation},\ and\ \citenamefont
  {Nori}}]{Johansson12}%
  \BibitemOpen
  \bibfield  {author} {\bibinfo {author} {\bibfnamefont {J.}~\bibnamefont
  {Johansson}}, \bibinfo {author} {\bibfnamefont {P.}~\bibnamefont {Nation}}, \
  and\ \bibinfo {author} {\bibfnamefont {F.}~\bibnamefont {Nori}},\ }\href@noop
  {} {\bibfield  {journal} {\bibinfo  {journal} {Computer Physics
  Communications}\ }\textbf {\bibinfo {volume} {183}},\ \bibinfo {pages} {1760
  } (\bibinfo {year} {2012})}\BibitemShut {NoStop}%
\bibitem [{\citenamefont {Johansson}\ \emph {et~al.}(2013)\citenamefont
  {Johansson}, \citenamefont {Nation},\ and\ \citenamefont
  {Nori}}]{Johansson13}%
  \BibitemOpen
  \bibfield  {author} {\bibinfo {author} {\bibfnamefont {J.}~\bibnamefont
  {Johansson}}, \bibinfo {author} {\bibfnamefont {P.}~\bibnamefont {Nation}}, \
  and\ \bibinfo {author} {\bibfnamefont {F.}~\bibnamefont {Nori}},\ }\href@noop
  {} {\bibfield  {journal} {\bibinfo  {journal} {Computer Physics
  Communications}\ }\textbf {\bibinfo {volume} {184}},\ \bibinfo {pages} {1234
  } (\bibinfo {year} {2013})}\BibitemShut {NoStop}%
\bibitem [{\citenamefont {Magesan}\ \emph {et~al.}(2012)\citenamefont
  {Magesan}, \citenamefont {Gambetta}, \citenamefont {Johnson}, \citenamefont
  {Ryan}, \citenamefont {Chow}, \citenamefont {Merkel}, \citenamefont
  {da~Silva}, \citenamefont {Keefe}, \citenamefont {Rothwell}, \citenamefont
  {Ohki}, \citenamefont {Ketchen},\ and\ \citenamefont {Steffen}}]{Magesan12b}%
  \BibitemOpen
  \bibfield  {author} {\bibinfo {author} {\bibfnamefont {E.}~\bibnamefont
  {Magesan}}, \bibinfo {author} {\bibfnamefont {J.~M.}\ \bibnamefont
  {Gambetta}}, \bibinfo {author} {\bibfnamefont {B.~R.}\ \bibnamefont
  {Johnson}}, \bibinfo {author} {\bibfnamefont {C.~A.}\ \bibnamefont {Ryan}},
  \bibinfo {author} {\bibfnamefont {J.~M.}\ \bibnamefont {Chow}}, \bibinfo
  {author} {\bibfnamefont {S.~T.}\ \bibnamefont {Merkel}}, \bibinfo {author}
  {\bibfnamefont {M.~P.}\ \bibnamefont {da~Silva}}, \bibinfo {author}
  {\bibfnamefont {G.~A.}\ \bibnamefont {Keefe}}, \bibinfo {author}
  {\bibfnamefont {M.~B.}\ \bibnamefont {Rothwell}}, \bibinfo {author}
  {\bibfnamefont {T.~A.}\ \bibnamefont {Ohki}}, \bibinfo {author}
  {\bibfnamefont {M.~B.}\ \bibnamefont {Ketchen}}, \ and\ \bibinfo {author}
  {\bibfnamefont {M.}~\bibnamefont {Steffen}},\ }\href@noop {} {\bibfield
  {journal} {\bibinfo  {journal} {Phys. Rev. Lett.}\ }\textbf {\bibinfo
  {volume} {109}},\ \bibinfo {pages} {080505} (\bibinfo {year}
  {2012})}\BibitemShut {NoStop}%
\end{thebibliography}%
\end{document}